\documentclass[useAMS]{mn2e}
\usepackage{longtable}
\usepackage{graphicx}
\usepackage{epsfig}
\usepackage{rotating}
\usepackage{lscape}

\def\lsim{\,\lower2truept\hbox{${<\atop\hbox{\raise4truept\hbox{$\sim$}}}$}\,}
\def\gsim{\,\lower2truept\hbox{${> \atop\hbox{\raise4truept\hbox{$\sim$}}}$}\,}

\title[Photo-$z$ accuracy in {\it AKARI} Deep Surveys]{Photometric redshift accuracy in {\it AKARI} Deep Surveys}
\author[M. Negrello et al. ]{
\parbox[t]{\textwidth}{
M. Negrello$^{1}$\thanks{E-mail: M.Negrello@open.ac.uk},
S. Serjeant$^{1}$, 
C. Pearson$^{2,3}$\thanks{visiting Research Fellow at the Open University}, 
T. Takagi$^{4}$, 
A. Efstathiou$^{5}$, 
T. Goto$^{4}$,
D. Burgarella$^{6}$,
W.-S. Jeong$^{4,7}$, 
M. Im$^{8,9}$, 
H. M. Lee$^{8}$, 
H. Matsuhara$^{4}$, 
S. Oyabu$^{4}$, 
T. Wada$^{4}$, 
G. White$^{1}$
} \\ \\
$^{1}$Department of Physics and Astronomy, Open University, Walton
Hall, Milton Keynes MK7 6AA, United Kingdom \\
$^{2}$Space Science and Technology Department, CCLRC Rutherford Appleton
Laboratory, Chilton, Didcot, Oxfordshire OX11 0QX, United Kingdom \\
$^{3}$Department of Physics, University of Lethbridge, 4401 University Drive,
Lethbridge, Alberta T1J 1B1, Canada \\
$^{4}$Institute of Space and Astronautical Science, Japan Aerospace Exploration Agency, Sagamihara, Kanagawa 229-8510, Japan \\
$^{5}$Department of Computer Science and Engineering, Cyprus College, 6 Diogenes Str, 1516 Nicosia, Cyprus \\
$^{6}$Observatoire Astronomique Marseille Provence, Laboratoire d’Astrophysique de Marseille, 13376 Marseille Cedex 12, France \\
$^{7}$Space Science Division, Korea Astronomy $\&$ Space Science Institute (KASI), 61-1, Whaam-dong, Yuseong-gu, Deajeon, 305-348, South Korea \\
$^{8}$Department of Physics and Astronomy, FPRD, Seoul National University, Seoul 151-742, Korea \\ 
$^{9}$Infrared Processing and Analysis Center, California Institute of Technology, Pasadena, CA91125, USA
}

\begin{document}

\date{....}

\pagerange{\pageref{firstpage}--\pageref{lastpage}} \pubyear{2008}

\input BoxedEPS.tex \SetEPSFDirectory{./}

\SetRokickiEPSFSpecial

\maketitle

\label{firstpage}

\begin{abstract}
  We investigate the photometric redshift accuracy achievable with the
  {\it AKARI} infrared data in deep multi-band surveys, such as in the
  North Ecliptic Pole field. We demonstrate that the passage of
  redshifted policyclic aromatic hydrocarbons and silicate features
  into the mid-infrared wavelength window covered by {\it AKARI} is a
  valuable means to recover the redshifts of starburst galaxies. To
  this end we have collected a sample of $\sim$60 galaxies drawn from
  the GOODS-North Field with spectroscopic redshift $0.5\lsim z_{\rm
    spec}\lsim1.5$ and photometry from 3.6 to 24 \,$\mu$m, provided by
  the {\it Spitzer}, {\it ISO} and {\it AKARI} satellites. The
  infrared spectra are fitted using synthetic galaxy Spectral Energy
  Distributions which account for starburst and active nuclei
  emission. For $\sim90\%$ of the sources in our sample the redshift
  is recovered with an accuracy $|z_{\rm phot}-z_{\rm spec}|/(1+z_{\rm
    spec})\lsim10\%$. A similar analysis performed on different sets of
  simulated spectra shows that the {\it AKARI} infrared data alone can
  provide photometric redshifts accurate to $|z_{\rm phot}-z_{\rm
    spec}|/(1+z_{\rm spec})\sim10\%$ (1$\,\sigma$) at $z\lsim2$.  At
  higher redshifts the PAH features are shifted outside the wavelength range covered by {\it
    AKARI} and the photo-z estimates rely on the
  less prominent 1.6$\,\mu$m stellar bump; the accuracy achievable in this case on $(1+z)$ is
  $\sim10-15\%$, provided that the AGN contribution to the
  infrared emission is subdominant.
  Our technique is no more prone to redshift aliasing than optical-uv
  photo-$z$, and it may be possible to reduce this aliasing further
  with the addition of submillimetre and/or radio data.
\end{abstract}


\section{Introduction}
\label{sec:intro}

The discovery that infrared luminous starbursting galaxies are
significant and possibly dominant contributors to the cosmic
starformation history of the Universe has had an enormous impact on
the understanding of galaxy evolution (Hughes et al. 1998; Lagache, Puget $\&$ Dole 2005). However, these galaxies are often too faint in the
optical for large optical redshift compaigns, or have ambiguous
optical identifications (Chapman et al. 2004, 2005). Attention therefore have been
focusing in the last years on infrared photometric redshift
estimators. Sawicki (2002) showed that the 1.6$\,\mu$m spectral
feature arising from the photospheric emission from evolved stars
could be used to obtain crude photometric redshift constraints with
the 3.6-8.0$\,\mu$m photometry from the IRAC instrument on {\it
  Spitzer}. Unfortunately, the 8-24$\,\mu$m wavelength gap in {\it
  Spitzer}'s wide-field survey capability prevented the use of longer
wavelength rest-frame features as redshift indicators, such as the much more prominent
Policyclic Aromatic Hydrocarbons (PAH) and $\sim10\,\mu$m silicate
absorption features. 

{\it AKARI} is a Japanese-led infrared space telescope, which was
launched successfully in Feb 2006. It has undertaken deep surveys near
the Ecliptic Poles, far deeper than its all-sky survey. The North
Ecliptic Pole (NEP) field is the largest deep-field legacy survey of
{\it AKARI}, covering $0.38$\,deg$^{2}$, and is its major deep-field
legacy (Wada et al. 2008). A wider and shallower survey at the NEP has
also been performed with {\it AKARI}, over a 5.8\,deg$^{2}$ area (Lee
et al. 2008), and further deep photometric surveys have been made of
well-studied {\it Spitzer} fields (Pearson et al. 2008).

{\it AKARI} observed in 9 bands from $\sim$2 up to
$\sim$24$\,\mu$m. The mid-infrared imaging spans the $8-24\,\mu$m gap
between {\it Spitzer}'s IRAC and MIPS instruments. The unique
diagnostic power of {\it AKARI}'s Spectral Energy Distribution (SED)
analysis has been confirmed by Takagi et al. (2007), who very
successfully traced several starburst PAH features and identified the
AGN dust torus excess over the starburst emission, obtaining at the
same time a good match between the photometric redshift derived from
HyperZ (Bolzonella et al. 2000) and the one obtained by fitting the
infrared data with reliable starburst models.

In the present work we investigate the photometric redshift accuracy
achievable with {\it AKARI} in deep multi-band surveys from infrared data alone.
Infrared spectra are fitted using reference SED templates derived from
the starburst model of Takagi, Arimoto $\&$ Hanami (2003, hereafter
TAH03; see also Takagi, Hanami $\&$ Arimoto 2004, hereafter THA04),
and already exploited in Takagi et al. (2007).  However in order to
deal with a possible excess over starburst emission due to an AGN
activity we have included a set of reference AGN spectra derived from
the model of Efstathiou $\&$ Rowan-Robinson (1995, hereafter ER95).
Our photometric redshift code combines the starburst and AGN
components to provide the best fit to observed or simulated spectra.
We fit the infrared SEDs of a sample of galaxies with spectroscopic
redshift drawn from the GOODS-North Field where deep AKARI observation
have been performed at 11 and 18$\,\mu$m, and of a set of simulated
spectra in the NEP Deep Field with {\it AKARI}. We aim to demonstrate
in this way the power of the PAH and silicate absorption feaures to
obtain reliable redshift estimates based on infrared photometry alone.

We start in Section~\ref{sec:ifr_models} describing the reference SED
templates.  Section~\ref{sec:data} presents the sample of infrared
sources drawn from the GOODS-N field and Section~\ref{sec:results}
provides our results on the comparison between the photometric and the
spectroscopic redshifts.  Section~\ref{sec:simulations} describes the
simulations and the results on the photometric redshift accuracy for
the {\it AKARI} NEP Deep field.  Conclusions are summarized in
Section~\ref{sec:conclusions}.

\section{Infrared SED templates}
\label{sec:ifr_models}

\subsection{Starburst component}

We adopt the StarBUrst with Radiative Transfer (SBURT) model of THA03
and TAH04 to deal with the infrared emission due to starburst
activity.

The model deals with the star formation and chemical evolution in the
starburst regions using the infall model of chemical evolution of
Arimoto et al. (1992). Under the assumption that the amount of gas in
the outflow from the starburst region has a negligible impact on the
chemical evolution as a whole, the starburst is characterized by the
rate of gas infall and that of the star-formation. However, for
simplicity, THA03 set the infall time-scale equal to star-formation
time-scale, $t_{0}$. The equations describing the time variation of
gas mass, total stellar mass and gas metallicity are solved
numerically using in input the unobscured stellar SEDs derived from
the evolutionary population synthesis code of Kodama $\&$ Arimoto
(1997). A top-heavy IMF is assumed with a slope $x=1.10$, flatter than
the Salpeter IMF ($x=1.35$). The initial metallicity of the gas cloud,
$Z_{i}$, is set to zero. According to TAH03, the chemical evolution as
a function of $t/t_{0}$ as well as the properties of the SED are
almost independent of the practical choice of $t_{0}$, specifically
when $t_{0}\gsim50$ Myr. Therefore, all the reference SED templates
used here for the starburst are calculated adopting $t_{0}=100$ Myr.

THA03 assumes a starburst region in which stars and dust are
distributed within a radius $R_{\rm t}$, and introduce the following
mass-radius relation
\begin{equation}
  \frac{R_{\rm t}}{1 \rm kpc}=\Theta\left(\frac{M_{\star}}{10^9 \rm M_{\odot}}\right)^{\gamma},
\end{equation}
where $\Theta$ is a compactness factor that expresss the matter
concentration, with the mean density being higher for smaller values
of $\Theta$, while $M_{\star}$ is the (time-dependent) total stellar
mass in the starburst region.  The exponent $\gamma$ is set to
$\frac{1}{2}$ which results in a constant surface brightness of the
straburst galaxy for constant $\Theta$. Dust is assumed to be
homogeneously distributed within $R_{\rm t}$ while a King profile is
adopted for the stellar density distribution. The optical depth of the
starburst regions is a function of the starburst age and of the
compactness factor, and increases for decreasing values of $\Theta$.

The amount of dust is given by the chemical evolution assuming a
constant dust-to-metal ratio, $\delta_{0}$, with three different dust
models, i.e. the model for dust in the Milky Way (MW), the Large
Magellanic Cloud (LMC) and the Small Magellanic Cloud (SMC). The
values of $\delta_{0}$ derived from the extinction curve and the
spectra of cirrus emission are 0.40, 0.55 and 0.75 for the MW, LMC and
SMC, respectively. The difference among the three extinction curves is
attributed to the variation of the ratio of carbonaceous dust
(graphite and PAHs) to silicate grain with the fraction of
  silicate dust grains increasing from MW to SMC model, i.e. as a
  function of metallicity; therefore the silicate absorption features
  are most prominent in the SEDs described by the SMC type dust
  model.  The SED from ultraviolet to submillimetre wavelengths of
starburst region is calculated for each starburst age, compactness
parameter and extinction curve using a radiative transfer code that
assumes isotropic multiple scattering and accounts for self-absorption
of re-emitted energy from dust.

The absorption/emission properties of the dust are responsible for the
infrared SED features of starforming galaxies: the PAHs emissions at
3.3, 6.2, 7.7, 8.6 and 11.3$\,\mu$m and the silicate resonances at 9.7
and 18$\,\mu$m. {\bf An example of starburst SED template drawn from
  the Takagi et al. model, showing the main infrared features, is
  presented in Fig.~\ref{fig:sed_vs_filters}.} The efficiency of these
features as redshift indicators is what we are going to put to the
test in the present work.

\setcounter{figure}{0}
\begin{figure}
  \hspace{-1.5cm}
  \includegraphics[height=8.2cm,width=11.0cm]
  {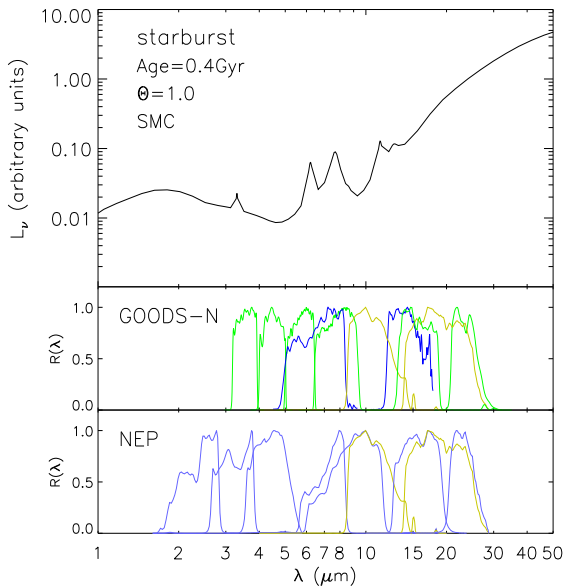} \vskip+0.2truecm\caption{Example of a
      starburst SED template (rest-frame) drawn from the Takagi et
      al. model, showing the main infrared features exploited as
      redshift indicators: PAH emissions at 3.3, 6.2, and 7.7$\,\mu$m,
      and silicate absorption at 9.7$\,\mu$m. The 1.6$\,\mu$m stellar
      bump is also visible. In the central and in the bottom panels
      are shown the transmission curves of the infrared filters
      exploited in the GOODS-N field and in the AKARI NEP deep field,
      respectively, normalized to a maximum value of one: {\it
        Spitzer} 3.6, 4.5, 5.8, 8.0, 16 and 24$\,\mu$m filters (
      green), {\it ISO} 6.5 and 15$\,\mu$m filters (blue), {\it AKARI}
      filters (the 11 and 18$\,\mu$m filters in yellow, the others in light blue).}
  \label{fig:sed_vs_filters}
\end{figure}

\subsection{AGN component}

The infrared spectrum of the AGN dusty torus is modelled as in
ER95. They combine an accurate solution for the axially symmetric
radiative transfer problem in dust clouds with the multigrain dust
model of Rowan-Robinson (1992) and three different geometries for the
dusty torus that surround the central supermassive black hole. Here we
assume their thick tapered disc model, which is a disc-like torus
whose height increases with the distance from the central source but
tapers to a constant height in the outer parts. This choice has been
found to provide the best agreement with the observational constraints
on AGN. The tapered disc is assumed to have a $r^{-1}$ density
distribution. The ratio between the outer and the inner radius of the
torus is set to 20, in the middle of the values considered by ER95,
while we assume a value of 45$^{\circ}$ for the half-opening angle of
the torus (the values assumed by ER95 for the nucleus of NGC
1068). The equatorial optical depth of the torus is fixed to 1000.
Therefore the SED of the AGN torus depends only on the viewing angle
of
the torus, $\theta_{view}$. \\
With the above choice of parameters the SED shows deep absorption
features at $\sim10\,\mu$m due to silicate dust for edge-on views of
the torus ($\theta_{view} = 0^{\circ}$), but it is rather featureless
when the torus is seen face-on ($\theta_{view} = 90^{\circ}$).

\subsection{Reference SED templates and fitting procedure}
\label{subsec:reference_templates}

Using the models described above we have created a set of reference
SED templates meant for fitting any set of infrared photometric data
(both real and simulated). These templates have been derived as
follows (see TAH04).
\begin{itemize}
\item {\sc STARBURST component}: for each type of extinction curve, we
  adopt 10 different starburst ages in the interval $t/t_{0}=0.1-6.0$
  (or equivalently $t=10-600$ Myr, being $t_{0}=100$ Myr), and 17
  different compactness factors in the range $\Theta=0.3-5$.
\item {\sc AGN component}: we vary the viewing angle between
  0$^{\circ}$ and 90$^{\circ}$, with 11 discrete values.
\end{itemize}
The resulting SED templates are convolved with the response functions
of the instruments for different values of the redshift in the range
$z=0-7$ and in steps of 0.02.  The fluxes derived for the two SED
components (i.e. starburst and AGN torus) are then linearly combined
to give the predicted flux at the passbands of the instruments. The
best fit parameters of the SED models, including the redshift, are
then obtained by minimizing the $\chi^2$
\begin{equation}
  \chi^{2} = \sum_{i=1}^{N_{\rm filters}}\left[\frac{F_{\rm obs}(i) - \sum_{k=1}^{2}f_{k}\times F_{k}(i)}{\sigma(i)}\right]^{2}, 
  \label{eq:chisq}
\end{equation}
where $F_{\rm obs}(i)$ is the measured flux in the passband $i$ and
$\sigma(i)$ is its corresponding uncertainty. $F_{k}(i)$ represents
the predicted flux in the passband $i$ contributed by the SED
component $k$, with $k=1$, 2 denoting ``Starburst'' and ``AGN'',
respectively. $f_{k}$ is therefore the relative contribution of the
$k$ component to the total bolometric luminosity of the galaxy. Note
that for each set of values of the SED model parameters, the values of
$f_{1}$ and $f_{2}$ are fixed by the condition of minimization with
the restriction that they must be all non-negative. When a negative
value is obtained for either $f_{1}$ or $f_{2}$, then that parameter
is set to zero and the process of minimization is performed on the
other SED component alone. We descard those sets of values of the SED
model parameters for which the bolometric luminosity\footnote{The
  bolometric luminosity is obtained by integrating the SED in the
  rest-frame wavelength interval 0.1-1000\,$\mu$m.} resulting from the
minimization over the parameters $f_{1}$ and $f_{2}$ lies above
$10^{14}$ L$_{\odot}$ or below 10$^{8}$ L$_{\odot}$.  Since we are
dealing with photometric data from different telescopes we have set a
minimum error of 10$\%$ on the measured fluxes before
performing the SED fitting. \\
The goodness of the fit is described by the probability for a
$\chi^{2}$-distribution, with the number of degree of freedom set by
the problem, to produce a value of $\chi^{2}$ higher than the one
obtained by the best SED-fit.  The number of degree of freedom is
given by $\nu=n_{\rm data}-n_{\rm par}$ where $n_{\rm data}$ is the
number of fitted data (ranging from 8 to 10 depending on the
availability of {\it ISO} and/or IRS photometry) while $n_{\rm par}$
is the number of parameters of the model, i.e. the redshift, the
normalization of the SED components and the SED model parameters. As a
result\footnote{The type of the extinction curve (i.e. MW, LMC, SMC)
  is not considered as a SED parameter and therefore it has not been
  included in the calculation of $n_{\rm par}$.} $n_{\rm par}=6$ if
both $f_{1}$ and $f_{2}$ are positive, otherwise $n_{\rm par}=4$
($f_{2}=0$) or $n_{\rm par}=3$ ($f_{1}=0$). A 99$\%$ confidence
interval on the estimated photometric redshift is derived by the
$\Delta\chi^{2}$ method (Avni 1976), being the number of ``interesting
parameters'' equal to 1, i.e. $z_{\rm phot}$. In this case the value
of $\Delta\chi^{2}$ defining the confidence interval at 99$\%$ is
6.63. A best SED-fit is considered ``good'' if $P_{\chi^{2}}>1\%$.

\section{Data sets}
\label{sec:data}

In order to show the reliability of the redshift estimates based on
the PAH and silicate infrared features, we have assembled a sample of
galaxies with flux measurements in the wavelength range 3.6-24
$\,\mu$m and with measured spectroscopic redshifts. The sample has
been selected within the northern field of the Great Observatories
Origins Deep Survey (GOODS, Dickinson et al. 2001) because of its
richness in multiwavelength photometric data and spectroscopic
redshifts. The GOODS observations covered two fields on the sky: a
northern target area (GOODS-N) coincident with (but significantly
larger than) the Hubble-Deep Field North (HDF-N, Williams et
al. 1996), and a similarly sized southern field (GOODS-S) coincident
with the Chandra Deep Field South (Giacconi et al. 2000). The GOODS-N
field has been imaged at infrared wavelengths by different
instruments: at 6.75$\,\mu$m and 15$\,\mu$m by {\it ISO}CAM on board
of the Infrared Space Observatory ({\it ISO}, Kessler et al. 1996), at
3.5, 4.5, 5.8, 8$\,\mu$ and 24 $\,\mu$m by {\it Spitzer}, and, more
recently, at 11 and 18$\,\mu$m by {\it AKARI} (Pearson et al. 2008).
Some subregions of the GOODS-N field (including part of the HDF-N)
were also imaged at 16 $\mu$m by the {\it Spitzer} Infra-Red
Spectrograph (IRS) blue peak-up filter (Teplitz et al. 2005). It is
these sets of infrared data we have exploited here and we provide
brief descriptions below.

\setcounter{table}{0}
\begin{table}
  \scriptsize
  \caption{Some information about the {\it Spitzer} and {\it AKARI} infrared catalogues used in 
    the present work: observing wavelength, Full With at Half Maximum (FWHM) of the instrument, 
    number of surces detected above 3$\sigma$, corresponding minimum source flux, and surveyed area.}
  \vspace{-0.5cm}
  \begin{center}
    \begin{tabular}{rrrrrr}
      \hline
      instrument  & waveband  & FWHM     & $\sharp$ sources  & S$_{min}$  & Area~~~~ \\ 
      & ($\mu$m)  & (arcsec) &                   & ($\mu$Jy)  & (arcmin$^{2}$) \\
      \hline
      {\it Spitzer}/IRAC &  3.6 &  1.6 & 5792 &  0.52 & 230 \\ 
      {\it Spitzer}/IRAC &  4.5 &  1.7 & 5576 &  0.53 & 230 \\ 
      {\it Spitzer}/IRAC &  5.8 &  1.9 & 2328 &  2.74 & 226 \\
      {\it Spitzer}/IRAC &  8.0 &  2.0 & 2186 &  1.80 & 249 \\
      {\it AKARI}/IRC    & 11.0 &  4.8 &  242 & 44.04 & 101 \\
      {\it AKARI}/IRC    & 18.0 &  5.7 &  233 & 96.29 & 115 \\
      {\it Spitzer}/MIPS & 24.0 &  6.0 & 1199 & 80.00 & 254 \\
      \hline
    \end{tabular}
  \end{center}
  \label{tab:catalogues_info}
\end{table}

\subsection{{\it ISO}}

We use the catalogue of {\it ISO} sources in the HDF-N produced by
Aussel et al. (1999).  The source catalogue is claimed to be $95\%$
complete at 200$\,\mu$Jy in the 15$\,\mu$m band and at 65$\,\mu$Jy in
the 6.5$\mu$m band. It includes 49 objects, 42 detected only at
15$\,\mu$m, 3 at only 6.5$\,\mu$m and 4 at both wavelengths. Aussel et
al. also provide an additional, less secure, list of 51 sources of
which 47 are detected at 15$\,\mu$m only, 4 at 6.5$\,\mu$m only, but
none in both filters. All together the two catalogues include a total
of 100 objects.

\subsection{{\it Spitzer}}

The {\it Spitzer} data for the HDF-N are part of the GOODS {\it
  Spitzer} Legacy Data Products\footnote{see
  http:$//$data.spitzer.caltech.edu$/$popular$/$goods$/$Documents$/$
  goods$\underline{}$dataproducts.html} and are in the public domain.
The {\it Spitzer} data sets include the images of both GOODS fields at
3.6, 4.5, 5.8, 8.0$\,\mu$m from the Infra-Red Array Camera (IRAC) and
at 24$\,\mu$m from the Multiband Imaging Photometer for {\it Spitzer}
(MIPS), plus a list of sources for the MIPS 24$\,\mu$m imaging.  The
MIPS catalogue consists of 1199 sources detected at $3\sigma$, with
flux densities greater than 80$\,\mu$Jy (see
Table~\ref{tab:catalogues_info}), a limit where the source extraction
is stated to be highly complete and reliable.  Source lists for the
IRAC imaging have not yet been released by the GOODS
consortium. Therefore we derived our own IRAC source catalogues in
GOODS-N using SExtractor (Bertin $\&$ Arnouts 1996) with the default
set of values for the configuration parameters.  The source fluxes and
the corresponding errors have been derived from the automatic aperture
magnitudes. Sources were extracted down to a minimum signal-to-noise
ratio of 3. The resulting catalogues consist of 5792, 5576, 2328, 2186
sources at 3.6, 4.5, 5.8 and 8.0$\,\mu$m, respectively (see
Table~\ref{tab:catalogues_info}). \\
Imaging at 16$\,\mu$m of selected areas within the GOODS-N field have
been obtained as a result of a pilot study with the {\it Spitzer} IRS
(Teplitz et al. 2005). The majority of the area (30 of 35 arcmin$^2$)
reaches a 3$\sigma$ sensitivity of $\sim$75$\,\mu$Jy. Teplitz et
al. (2005) detected 149 objects at 16$\,\mu$m, with flux ranging from
21$\,\mu$Jy to 1.24 mJy, and with photometry in good agreement with
the 15$\,\mu$m {\it ISO} survey of the same area.

\subsection{{\it AKARI}}

The GOODS-N region has recently been imaged at 11$\,\mu$m and
18$\,\mu$m with the {\it AKARI} Infra-Red Camera (IRC) within the
FUHYU program. FUYHU is an {\it AKARI} mission program to follow-up
well-studied {\it Spitzer} fields (Pearson et al. 2008) including the
GOODS-N field. This survey has mapped the Lockman-Hole and ELAIS N1
fields, other well-studied fields with sufficiently high $|\beta|$
ecliptic latitudes and consequently high {\it AKARI} visibility. The
source extraction and flux calibration are described in Pearson et
al. (2008), and are based partly on the source extraction methodology
developed for sub-millimeter surveys (Serjeant et al. 2003).  The
samples comprises 242 detections at 11$\mu$m and 233 at 18$\,\mu$m
respectively, with a signal-to-noise ratio higher than 3 (see
Table~\ref{tab:catalogues_info}).

\subsection{Spectroscopic redshifts}

The region around the HDF-N had been the subject of intensive
spectroscopic campaigns in the 90s' by a variety of groups using the
Low Resolution Imaging Spectrograph (LRIS; Oke et al. 1995) on the
Keck 10 meter telescopes.  A compilation of the LRIS-spectra for 671
sources is presented in Cohen et al. (2000). Subsequently, two
parallel spectroscopic projects were carried out in the GOODS area
using the Deep Imaging Multi-Object Spectrograph (DEIMOS) on the Keck
II telescope.  The Team Keck Treasury Survey (TKTS) of the GOODS-N
field (Wirth et al. 2004) has focused on an $R_{AB}<24.4$ magnitude
selected sample while the survey by Cowie et al. (2004) was embedded
within observations targeted at high-redshift galaxies and X-ray and
radio selected galaxies. More recently, Steidel and collaborators
  have conducted a spectroscopic survey in the GOODS-N with the blue
  arm of LRIS (LRIS-B) targeting several hundreds of star-forming
  galaxies and AGNs at redshifts $1.4\lsim z\lsim3.0$. The source
  candidates for spectroscopic follow-up were pre-selected using
  different criteria based on their optical colours (Steidel et
  al. 2003, 2004; Adelberger et al. 2004).  The resulting
  spectroscopic catalogue, presented by Reddy et al. (2006) includes
  342 objects and provide for each of them both optical photometry and
  infrared photometry at 3.6, 4.5, 5.8, 8.0 and 24$\,\mu$m from {\it
    Spitzer}. We use here the
  compilation\footnote{http://www2.keck.hawaii.edu/tksurvey/} of
  redshifts assembled by the Keck Team combining all the spectroscopic
  surveys within the GOODS-N field produced up to the 2004, and we add
  to it the spectroscopic sample of Reddy et al. (2006). Note that the
  TKTS is based primarly on observations with the DEIMOS instrument on
  Keck II, whose spectral range only allows the detection of emission
  and absorption features from objects at $z\lsim1.2$. Conversely, the
  selection criteria used in the spectroscopic survey of Reddy et
  al. is better at identifying galaxies at $z\gsim1.4$. We found indeed that only 59 out
  of the 342 spectroscopic sources listed in the Reddy et
  al. catalogue are already included in the TKTS compendium redshift
  catalogue.

\subsection{Optical data}

Optical imaging of the GOODS-N field was obtained with the Avanced
Camera for Survey (ACS) on board the {\it Hubble Space Telescope}
(HST) at B, V, i$^{\prime}$, z$^{\prime}$ bands. The ACS images and
the source catalogues extracted by the GOODS Team are of public
domain\footnote{http://archive.stsci.edu/prepds/goods/}.

Optical/near-infrared ground-based images are also available for the
GOODS-N field (Capak et al. 2004). An intensive multi-color imaging
survey of $\sim0.2$ deg$^{2}$ centered on the HDF-N have been carried
out using different instruments: the Kitt Peak National Observatory
(KPNO) 4 meter telescope with the MOSAIC prime focus camera, the
Subaru 8.2 meter telescope with the Suprime-Cam instrument, and the
QUIRC camera on the Hawaii 2.2 meter telescope. The surveyed area is
referred to as the Hawaii HDF-N (Capak et al. 2004).  Data were
collected in $U$, $B$, $V$, $R$, $I$ and $z^{\prime}$ bands over the
whole field and in $HK^{\prime}$ over a smaller region covering the
Chandra Deep Field South, down to 5$\sigma$ (AB magnitudes) limits of
$27.1$, $26.9$, $26.8$, $26.6$, $25.6$, $25.4$ and $22.1$,
respectively.  The images and the corresponding catalogues are
available on the World Wide
Web\footnote{http://www.ifa.hawaii.edu/~capak/hdf/index.html}.

Here, optical data are used just for comparison with the infrared
imaging, but they are not exploited in the SED fit.


\setcounter{table}{1}
\begin{landscape}
  \begin{table}
    \begin{center}
      \scriptsize
      \caption{ \label{tab:realcat_info} List of infrared sources with
        spectroscopic redshift drawn from the GOODS-N field, by
        cross-matching the source lists of Aussel et al. (1999) and of
        Teplitz et al. (2005) with the {\it Spitzer} and {\it AKARI}
        catalogues in the same field. The source positions reported
        here are those of the spectroscopic
        counterpart.} 
      \begin{tabular}{lcccrrrrrrrrrrl}
        \hline
        \hline
        ID & \multicolumn{2}{c}{Spec. position} & $z_{\rm spec}$ & $F_{\rm 3.6\mu m}$ &
        $F_{\rm 4.5\mu m}$ & $F_{\rm 5.8\mu m}$ & $F_{\rm 8\mu m}$ & $F_{\rm 24\mu m}$ & $F_{\rm 6.5\mu m}$ & $F_{\rm 15\mu m}$ & 
        $F_{\rm 11\mu m}$ & $F_{\rm 18\mu m}$ &  $F_{\rm 16\mu m}$ & {\it ISO} name \\
        & $\alpha_{\rm J2000}$ & $\delta_{\rm J2000}$ & & & & & & & & & & & & \\
        & ($^{\rm h\ m\ s}$) & ($^{\circ}\ '\ ''$) & & ($\mu$Jy) & ($\mu$Jy) & ($\mu$Jy) & ($\mu$Jy) & ($\mu$Jy) & ($\mu$Jy) & ($\mu$Jy) & ($\mu$Jy) & ($\mu$Jy) & ($\mu$Jy) & \\
        \hline
        \hline
        ID1  &  12 36 48.30  &  +62 14 26.86  &  0.1389  &  86.3$^{+2.2}_{-2.2}$  &  60.1$^{+1.7}_{-1.7}$  &  54.8$^{+3.2}_{-3.2}$  &  386.6$^{+4.0}_{-4.0}$  &  460.0$^{+6.1}_{-6.1}$  & 254$^{+71}_{-73}$  &  307$^{+62}_{-67}$  &  364$^{+15}_{-15}$  &  291$^{+32}_{-32}$  &  283$^{+20}_{-20}$  &  HDF$\underline{~}$PM3$\underline{~}$24  \\
        ID2  &  12 37 23.79  &  +62 10 46.35  &  0.1133  &  92.9$^{+2.9}_{-2.9}$  &  62.1$^{+2.3}_{-2.3}$  &  54.9$^{+4.2}_{-4.2}$  &  242.3$^{+4.2}_{-4.2}$  &  198$^{+6.6}_{-6.6}$  &  -  &  -  & 184$^{+16}_{-16}$  &  129$^{+33}_{-33}$  &  141$^{+24}_{-24}$  &  - \\
        ID3  &  12 36 12.48  &  +62 11 40.79  &  0.2759  &  136.9$^{+4.1}_{-4.1}$  &  127.5$^{+3.5}_{-3.5}$  &  99.9$^{+6.4}_{-6.4}$  &  401.9$^{+6.1}_{-6.1}$  &  1240$^{+13}_{-13}$  &  -  &  -  & 900$^{+15}_{-15}$  &  1069$^{+32}_{-32}$  &  973$^{+27}_{-27}$  &  - \\
        ID4  &  12 36 34.47  &  +62 12 13.45  &  0.4573  &  316.3$^{+5.7}_{-5.7}$  &  244.5$^{+4.5}_{-4.5}$  &  196.0$^{+7.9}_{-7.9}$  &  338.4$^{+5.0}_{-5.0}$  &  1290.0$^{+8.8}_{-8.8}$  &  -  &  448$^{+68}_{-59}$  &  858$^{+18}_{-18}$  &  897$^{+32}_{-32}$  &  853$^{+32}_{-32}$  &  HDF$\underline{~}$PM3$\underline{~}$2  \\ 
        ID5  &  12 36 22.94  &  +62 15 26.97  &  2.5920  &  46.4$^{+2.2}_{-2.2}$  &  51.1$^{+1.9}_{-1.9}$  &  72.8$^{+4.8}_{-4.8}$  &  130.1$^{+3.0}_{-3.0}$  &  529.0$^{+6.2}_{-6.2}$  &  -  &  -  & 159$^{+15}_{-15}$  &  348$^{+32}_{-32}$  &  335$^{+29}_{-29}$  &  - \\
        ID6  &  12 37 08.32  &  +62 10 56.41  &  0.4225  &  61.9$^{+1.9}_{-1.9}$  &  64.5$^{+1.9}_{-1.9}$  &  50.0$^{+3.2}_{-3.2}$  &  183.3$^{+3.0}_{-3.0}$  &  648.0$^{+7.4}_{-7.4}$  &  -  &  -  &  576$^{+15}_{-15}$  &  494$^{+33}_{-33}$  &  423$^{+26}_{-26}$  &  - \\
        ID7  &  12 37 19.14  &  +62 11 31.58  &  0.5560  &  74.1$^{+2.1}_{-2.1}$  &  52.3$^{+1.7}_{-1.7}$  &  48.5$^{+3.2}_{-3.2}$  &  50.7$^{+1.7}_{-1.7}$  &  190.0$^{+5.9}_{-5.9}$  &  -  &  -  &  171$^{+15}_{-15}$  &  200$^{+32}_{-32}$  &  213$^{+24}_{-24}$  &  - \\
        ID8  &  12 36 39.71  &  +62 15 26.68  &  0.3765  &  46.6$^{+1.7}_{-1.7}$  &  40.3$^{+1.5}_{-1.5}$  &  27.9$^{+2.4}_{-2.4}$  &  66.7$^{+1.7}_{-1.7}$  &  161.0$^{+5.7}_{-5.7}$  &  -  &  -  &  121$^{+15}_{-15}$  &  136$^{+32}_{-32}$  &  121$^{+19}_{-19}$  &  - \\
        ID9  &  12 36 51.12  &  +62 10 31.23  &  0.4099  &  90.7$^{+3.3}_{-3.3}$  &  91.6$^{+2.9}_{-2.9}$  &  74.0$^{+5.0}_{-5.0}$  &  261.1$^{+4.6}_{-4.6}$  &  984$^{+9.1}_{-9.1}$  &  -  &  341$^{+40}_{-65}$  &  713$^{+15}_{-15}$  &  657$^{+33}_{-33}$  &  745$^{+29}_{-29}$  &  HDF$\underline{~}$PM3$\underline{~}$28  \\
        ID10  &  12 36 03.26  &  +62 11 11.27  &  0.6382  &  135.2$^{+4.1}_{-4.1}$  &  83.6$^{+2.8}_{-2.8}$  &  97.8$^{+6.2}_{-6.2}$  &  92.2$^{+3.0}_{-3.0}$  &  1210.0$^{+9.5}_{-9.5}$  &  -  &  -  &  608$^{+22}_{-22}$  &  733$^{+144}_{-144}$  &  655$^{+33}_{-33}$  &  - \\
        ID11  &  12 36 22.50  &  +62 15 44.78  &  0.6393  &  71.4$^{+2.6}_{-2.6}$  &  48.5$^{+1.8}_{-1.8}$  &  57.1$^{+4.1}_{-4.1}$  &  67.3$^{+2.1}_{-2.1}$  &  721.0$^{+6.6}_{-6.6}$  &  -  &  -  &  294$^{+15}_{-15}$  &  346$^{+32}_{-32}$  &  390$^{+28}_{-28}$  &  - \\
        ID12  &  12 36 50.20  &  +62 08 45.09  &  0.4335  &  101.3$^{+3.3}_{-3.3}$  &  86.3$^{+2.9}_{-2.9}$  &  79.8$^{+5.5}_{-5.5}$  &  129.9$^{+3.6}_{-3.6}$  &  585.0$^{+7.8}_{-7.8}$  &  -  &  -  &  323$^{+15}_{-15}$  &  293$^{+33}_{-33}$  &  348$^{+20}_{-20}$  &  - \\
        ID13  &  12 36 41.56  &  +62 09 48.54  &  0.5186  &  204.8$^{+4.9}_{-4.9}$  &  151.7$^{+3.8}_{-3.8}$  &  128.6$^{+7.3}_{-7.3}$  &  152.5$^{+4.1}_{-4.1}$  &  433.0$^{+6.9}_{-6.9}$  &  -  &  -  &  355$^{+15}_{-15}$  &  322$^{+34}_{-34}$  &  327$^{+23}_{-23}$  &  - \\
        ID14  &  12 36 55.75  &  +62 09 17.80  &  0.4191  &  72.9$^{+2.8}_{-2.8}$  &  71.7$^{+2.6}_{-2.6}$  &  67.4$^{+4.9}_{-4.9}$  &  150.6$^{+3.7}_{-3.7}$  &  846.0$^{+9.9}_{-9.9}$  &  -  &  -  &  411$^{+15}_{-15}$  &  426$^{+32}_{-32}$  &  408$^{+21}_{-21}$  &  - \\
        ID15  &  12 36 43.98  &  +62 12 50.44  &  0.5560  &  58.9$^{+2.1}_{-2.1}$  &  46.9$^{+1.8}_{-1.8}$  &  52.6$^{+3.7}_{-3.7}$  &  64.4$^{+2.1}_{-2.1}$  &  424$^{+4.6}_{-4.6}$  &  $<$50  &  282$^{+60}_{-64}$  &  281$^{+15}_{-15}$  &  317$^{+34}_{-34}$  &  343$^{+22}_{-22}$  &  HDF$\underline{~}$PM3$\underline{~}$17  \\
        ID16  &  12 36 48.63  &  +62 09 32.55  &  0.5174  &  31.6$^{+1.9}_{-1.9}$  &  25.7$^{+1.6}_{-1.6}$  &  20.1$^{+2.7}_{-2.7}$  &  29.8$^{+1.8}_{-1.8}$  &  127.0$^{+7.3}_{-7.3}$  &  -  &  -  &  86$^{+15}_{-15}$  &  131$^{+32}_{-32}$  &  104$^{+16}_{-16}$  &  - \\
        ID17  &  12 36 53.89  &  +62 12 54.40  &  0.6419  &  58.4$^{+1.9}_{-1.9}$  &  37.3$^{+1.4}_{-1.4}$  &  36.5$^{+2.7}_{-2.7}$  &  25.3$^{+1.2}_{-1.2}$  &  200.0$^{+6.3}_{-6.3}$  &  $<$36  &  179$^{+60}_{-43}$  &  92$^{+15}_{-15}$  &  181$^{+32}_{-32}$  &  207$^{+22}_{-22}$  &  HDF$\underline{~}$PM3$\underline{~}$33 \\
        ID18  &  12 36 36.80  &  +62 12 13.46  &  0.8477  &  126.4$^{+3.5}_{-3.5}$  &  83.7$^{+2.7}_{-2.7}$  &  68.9$^{+4.8}_{-4.8}$  &  50.4$^{+2.1}_{-2.1}$  &  379.0$^{+5.2}_{-5.2}$  &  $<$113  &  202$^{+58}_{-50}$  &  101$^{+15}_{-15}$  &  261$^{+32}_{-32}$  &  300$^{+22}_{-22}$  &  HDF$\underline{~}$PM3$\underline{~}$8 \\
        ID19  &  12 36 51.79  &  +62 13 54.19  &  0.5561  &  46.7$^{+1.6}_{-1.6}$  &  34.8$^{+1.3}_{-1.3}$  &  36.2$^{+2.6}_{-2.6}$  &  42.2$^{+1.4}_{-1.4}$  &  203.0$^{+6.3}_{-6.3}$  &  $<$39  &  151$^{+74}_{-68}$  &  195$^{+15}_{-15}$  &  176$^{+32}_{-32}$  &  185$^{+30}_{-30}$  &  HDF$\underline{~}$PM3$\underline{~}$30 \\
        ID20  &  12 36 17.44  &  +62 15 51.58  &  0.3758  &  22.7$^{+1.6}_{-1.6}$  &  25.9$^{+1.5}_{-1.5}$  &  17.5$^{+2.7}_{-2.7}$  &  54.9$^{+2.2}_{-2.2}$  &  145.0$^{+9.7}_{-9.7}$  &  -  &  -  &  119$^{+25}_{-25}$  &  116$^{+32}_{-32}$  &  95$^{+23}_{-23}$  &  - \\
        ID21  &  12 36 46.18  &  +62 11 42.41  &  1.0164  &  108.2$^{+3.4}_{-3.4}$  &  79.4$^{+2.7}_{-2.7}$  &  49.4$^{+4.2}_{-4.2}$  &  41.5$^{+2.0}_{-2.0}$  &  290.0$^{+4.9}_{-4.9}$  &  88$^{+45}_{-80}$  &  170$^{+59}_{-42}$  &  61$^{+15}_{-15}$  &  212$^{+37}_{-37}$  &  275$^{+18}_{-18}$  &  HDF$\underline{~}$PM3$\underline{~}$19 \\
        ID22  &  12 36 46.89  &  +62 14 47.78  &  0.5560  &  44.1$^{+1.6}_{-1.6}$  &  32.2$^{+1.3}_{-1.3}$  &  30.2$^{+2.4}_{-2.4}$  &  33.9$^{+1.3}_{-1.3}$  &  277$^{+11}_{-11}$  &  $<$179  &  144$^{+72}_{-47}$  &  66$^{+15}_{-15}$  &  266$^{+32}_{-32}$  &  193$^{+17}_{-17}$  &  HDF$\underline{~}$PM3$\underline{~}$23 \\
        ID23  &  12 36 31.65  &  +62 16 04.41  &  0.7840  &  44.9$^{+1.7}_{-1.7}$  &  30.0$^{+1.3}_{-1.3}$  &  30.9$^{+2.6}_{-2.6}$  &  23.8$^{+1.2}_{-1.2}$  &  301$^{+5.3}_{-5.3}$  &  -  &  -  &  87$^{+15}_{-15}$  &  239$^{+32}_{-32}$  &  245$^{+22}_{-22}$  &  - \\
        ID24  &  12 37 01.49  &  +62 08 42.37  &  0.7038  &  37.7$^{+2.3}_{-2.3}$  &  20.7$^{+1.6}_{-1.6}$  &  23.1$^{+3.5}_{-3.5}$  &  21.3$^{+1.9}_{-1.9}$  &  185.0$^{+7.2}_{-7.2}$  &  -  &  -  &  70$^{+15}_{-15}$  &  123$^{+33}_{-33}$  &  130$^{+24}_{-24}$  &  - \\
        ID25  &  12 36 49.72  &  +62 13 13.39  &  0.4745  &  53.5$^{+1.8}_{-1.8}$  &  49.0$^{+1.6}_{-1.6}$  &  47.4$^{+3.1}_{-3.1}$  &  113.0$^{+2.4}_{-2.4}$  &  371$^{+10}_{-10}$  &  136$^{+68}_{-57}$  &  320$^{+39}_{-62}$  &  356$^{+15}_{-15}$  &  410$^{+33}_{-33}$  &  370$^{+27}_{-27}$  &  HDF$\underline{~}$PM3$\underline{~}$27 \\
        ID26  &  12 36 39.93  &  +62 12 50.38  &  0.8462  &  64.1$^{+2.5}_{-2.5}$  &  38.9$^{+1.7}_{-1.7}$  &  40.0$^{+3.5}_{-3.5}$  &  30.9$^{+1.6}_{-1.6}$  &  493.0$^{+6.1}_{-6.1}$  &  $<$64  &  302$^{+67}_{-55}$  &  120$^{+15}_{-15}$  &  288$^{+32}_{-32}$  &  425$^{+20}_{-20}$  & HDF$\underline{~}$PM3$\underline{~}$11 \\
        ID27  &  12 36 17.33  &  +62 15 29.95  &  0.8497  &  79.4$^{+3.1}_{-3.1}$  &  55.8$^{+2.2}_{-2.2}$  &  52.9$^{+4.6}_{-4.6}$  &  31.6$^{+1.7}_{-1.7}$  &  499.0$^{+7.2}_{-7.2}$  &  -  &  -  &  93$^{+15}_{-15}$  &  354$^{+32}_{-32}$  &  448$^{+27}_{-27}$  &  - \\
        ID28  &  12 36 33.59  &  +62 13 20.24  &  0.8446  &  49.9$^{+2.3}_{-2.3}$  &  32.9$^{+1.7}_{-1.7}$  &  34.6$^{+3.6}_{-3.6}$  &  23.1$^{+1.5}_{-1.5}$  &  323.0$^{+6.8}_{-6.8}$  &  -  &  122$^{+54}_{-40}$  &  68$^{+15}_{-15}$  &  192$^{+35}_{-35}$  &  320$^{+29}_{-29}$  &  HDF$\underline{~}$PS3$\underline{~}$3  \\
        ID29  &  12 36 49.50  &  +62 14 07.13  &  0.7517  &  32.6$^{+1.3}_{-1.3}$  &  21.4$^{+1.0}_{-1.0}$  &  21.7$^{+2.0}_{-2.0}$  &  17.6$^{+1.0}_{-1.0}$  &  186.0$^{+6.4}_{-6.4}$  &  $<$40  &  150$^{+74}_{-48}$  &  69$^{+15}_{-15}$  &  143$^{+33}_{-33}$  &  130$^{+20}_{-20}$  &  HDF$\underline{~}$PM3$\underline{~}$26 \\
        ID30  &  12 36 33.66  &  +62 10 06.20  &  1.0156  &  72.1$^{+3.2}_{-3.2}$  &  56.2$^{+2.4}_{-2.4}$  &  44.0$^{+4.1}_{-4.1}$  &  45.4$^{+2.2}_{-2.2}$  &  581.0$^{+9.0}_{-9.0}$  &  -  &  -  &  94$^{+15}_{-15}$  &  398$^{+32}_{-32}$  &  438$^{+25}_{-25}$  &  - \\
        ID31  &  12 36 33.14  &  +62 15 14.01  &  0.5196  &  16.7$^{+1.1}_{-1.1}$  &  13.9$^{+0.9}_{-0.9}$  &  13.5$^{+1.9}_{-1.9}$  &  17.8$^{+1.1}_{-1.1}$  &  142.0$^{+7.0}_{-7.0}$  &  -  &  -  &  112$^{+15}_{-15}$  &  182$^{+32}_{-32}$  &  41$^{+17}_{-17}$  &  - \\
        ID32  &  12 36 54.81  &  +62 08 47.67  &  0.7913  &  64.1$^{+2.7}_{-2.7}$  &  44.1$^{+2.1}_{-2.1}$  &  41.9$^{+4.0}_{-4.0}$  &  42.9$^{+2.1}_{-2.1}$  &  246.0$^{+5.6}_{-5.6}$  &  -  &  -  &  75$^{+15}_{-15}$  &  102$^{+33}_{-33}$  &  134$^{+17}_{-17}$  &  - \\
        ID33  &  12 36 46.23  &  +62 15 27.64  &  0.8510  &  84.5$^{2.2}_{-2.2}$  &  56.7$^{+1.7}_{-1.7}$  &  54.2$^{+3.3}_{-3.3}$  &  42.3$^{+1.5}_{-1.5}$  &  544.0$^{+7.6}_{-7.6}$  &  -  &  418$^{+91}_{-94}$  &  119$^{+15}_{-15}$  &  350$^{+32}_{-32}$  &  433$^{+27}_{-27}$  &  HDF$\underline{~}$PM3$\underline{~}$21 \\
        ID34  &  12 36 55.94  &  +62 08 08.58  &  0.7920  &  108.5$^{+3.9}_{-3.9}$  &  76.0$^{+3.2}_{-3.2}$  &  73.2$^{+6.1}_{-6.1}$  &  73.9$^{+3.3}_{-3.3}$  &  832.0$^{+7.5}_{-7.5}$  &  -  &  -  &  213$^{+15}_{-15}$  &  384$^{+32}_{-32}$  &  568$^{+20}_{-20}$  &  - \\
        ID35  &  12 36 32.48  &  +62 15 13.57  &  0.6827  &  34.4$^{+1.6}_{-1.6}$  &  23.0$^{+1.2}_{-1.2}$  &  22.8$^{+2.3}_{-2.3}$  &  18.8$^{+1.1}_{-1.1}$  &  142.0$^{+6.7}_{-6.7}$  &  -  &  -  &  112$^{+15}_{-15}$  &  182$^{+32}_{-32}$  &  143$^{+17}_{-17}$  &  - \\
        ID36  &  12 36 31.50  &  +62 11 14.52  &  1.0124  &  42.8$^{+2.1}_{-2.1}$  &  32.3$^{+1.7}_{-1.7}$  &  27.9$^{+3.2}_{-3.2}$  &  39.7$^{+2.1}_{-2.1}$  &  480.0$^{+5.9}_{-5.9}$  &  -  &  355$^{+40}_{-60}$  &  75$^{+15}_{-15}$  &  287$^{+34}_{-34}$  &  430$^{+20}_{-20}$  &HDF$\underline{~}$PM3$\underline{~}$1 \\
        ID37  &  12 36 37.80  &  +62 11 49.70  &  0.8380  &  37.3$^{+2.0}_{-2.0}$  &  31.6$^{+1.6}_{-1.6}$  &  29.2$^{+3.2}_{-3.2}$  &  24.6$^{+1.8}_{-1.8}$ & 173.0$^{+3.6}_{-3.6}$  &  $<$61  &  212$^{+58}_{-55}$  &  81.4$^{+15}_{-15}$  &  238$^{+33}_{-33}$  &  142$^{+20}_{-20}$  &   HDF$\underline{~}$PM3$\underline{~}$10 \\
        ID38  &  12 36 38.30  &  +62 11 51.16  &  0.8410  &  36.9$^{+1.9}_{-1.9}$  &  23.3$^{+1.4}_{-1.4}$  &  29.3$^{+3.2}_{-3.2}$  &  16.9$^{+1.5}_{-1.5}$ & 230.0$^{+3.3}_{-3.3}$  &  $<$61  &  212$^{+58}_{-55}$  &  81.4$^{+15}_{-15}$  &  238$^{+33}_{-33}$  &  189$^{+19}_{-19}$  &  HDF$\underline{~}$PM3$\underline{~}$10 \\
        ID39  &  12 36 34.53  &  +62 12 41.34  &  1.2190  &  66.1$^{+2.6}_{-2.6}$  &  71.9$^{+2.4}_{-2.4}$  &  58.1$^{+4.4}_{-4.4}$  &  72.6$^{+2.4}_{-2.4}$  &  446.0$^{+5.1}_{-5.1}$  &  -  &  363$^{+79}_{-39}$  &  104$^{+15}_{-15}$  &  705$^{+32}_{-32}$  &  923$^{+32}_{-32}$  &  HDF$\underline{~}$PM3$\underline{~}$3 \\
        ID40  &  12 36 34.86  &  +62 16 28.62  &  0.8470  &  58.9$^{+1.9}_{-1.9}$  &  40.1$^{+1.5}_{-1.5}$  &  38.5$^{+2.8}_{-2.8}$  &  34.8$^{+1.3}_{-1.3}$  &  482.0$^{+4.6}_{-4.6}$  &  -  &  -  &  104$^{+15}_{-15}$  &  300$^{+32}_{-32}$  &  368$^{+21}_{-21}$  &  - \\
        ID41  &  12 36 35.59  &  +62 14 24.32  &  2.0050  &  71.6$^{+2.4}_{-2.4}$  &  99.9$^{+2.5}_{-2.5}$  &  163.4$^{+6.3}_{-6.3}$  &  282.6$^{+4.0}_{-4.0}$  &  1480$^{+10}_{-10}$  &  -  &  441$^{+43}_{-82}$  &  369$^{+15}_{-15}$  &  756$^{+32}_{-32}$  & 615$^{+32}_{-32}$  &  HDF$\underline{~}$PM3$\underline{~}$5 \\
        ID42  &  12 36 41.43  &  +62 11 42.81  &  0.5480  &  59.9$^{+2.5}_{-2.5}$  &  61.2$^{+2.3}_{-2.3}$  &  46.7$^{+4.0}_{-4.0}$  &  39.7$^{+2.0}_{-2.0}$  &  225.0$^{+5.0}_{-5.0}$  &  127$^{+99}_{-61}$  &  $<$72  &  55$^{+15}_{-15}$  &  137$^{+33}_{-33}$  &  134$^{+15}_{-15}$  & HDF$\underline{~}$PM2$\underline{~}$1 \\
        ID43  &  12 35 56.10  &  +62 12 19.57  &  0.9585  &  80.4$^{+3.1}_{-3.1}$  &  60.5$^{+2.4}_{-2.4}$  &  53.8$^{+4.7}_{-4.7}$  &  79.4$^{+2.9}_{-2.9}$  &  430.0$^{+6.1}_{-6.1}$  &  -  &  -  &  119$^{+15}_{-15}$  &  249$^{+34}_{-34}$  &  295$^{+25}_{-25}$  &  - \\
        ID44  &  12 36 44.35  &  +62 14 53.34  &  1.4865  &  13.8$^{+0.9}_{-0.9}$  &  14.1$^{+0.9}_{-0.9}$  &  16.8$^{+1.9}_{-1.9}$  &  21.7$^{+1.1}_{-1.1}$  &  81.7$^{+6.4}_{-6.4}$  &  $<$329  &  105$^{+94}_{-21}$  &  47$^{+15}_{-15}$  &  133$^{+32}_{-32}$  &  37$^{+16}_{-16}$  &  HDF$\underline{~}$PM3$\underline{~}$18 \\
        \multicolumn{10}{r}{\emph{continued on next page}}
      \end{tabular}
    \end{center}
  \end{table}
\end{landscape}


\setcounter{table}{1}
\begin{landscape}
  \begin{table}
    \scriptsize
    \begin{center}
      \caption{{\it Continued}}
      \begin{tabular}{lcccrrrrrrrrrrl}
        \hline
        \hline
        ID & \multicolumn{2}{c}{Spec. position} & $z_{\rm spec}$ & $F_{\rm 3.6\mu m}$ &
        $F_{\rm 4.5\mu m}$ & $F_{\rm 5.8\mu m}$ & $F_{\rm 8\mu m}$ & $F_{\rm 24\mu m}$ & $F_{\rm 6.5\mu m}$ & $F_{\rm 15\mu m}$ & 
        $F_{\rm 11\mu m}$ & $F_{\rm 18\mu m}$ &  $F_{\rm 16\mu m}$ & {\it ISO} name \\
        & $\alpha_{\rm J2000}$ & $\delta_{\rm J2000}$ & & & & & & & & & & & & \\
        & ($^{\rm h\ m\ s}$) & ($^{\circ}\ '\ ''$) & & ($\mu$Jy) & ($\mu$Jy) & ($\mu$Jy) & ($\mu$Jy) & ($\mu$Jy) & ($\mu$Jy) & ($\mu$Jy) & ($\mu$Jy) & ($\mu$Jy) & ($\mu$Jy) & \\
        \hline
        \hline
        ID45  &  12 36 29.16  &  +62 10 46.46  &  1.0130  &  98.1$^{+3.5}_{-3.5}$  &  88.3$^{+3.1}_{-3.1}$  &  76.0$^{+5.3}_{-5.3}$  &  70.6$^{+2.7}_{-2.7}$  &  724.0$^{+12}_{-12}$  &  -  &  -  &  75$^{+15}_{-15}$  &  434$^{+33}_{-33}$  &  465$^{+25}_{-25}$  &  - \\
        ID46  &  12 36 46.60  &  +62 10 49.36  &  0.9399  &  32.6$^{+1.9}_{-1.9}$  &  24.3$^{+1.5}_{-1.5}$  &  24.3$^{+3.0}_{-3.0}$  &  18.2$^{+1.4}_{-1.4}$  &  354$^{+6.5}_{-6.5}$  &  -  &  327$^{+39}_{-63}$  &  71$^{+15}_{-15}$  &  259$^{+32}_{-32}$  &  315$^{+29}_{-29}$  &  HDF$\underline{~}$PM3$\underline{~}$22 \\
        ID47  &  12 36 36.86  &  +62 11 35.17  &  0.07860  &  170.6$^{+4.1}_{-4.1}$  &  114.9$^{+3.1}_{-3.1}$  &  113.5$^{+6.0}_{-6.0}$  &  545.2$^{+6.6}_{-6.6}$  &  732.0$^{+9.1}_{-9.1}$  &  $<$135  &  300$^{+62}_{-67}$  &  370$^{+15}_{-15}$  &  377$^{+32}_{-32}$  &  -  &  HDF$\underline{~}$PM3$\underline{~}$7  \\
        ID48  &  12 36 36.64  &  +62 13 47.12  &  0.9590  &  78.4$^{+2.6}_{-2.6}$  &  77.3$^{+2.3}_{-2.3}$  &  91.1$^{+5.0}_{-5.0}$  &  110.6$^{+2.7}_{-2.7}$  &  474.0$^{+5.9}_{-5.9}$  &  -  &  353$^{+40}_{-66}$  &  188$^{+15}_{-15}$  &  346$^{+33}_{-33}$  &  -  &  HDF$\underline{~}$PM3$\underline{~}$6  \\
        ID49  &  12 37 05.87  &  +62 11 54.03  &  0.9032  &  83.5$^{+2.3}_{-2.3}$  &  55.8$^{+1.7}_{-1.7}$  &  50.4$^{+3.1}_{-3.1}$  &  39.7$^{+1.4}_{-1.4}$  &  655.0$^{+ 8.1}_{- 8.1}$  &  -  &  431$^{+43}_{-80}$  &  100$^{+15}_{-15}$  &  397$^{+33}_{-33}$  &  -  &  HDF$\underline{~}$PM3$\underline{~}$45 \\
        ID50  &  12 36 59.92  &  +62 14 50.28  &  0.7610  &  50.5$^{+1.7}_{-1.7}$  &  33.6$^{+1.3}_{-1.3}$  &  39.3$^{+2.7}_{-2.7}$  &  33.0$^{+1.2}_{-1.2}$  &  466.0$^{+5.5}_{-5.5}$  &  -  &  295$^{+61}_{-66}$  &  166$^{+15}_{-15}$  &  310$^{+34}_{-34}$  &  -  &  HDF$\underline{~}$PM3$\underline{~}$40 \\
        ID51  &  12 36 53.21  &  +62 11 17.12  &  0.9350  &  83.4$^{+2.8}_{-2.8}$  &  58.4$^{+2.2}_{-2.2}$  &  51.5$^{+4.1}_{-4.1}$  &  47.1$^{+2.2}_{-2.2}$  &  367.0$^{+6.4}_{-6.4}$  &  -  &  174$^{+59}_{-43}$  &  65$^{+15}_{-15}$  &  214$^{+33}_{-33}$  &  -  &  HDF$\underline{~}$PM3$\underline{~}$31 \\ 
        ID52  &  12 36 53.37  &  +62 11 39.97  &  1.2698  &  32.8$^{+1.6}_{-1.6}$  &  34.6$^{+1.6}_{-1.6}$  &  26.4$^{+2.7}_{-2.7}$  &  40.3$^{+1.7}_{-1.7}$  &  322.0$^{+6.2}_{-6.2}$  &  -  &  180$^{+60}_{-43}$  &  57$^{+15}_{-15}$  &  393$^{+33}_{-33}$  &  -  &  HDF$\underline{~}$PM3$\underline{~}$32 \\
        ID53  &  12 36 58.99  &  +62 12 09.20  &  0.8517  &  62.7$^{+2.0}_{-2.0}$  &  42.6$^{+1.6}_{-1.6}$  &  35.1$^{+2.7}_{-2.7}$  &  24.9$^{+1.2}_{-1.2}$  &  269.0$^{+6.0}_{-6.0}$  &  $<$66  &  157$^{+75}_{-49}$  &  85$^{+15}_{-15}$  &  179$^{+33}_{-33}$  &  -  &  HDF$\underline{~}$PM3$\underline{~}$39 \\
        ID54  &  12 37 02.74  &  +62 14 02.02  &  1.2463  &  62.6$^{+1.9}_{-1.9}$  &  60.0$^{+1.7}_{-1.7}$  &  40.3$^{+2.8}_{-2.8}$  &  49.5$^{+1.6}_{-1.6}$  &  334.0$^{+7.6}_{-7.6}$  &  -  &  144$^{+73}_{-47}$  &  61$^{+15}_{-15}$  &  417$^{+33}_{-33}$  &  -  &  HDF$\underline{~}$PM3$\underline{~}$44 \\
        ID55  &  12 36 57.79  &  +62 14 55.28  &  0.8493  &  45.9$^{+1.6}_{-1.6}$  &  31.8$^{+1.2}_{-1.2}$  &  33.4$^{+2.5}_{-2.5}$  &  31.0$^{+1.2}_{-1.2}$  &  366.0$^{+7.8}_{-7.8}$  &  -  &  225$^{+60}_{-56}$  &  79$^{+15}_{-15}$  &  258$^{+32}_{-32}$  &  -  &  HDF$\underline{~}$PM3$\underline{~}$37 \\
        ID56  &  12 36 38.13  &  +62 11 16.44  &  1.0174  &  52.2$^{+2.3}_{-2.3}$  &  39.6$^{+1.9}_{-1.9}$  &  32.7$^{+3.4}_{-3.4}$  &  31.9$^{+1.9}_{-1.9}$  &  291.0$^{+4.9}_{-4.9}$  &  -  &  212$^{+58}_{-55}$  &  56$^{+15}_{-15}$  &  287$^{+39}_{-39}$  &  -  &  HDF$\underline{~}$PM3$\underline{~}$9 \\
        ID57  &  12 36 54.64  &  +62 11 27.43  &  0.2542  &  50.1$^{+2.0}_{-2.0}$  &  43.7$^{+1.8}_{-1.8}$  &  27.1$^{+2.8}_{-2.8}$  &  118.2$^{+2.9}_{-2.9}$  &  173.0$^{+6.0}_{-6.0}$  &  -  &  42$^{+29}_{-09}$  &  180$^{+15}_{-15}$  &  103$^{+33}_{-33}$  &  -  &  HDF$\underline{~}$PS3$\underline{~}$23  \\
        ID58  &  12 36 42.19  &  +62 15 45.77 &  0.8572  &  126.6$^{+2.7}_{-2.7}$  &  102.6$^{+2.3}_{-2.3}$  &  106.1$^{+4.5}_{-4.5}$  &  121.0$^{+2.3}_{-2.3}$  &  850.0$^{+7.5}_{-7.5}$  &  -  &  459$^{+46}_{-86}$  &  190$^{+15}_{-15}$  &  457$^{+32}_{-32}$  &  -  &  HDF$\underline{~}$PS3$\underline{~}$10 \\
        ID59  &  12 37 04.33  &  +62 14 46.58  &  2.2110  &  5.6$^{+0.6}_{-0.6}$  &  8.5$^{+0.7}_{-0.7}$  &  15.7$^{+1.8}_{-1.8}$  &  39.5$^{+1.4}_{-1.4}$  &  376.0$^{+4.1}_{-4.1}$  &  -  &  80$^{+74}_{-19}$  &  97$^{+15}_{-15}$  &  187$^{+32}_{-32}$  &  -  &  HDF$\underline{~}$PS3$\underline{~}$36 \\
        \hline
        \hline
      \end{tabular}
    \end{center}
  \end{table}
\end{landscape}

\setcounter{figure}{1}
\begin{figure*}
  \HideDisplacementBoxes
  \hSlide{-5.5cm}\vSlide{+0.0cm}\ForceHeight{21.0cm}\BoxedEPSF{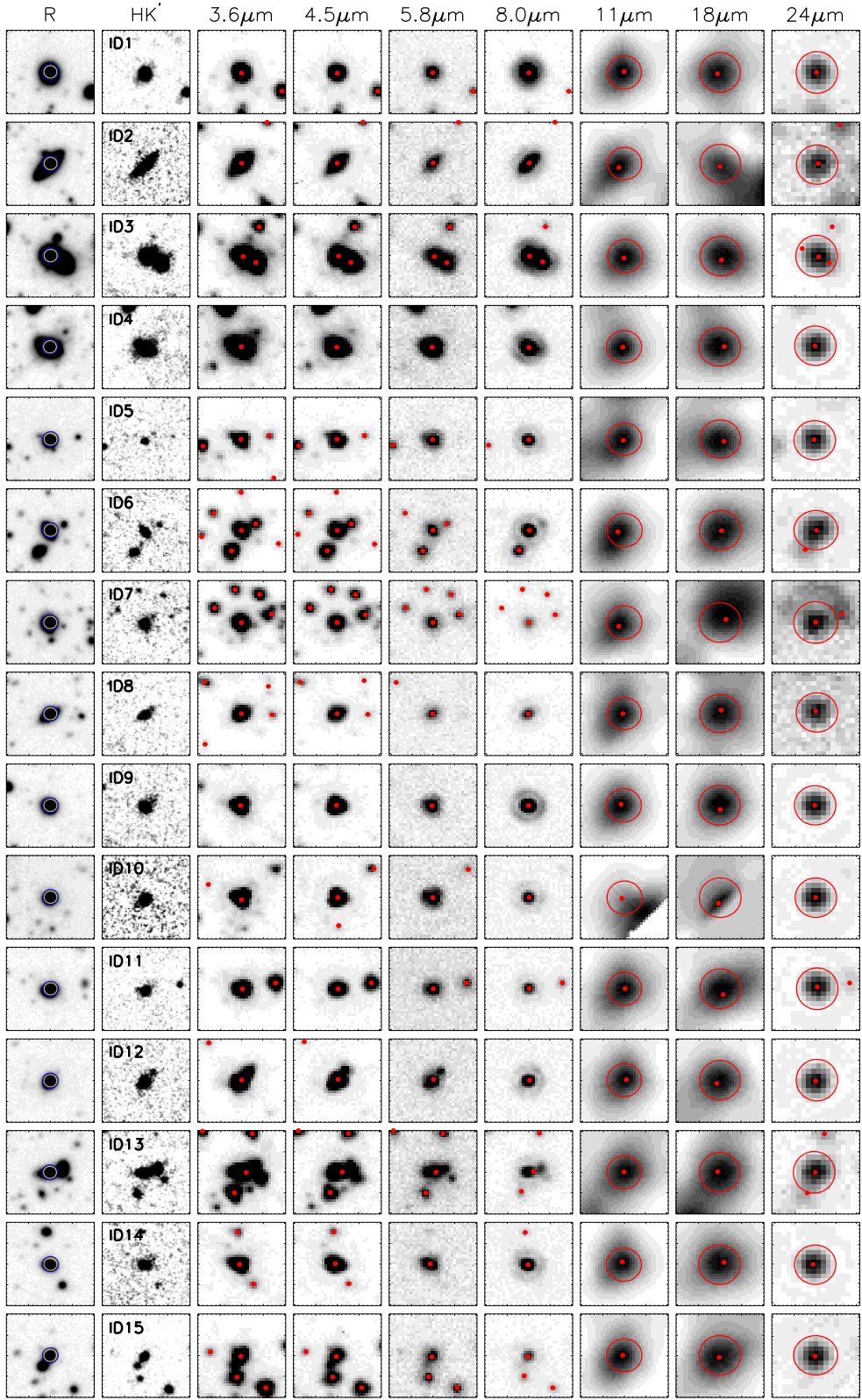}
  \vskip0.0truecm\caption{20$^{\prime\prime}\times20^{\prime\prime}$
    postage stamp images centered on the infrared sources listed in
    Table~\ref{tab:realcat_info}. From left to right: R and
    HK$^{\prime}$ optical images, IRAC images at 3.6, 4.5, 5.8
    8.0$\,\mu$m, {\it AKARI} images at 11 and 18$\,\mu$m, and MIPS
    image at 24$\,\mu$m. Red dots in the {\it Spitzer} and {\it AKARI}
    images represent 3$\sigma$ detections at the corresponding
    wavebands while the circle is the ``error''-circle of
    2$\sigma_{\rm FWHM}$-radius used for cross-matching the
    spectroscopic catalogue with the infrared catalogues,
    i.e. 2$\sigma_{\rm FWHM}$ = 4.08$^{\prime\prime}$,
    4.84$^{\prime\prime}$ and 5.10$^{\prime\prime}$ at 11, 18 and
    24$\,\mu$m, respectively. A circle of $1.5^{\prime\prime}$ in the
    $R$ band image indicates the spectroscopic counterpart.}
  \label{fig:PostageStamps}
\end{figure*}

\setcounter{figure}{1}
\begin{figure*}
  \HideDisplacementBoxes
  \hSlide{-5.5cm}\vSlide{+0.0cm}\ForceHeight{21.0cm}\BoxedEPSF{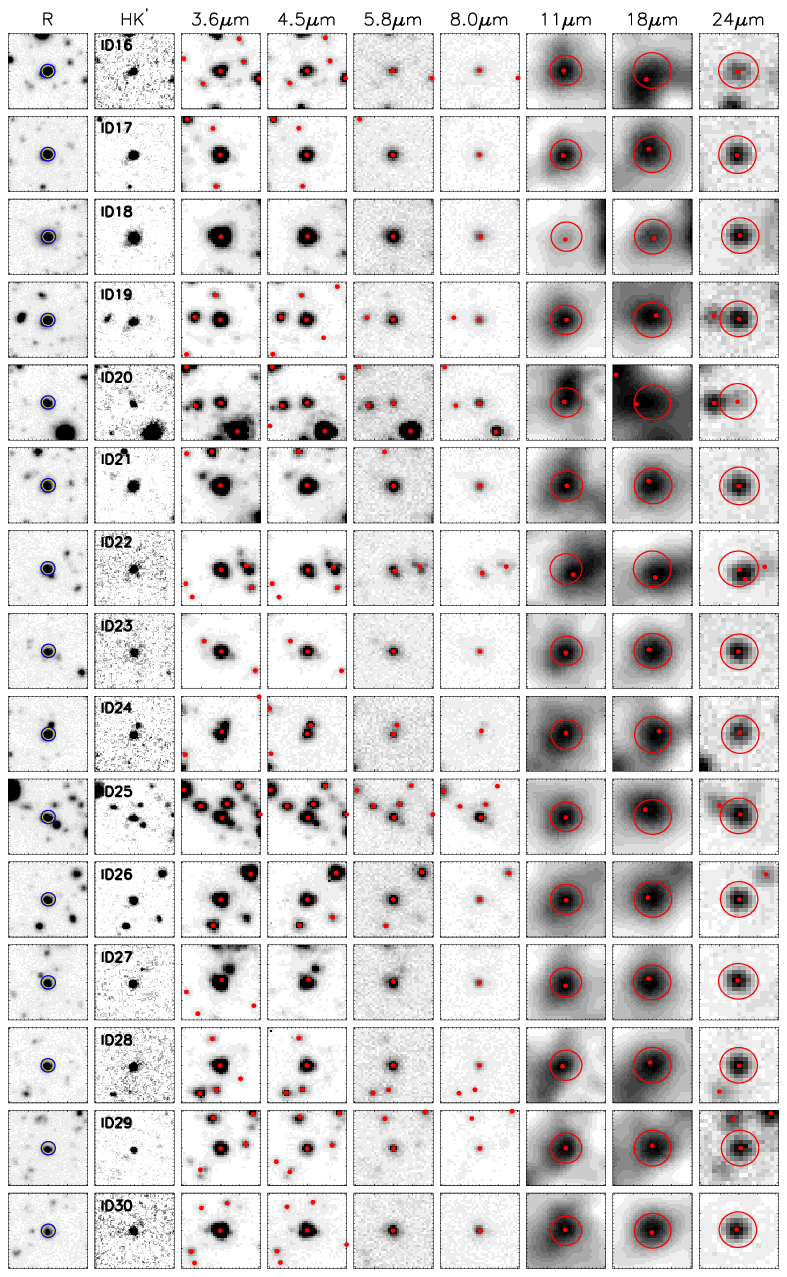}
  \vskip0.0truecm\caption{{\it Continued.}}
\end{figure*}

\setcounter{figure}{1}
\begin{figure*}
  \HideDisplacementBoxes
  \hSlide{-5.5cm}\vSlide{+0.0cm}\ForceHeight{21.0cm}\BoxedEPSF{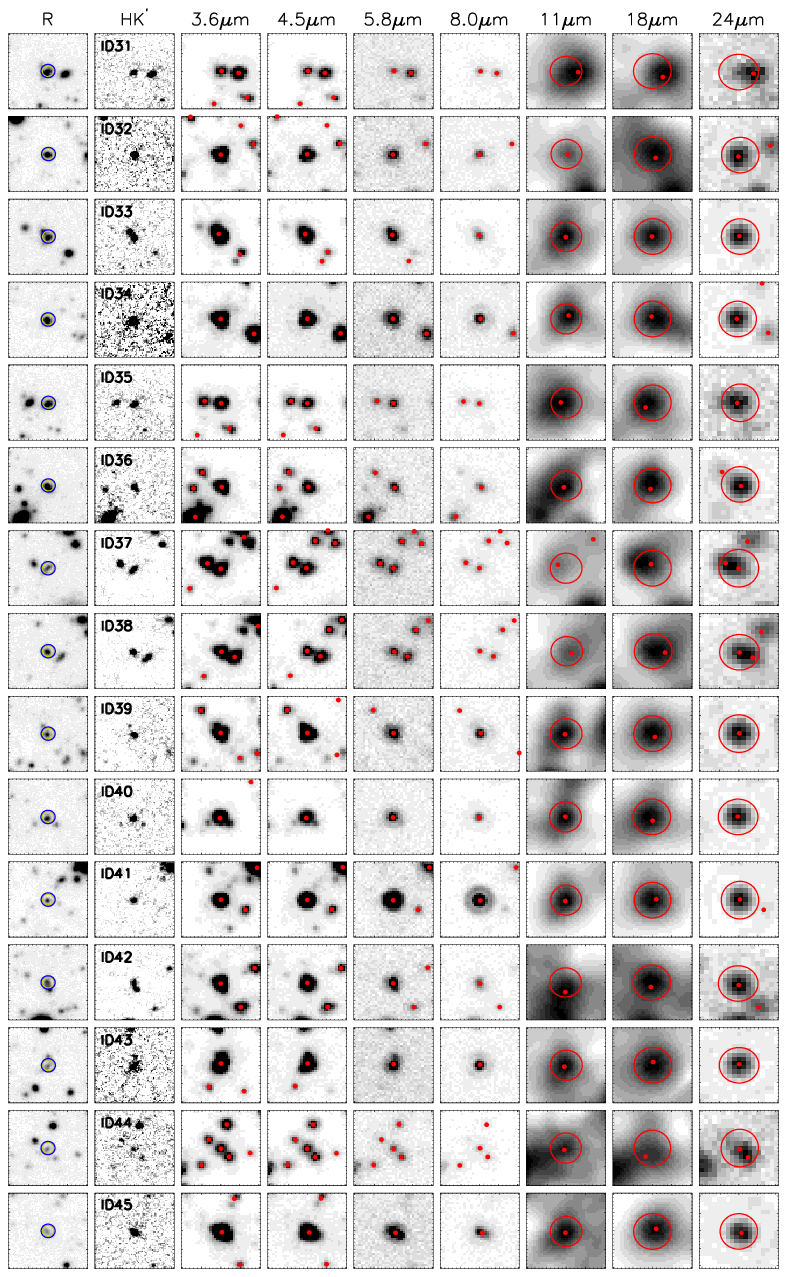}
  \vskip0.0truecm\caption{{\it Continued.}}
\end{figure*}

\setcounter{figure}{1}
\begin{figure*}
  \HideDisplacementBoxes
  \hSlide{-5.5cm}\vSlide{+0.0cm}\ForceHeight{21.0cm}\BoxedEPSF{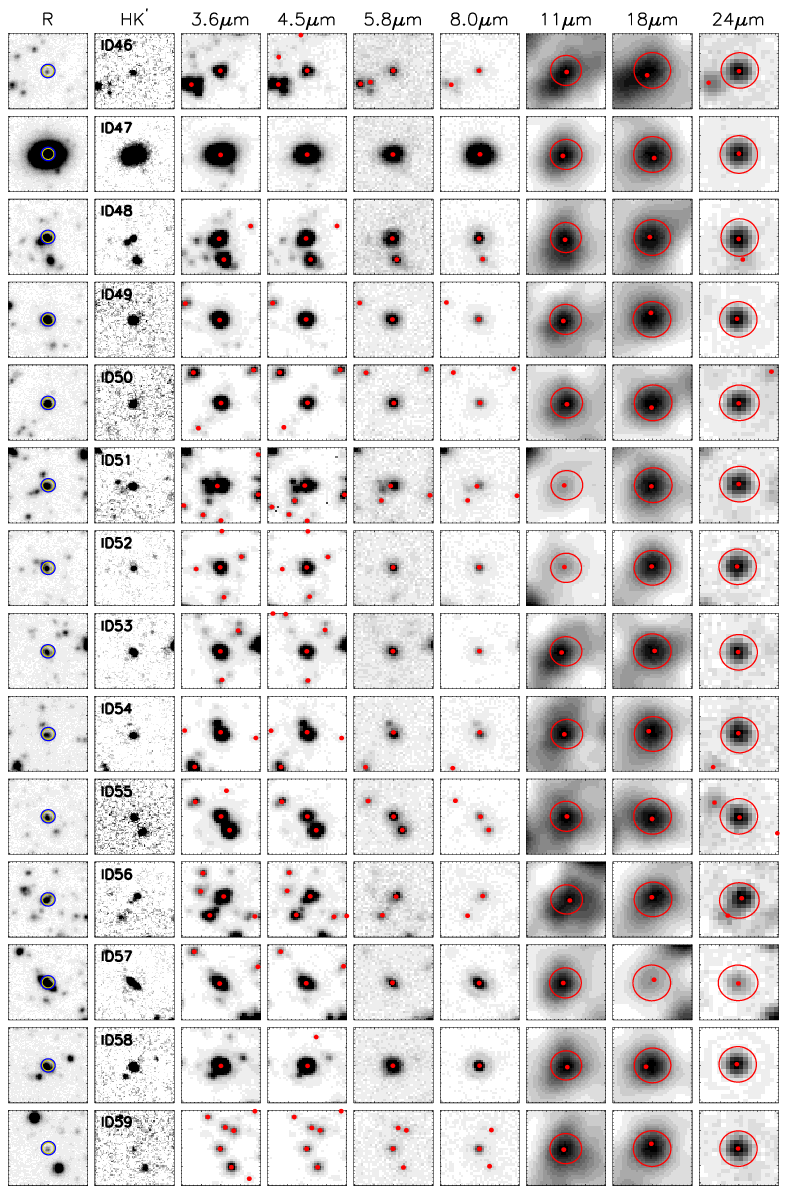}
  \vskip0.0truecm\caption{{\it Continued}.}
\end{figure*}

\subsection{Cross-matching}

We have cross-matched the spectroscopic catalogues with each of the
infrared catalogues mentioned above by using for the matching radius a
value equal to twice the Gaussian rms width $\sigma_{\rm FWHM}$
(=FWHM/2$\sqrt{2\ln{2}}$) of the instrument beam, where $\sigma_{\rm
  FWHM}$ is 6.4$^{\prime\prime}$ for {\it ISO}CAM at 15$\mu$m,
0.68$^{\prime\prime}$, 0.72$^{\prime\prime}$, 0.81$^{\prime\prime}$,
0.85$^{\prime\prime}$ for IRAC at 3.6, 4.5, 5.8, 8.0$\mu$m
respectively, 2.55$^{\prime\prime}$ for MIPS at 24$\mu$m,
1.53$^{\prime\prime}$ for IRS at 16 $\mu$m, 2.04$^{\prime\prime}$ and
2.42$^{\prime\prime}$ for the {\it AKARI} at 11 and 18$\mu$m,
respectively. In order to ensure full spectral coverage from mid-
  to far-IR wavelengths we kept only spectroscopic sources with a
counterpart in {\it all} of the {\it Spitzer} and {\it AKARI} infrared
catalogues considered here plus a counterpart in either the {\it ISO}
15$\,\mu$m or the IRS 16$\,\mu$m catalogues. The advantage of
  covering the whole mid- to far-IR spectral range is made clear by
  looking at the middle panel of Fig.~\ref{fig:sed_vs_filters}, where
  the {\it Spitzer}, {\it AKARI} and {\it ISO} filters are shown: all
  the main PAH and silicate features are sampled up to $z\sim2$ with
  the photometric data exploited here. At $z\gsim2$ the PAH features
are shifted outside the wavelength range covered by {\it Spitzer} and
{\it AKARI} so that they can no longer be exploited as redshift
indicators. In this case, the rest-frame 1.6$\,\mu$m bump can be used
instead for photometric redshift estimates (Sawicki 2002).

We ended up with a sample of 59 objects; 22 of them have flux
measurements at both 15$\,\mu$m and 16$\,\mu$m.  {\it ISO} 6.5$\,\mu$m
flux measurements (or upper limits) have been included, when
available, in the SED fitting process.  The redshifts of the sources
and their infrared fluxes are listed in
Table~\ref{tab:realcat_info}. We also provide in
Fig.~\ref{fig:PostageStamps} postage stamp images for all the sources
in our sample at $R$, $HK^{\prime}$, {\it Spitzer} and {\it AKARI}
wavebands.

For almost all of the sources in the sample we found a single
counterpart at each infrared waveband. However few of them appear to
lie in crowded regions, possibly affecting the infrared flux at the
longest wavebands, see e.g. ID3, ID20, ID25, ID31, ID35, ID37, ID38
and ID44. There are in particular pairs of objects ``sharing'' the
same {\it AKARI} (and {\it ISO}) fluxes, i.e. ID31-ID35 and
ID37-ID38. In this cases the {\it AKARI} (and {\it ISO}) photometry
should be taken as an upper limit to the intrinsic flux of the source
at those wavelengths.

Most of the sources in the sample lie at $z\lsim1.5$. Only 3 out of 59
have $z_{\rm spec}>1.5$: ID5 ($z_{\rm spec}=2.5920$), ID41 ($z_{\rm
  spec}=2.0050$) and ID59 ($z_{\rm spec}=2.2110$). We found that
  just five objects in the Reddy et al. catalogue have a counterpart
  at both 11 and 18$\,\mu$m {\it and} either at 15 or 16$\,\mu$m. They
  are (according to the names they have in the Reddy et al. source
  list): BX1321 (at z=0.139), BM1156 (at 2.211), BM1299 (at z=1.595),
  BM1326 (at 1.268), and MD39 (at 2.583). Four of them are in common
  with the TKTS spectroscopic catalogue and fall into the final sample
  of 59 infrared sources presented here, corresponding to ID1, ID5,
  ID52 and ID59. MD39 is instead missing from our final sample because
  it lacks a counterpart in our IRAC catalogues at both 3.6 and
  8.0$\,\mu$m. MD39 appears to be the source very close to ID24 in the
  corresponding postage stamp image of Fig.~\ref{fig:PostageStamps},
  at wavelength $\lambda\lsim5.8\,\mu$m. The pair of objects is not
  resolved at longer wavelengths and indeed Reddy et al. do not
  provide a 24$\,\mu$m flux estimate for it, probably because it
  cannot easily be deblended from the nearby ID24 source (although
  they provide for it flux estimates at all the four IRAC
  wavebands). For this reason we have decided not to include MD39 in
  our final catalogue and to add instead ID24 to the list of our
  infrared sources with uncertain far-IR photometry.

\setcounter{figure}{2}
\begin{figure*}
  \vspace{-2.5cm} \hspace{+3.5cm}
  \includegraphics[height=12.0cm,width=17.5cm]
  {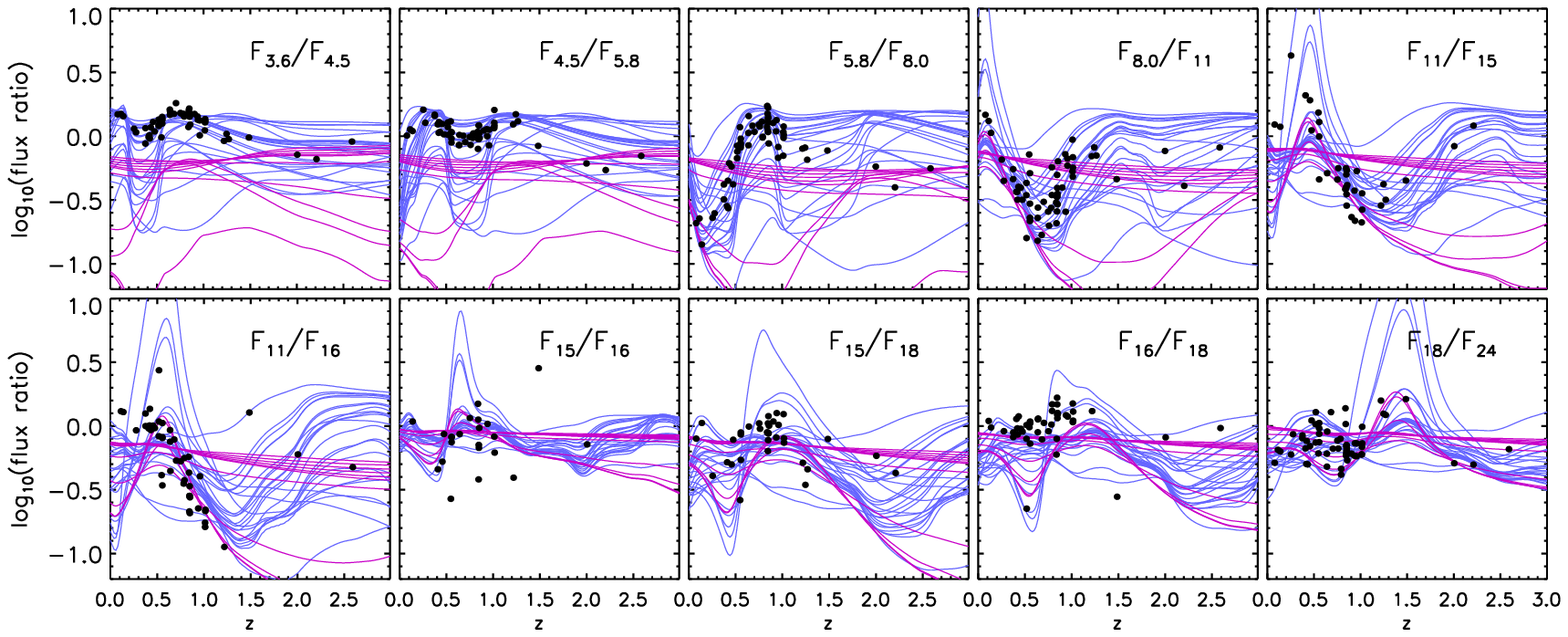}
  \vskip-2.3truecm\caption{Observed flux ratios as a function of
    spectroscopic redshift (black dots) for the sample of infrared
    sources extracted from the GOODS-N field. Blue lines show the
    predicted colors from the starburst models of Takagi et al. for a
    subsample of representative values of the sed model parameters
    (see text). The predicted colors from the AGN templates exploited
    in the SED fitting are shown in magenta.}
  \label{fig:colors_vs_zspec_realcat}
\end{figure*}

Fig.~\ref{fig:colors_vs_zspec_realcat} shows the observed flux
  ratios as a function of the spectroscopic redshifts (black dots),
  and compare them to the predictions from the starburst reference templates
  for a representative set of SED model
  parameters (blue curves), and from the AGN reference templates (megenta
  curves). The effects of the PAH 6.2$\,\mu$m emission feature and of the
  9.7$\,\mu$m silicate absorption are clearly manifest in the data. The
  deep ``well'' observed in the F$_{5.8}$/F$_{8.0}$,
  F$_{8.0}$/F$_{11}$, F$_{11}$/F$_{15}$ and F$_{11}$/F$_{16}$ diagrams
  at $z\sim0.2$, $z\sim0.6$, $z\sim1.2$ and $z\sim1.4$ is due to the
  passage of the PAH 6.2$\,\mu$m feature through the 8.0, 11, 15 and
  16$\,\mu$m wavebands, respectively.
The 9.7$\,\mu$m silicate absorption manifests itself as a
  bump in the 8.0/11$\,\mu$m flux ratios around $z\sim0$ where the
  feature falls into the AKARI 11$\,\mu$m passband. As the same
  feature enters the 15 and 16$\,\mu$m wavebands, which occurs at
  $z\sim0.5$, it produces a significant corresponding peak in the
  F$_{11}/$F$_{15}$ and F$_{11}/$F$_{16}$ diagrams. A hint of a bump
  due to the silicate absorption is seen also in the 15/18$\,\mu$m and
  16/18$\,\mu$m flux ratios at $z\sim0.7$, although it is not
  particularly prominent due to the wider wavelength coverage of the
  AKARI 18$\,\mu$m filter compared to the 11$\,\mu$m band. Finally, by
  entering the MIPS 24$\,\mu$m band, the silicate absorption induces
  another significant bump in the 18/24$\,\mu$m flux ratio at
  $z\sim1.5$. Note that such a bump can be used for efficiently
  selecting ultraluminous infrared galaxies in the redshift range
  $1\lsim z\lsim2$ with {\it AKARI} (see Takagi $\&$ Pearson 2005).
  The $\sim10\,\mu$m silicate absorption is also responsible for the
  behaviour in the predicted flux ratios of the AGN templates when the
  torus is seen edge-on. For a face-on AGN the corresponding flux
  ratios are almost independent of redshift, implying that for a
  power-law infrared spectrum the recovery of the redshift from
  mid-/far-IR photometry alone is extremely challenging, if not
  impossible.

  On average the range of flux ratios spanned by the data is accounted
  for by the model with the sole exception of the 15/16$\,\mu$m flux
  ratios. Indeed we observe for few objects a significant steep
  increase of the flux from 15 to 16$\,\mu$m (ID28, ID39, ID42, which
  lie well below the theoretical expectations) or, conversely, a
  notable decrease of the flux when moving from 15 to 16$\,\mu$m
  (ID44).

\section{Results of SED fitting}
\label{sec:results}

The results of the SED fitting are listed in
Table~\ref{tab:SEDfit_results}. A comparison between the photometric
data and the best-fit SED model is shown in Fig.~\ref{fig:SEDs}. The
derived photometric redshifts versus the spectroscopic redshifts are
presented in Fig.~\ref{fig:zphot_vs_zspec_realcat} where sources with
$P_{\chi^2}>1\%$ are indicated by filled circles (42 in total,
i.e. $\sim70\%$ of the whole sample), and those with $P_{\chi^2}<1\%$
are identified by open circles. In the same figure the solid line
marks the ideal case in which $z_{\rm phot}=z_{\rm spec}$, while the
dot-dashed lines delimit the region where $|z_{\rm phot}-z_{\rm
  spec}|/(1+z_{\rm spec})<10\%$. Error bars corresponding to a 99$\%$
confidence limit have been drawn, for reason of clarity, only for
objects with $P_{\chi^2}>1\%$ which lie within the 10$\%$ confidence
region.

For almost all of the sources with $z_{\rm spec}\lsim1.5$ and
$P_{\chi^2}>1\%$ the redshift is recovered with an accuracy $|z_{\rm
  phot}-z_{\rm spec}|/(1+z_{\rm spec})\lsim10\%$. The only exceptions
are ID22 (at $z_{\rm spec}=0.5560$), ID42 (at $z_{\rm spec}=0.5480$)
and ID44 (at $z_{\rm spec}=1.4865$).

For ID22, we would expect the PAH 6.2 and 7.7$\,\mu$m features to enter the
11$\,\mu$m AKARI filter at $z_{\rm spec}\sim0.5$ and therefore to
produce a bump in the measured flux at that waveband. This is for
example what we observe in the spectrum of ID7 and ID15 whose
redshifts are very close to that of ID22: a bump at 11$\mu$m followed
by a plateau at longer wavelengths as a result of the convolution of
the instruments filters whith the redshifted silicate absorption
features at 9.7 and 18$\,\mu$m and with the PAH 11.3$\,\mu$m mission
feature. On the contrary, the SED of ID22 increases monotonically in
the interval 8-18$\,\mu$m. Interestingly, we found that at $z\sim0.5$
the highest contribution to the value of the $\chi^2$ in
Eq.~\ref{eq:chisq} (after minimization over the other SED model
parameters) comes from the {\it AKARI} 11$\,\mu$m waveband. By
removing the data point at 11$\,\mu$m from the measured spectrum and
performing the fit on the other photometric data alone we get $z_{\rm
  phot} = 0.56$, in excellent agreement with the spectroscopic
value. We also tested that none of the flux measurements at
wavelengths $>11\,\mu$m have such an effect on the photo-$z$
estimate. Therefore the failure in the recovery of the redshift for
ID22 seems to be determined exclusively by the AKARI photometry at
11$\,\mu$m.  This result points to the conclusion that the source has
an intrinsically low PAH 7.7$\,\mu$m emission that our reference SED
templates are not able to account for that.


\setcounter{figure}{3}
\begin{figure*}
  \HideDisplacementBoxes
  \hSlide{-3.0cm}\vSlide{+0.0cm}\ForceHeight{4.8cm}\BoxedEPSF{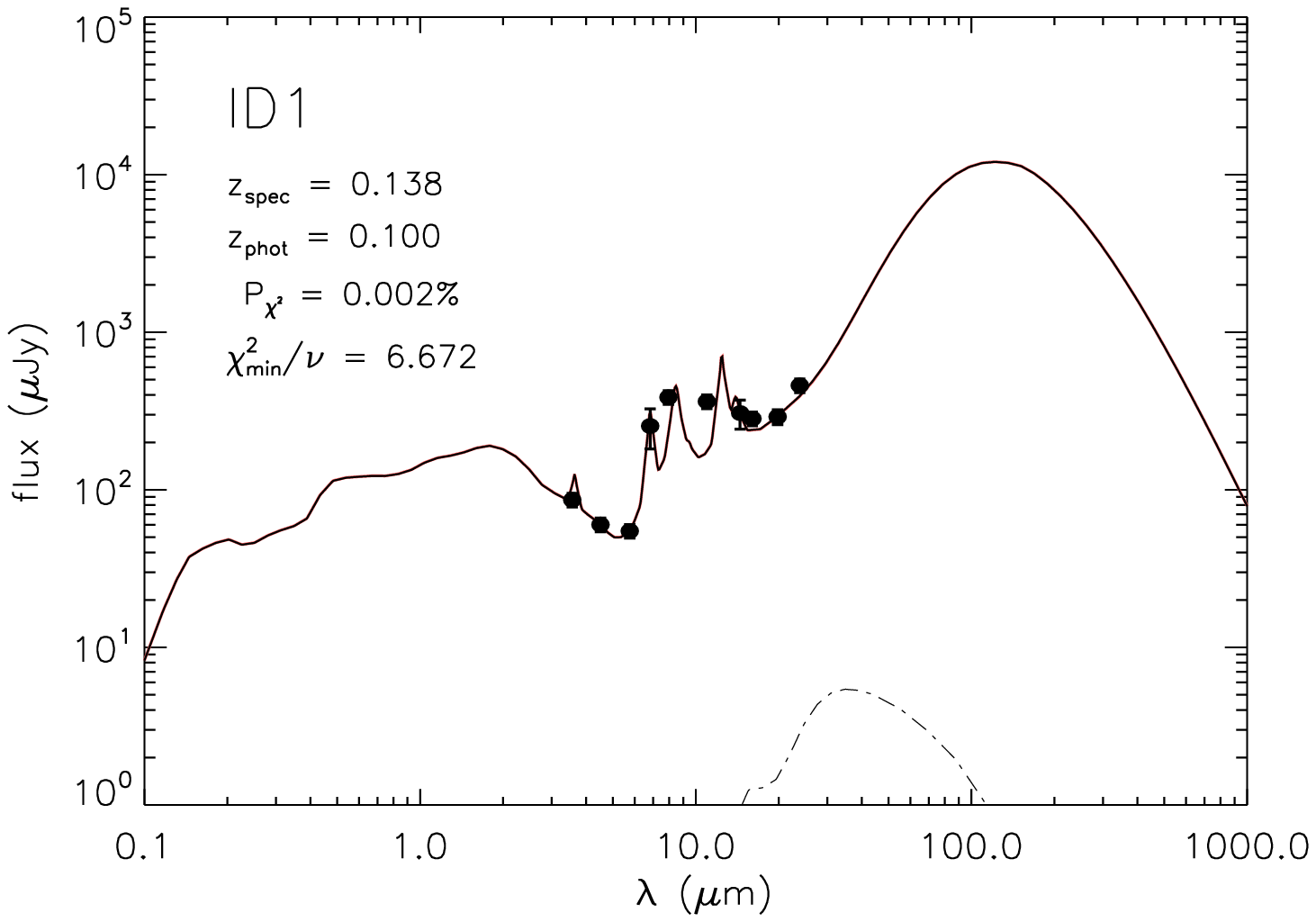}
  \hSlide{+2.0cm}\vSlide{+0.0cm}\ForceHeight{4.8cm}\BoxedEPSF{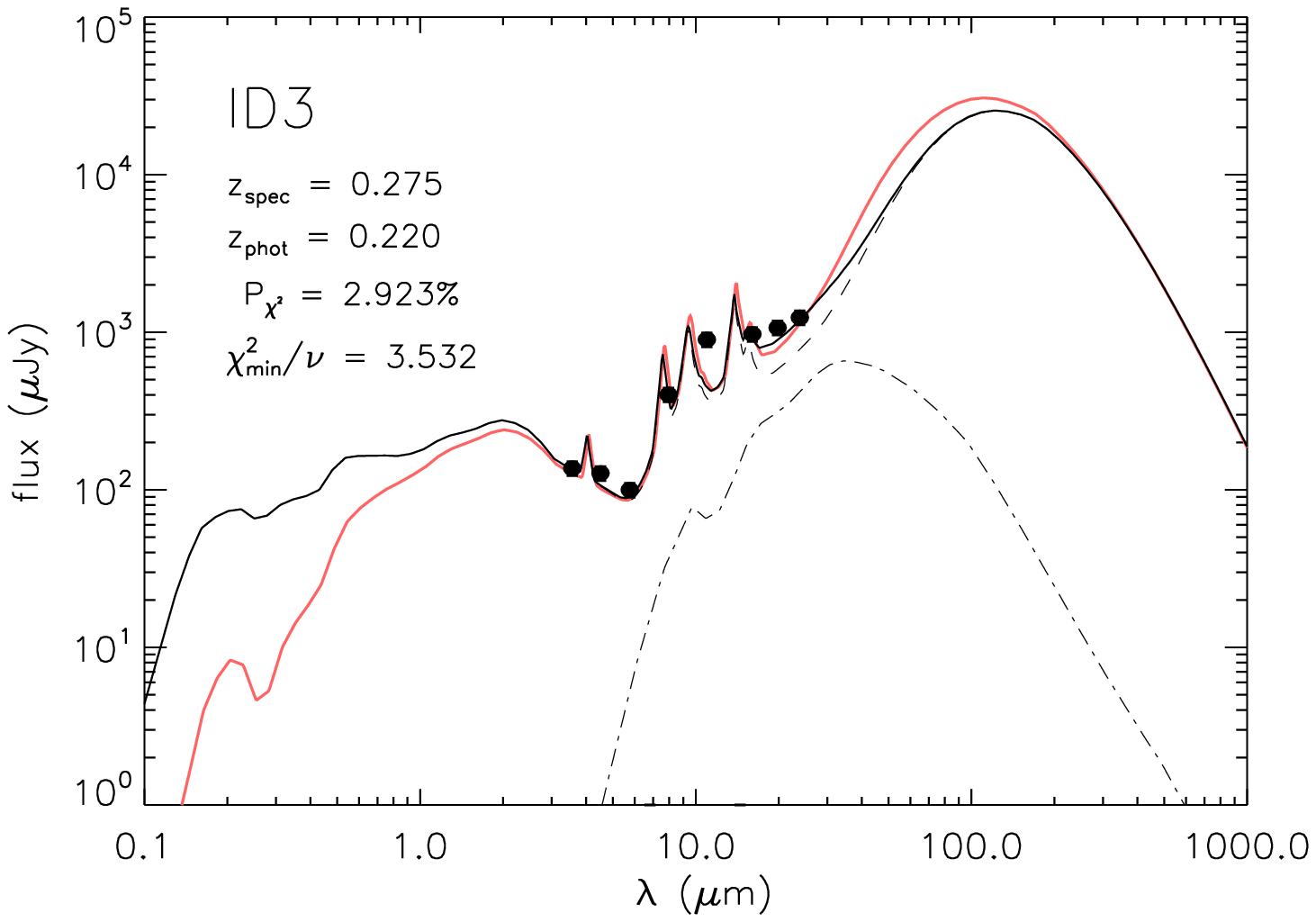}
  \vspace{-5.5cm}\hSlide{-0.5cm}\vSlide{+0.7cm}\ForceHeight{4.8cm}\BoxedEPSF{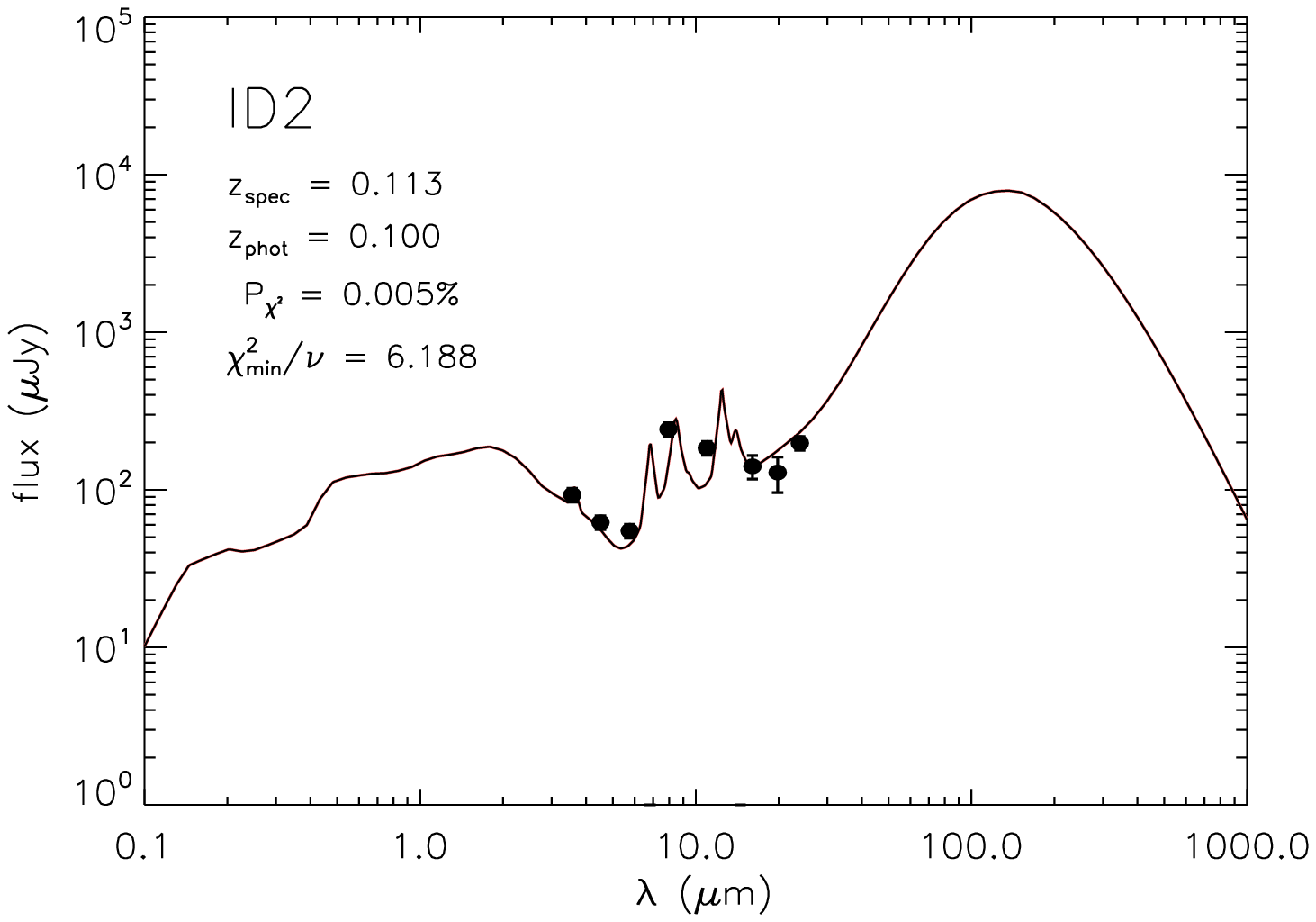} \\
  \hSlide{-3.0cm}\vSlide{+0.0cm}\ForceHeight{4.8cm}\BoxedEPSF{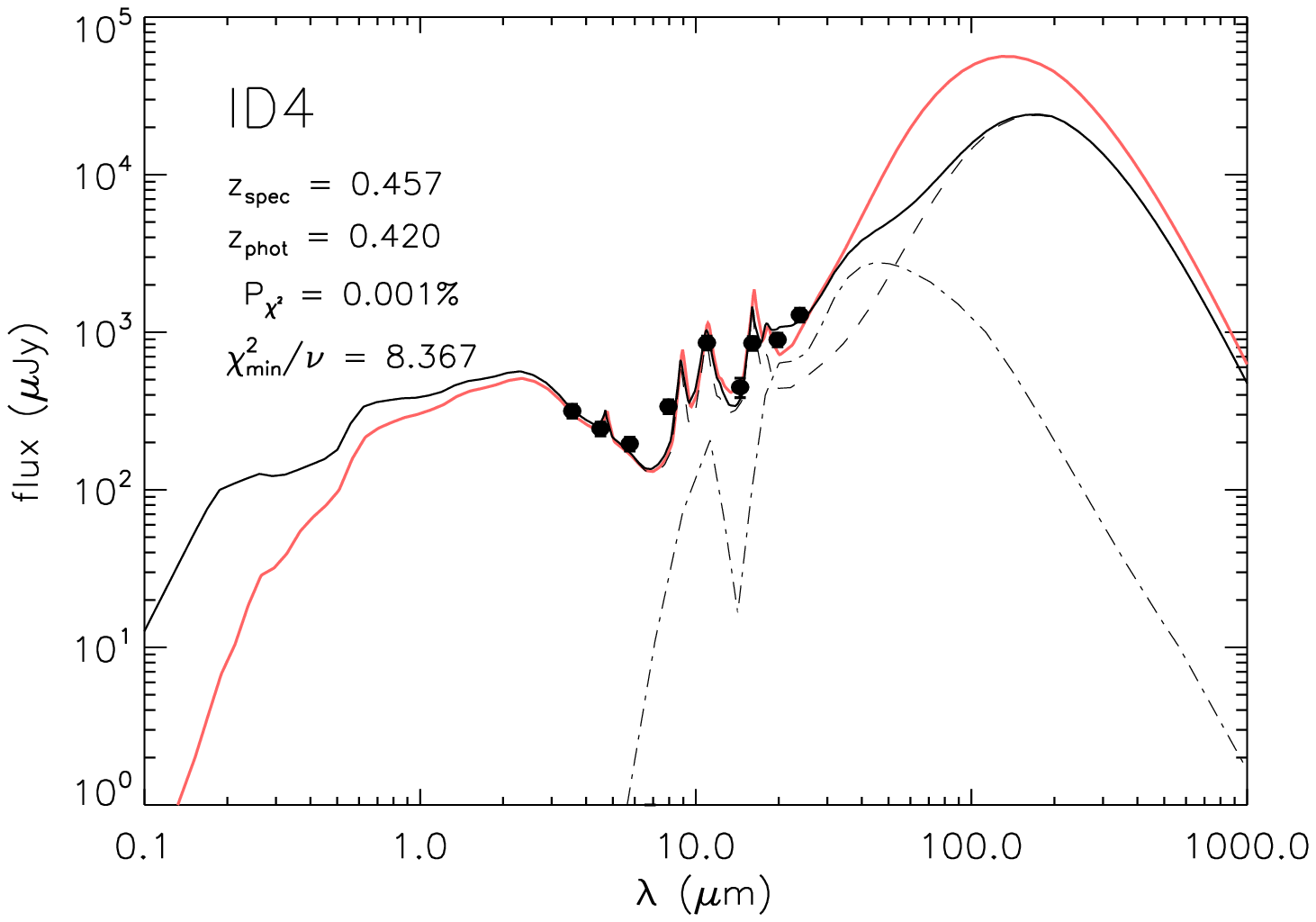}
  \hSlide{+2.0cm}\vSlide{+0.0cm}\ForceHeight{4.8cm}\BoxedEPSF{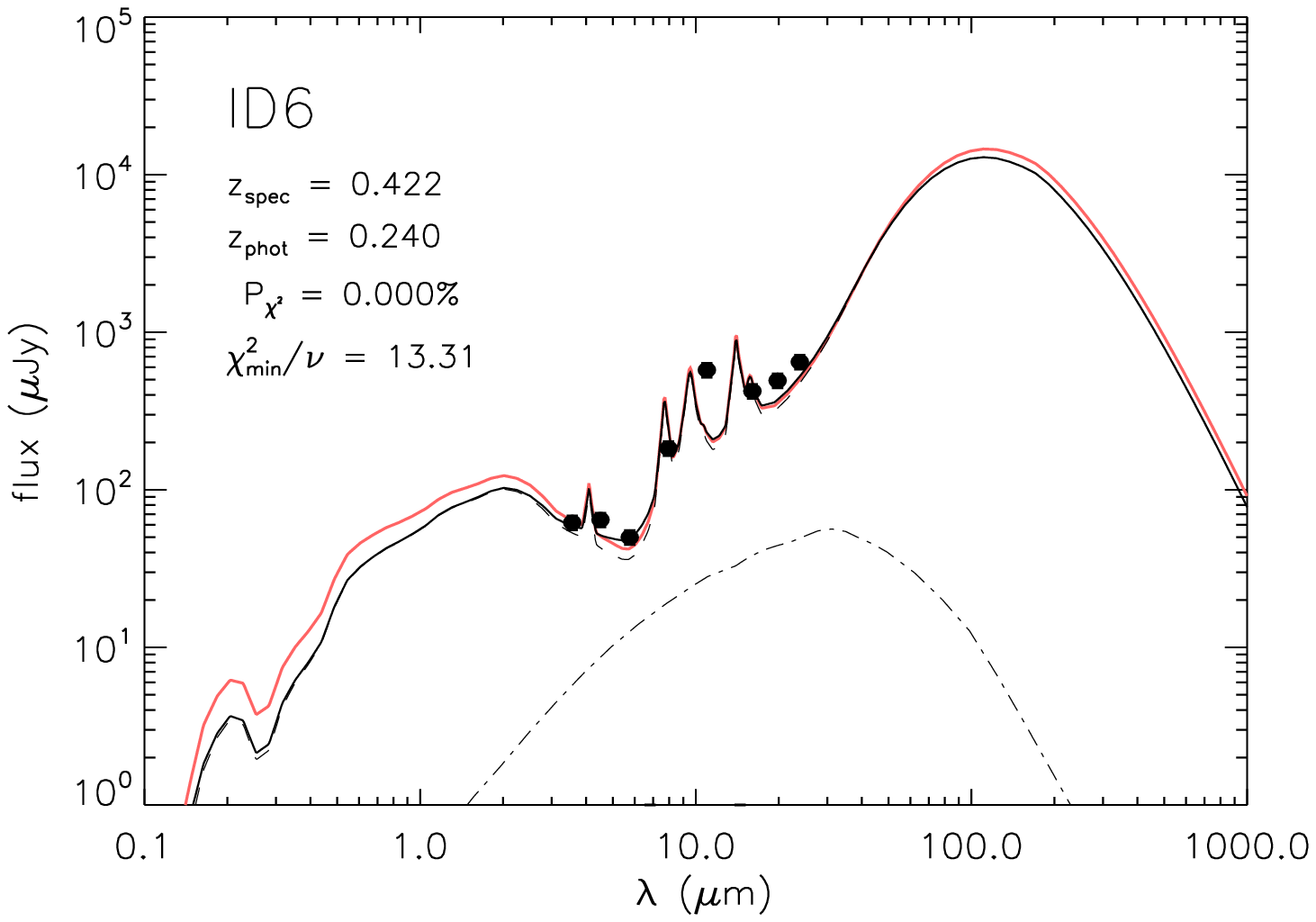}
  \vspace{-5.5cm}\hSlide{-0.5cm}\vSlide{+0.7cm}\ForceHeight{4.8cm}\BoxedEPSF{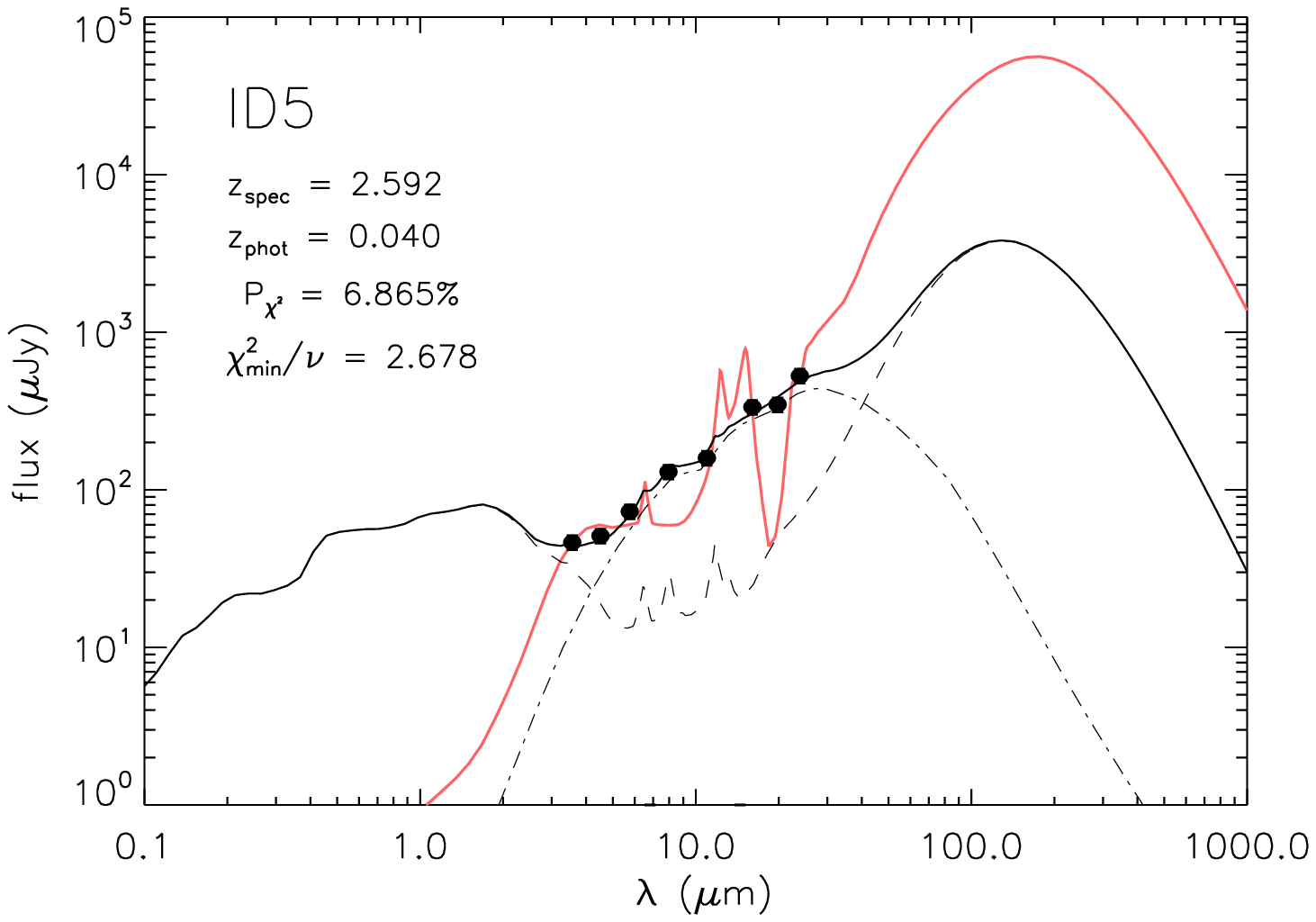} \\
  \hSlide{-3.0cm}\vSlide{+0.0cm}\ForceHeight{4.8cm}\BoxedEPSF{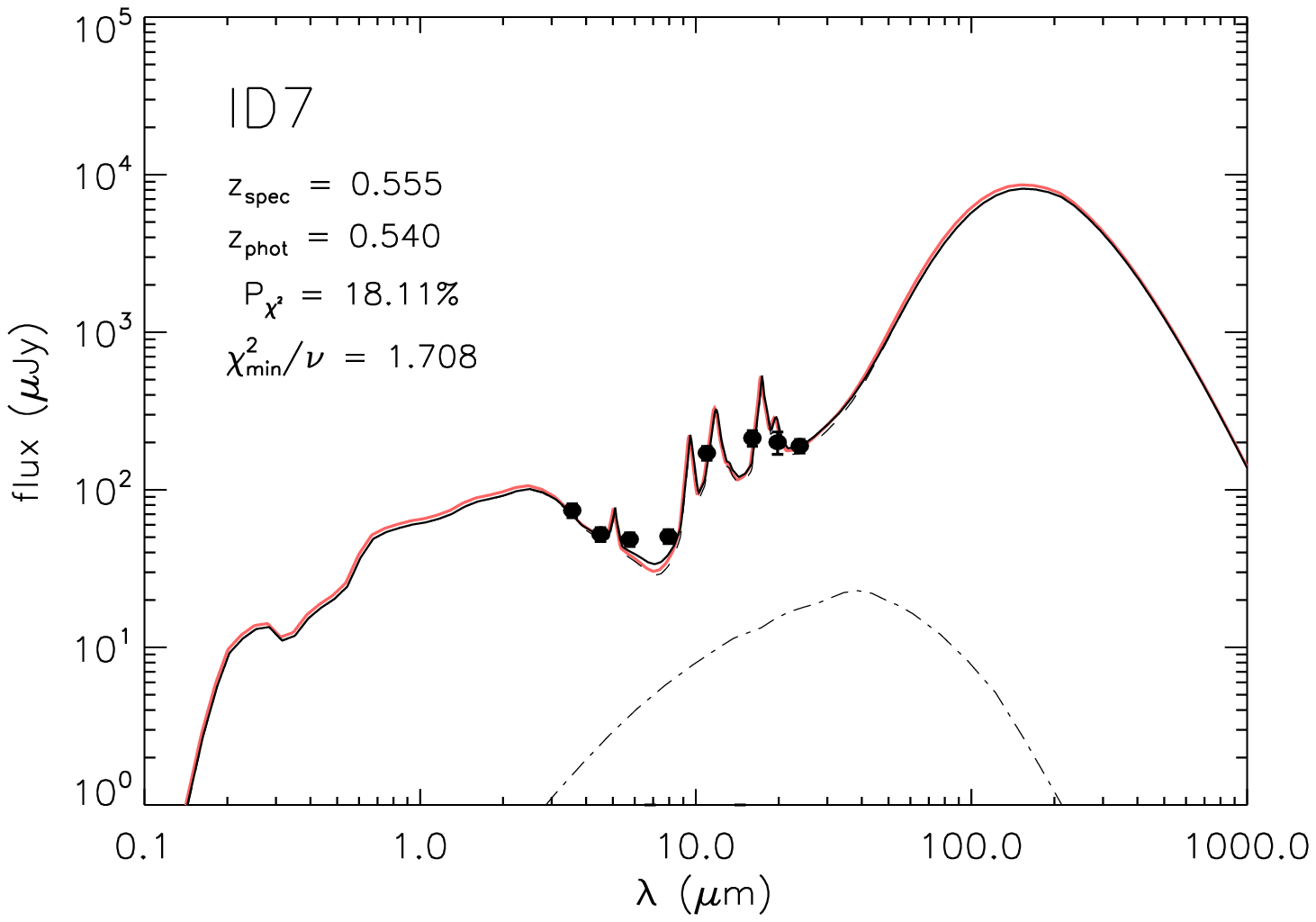}
  \hSlide{+2.0cm}\vSlide{+0.0cm}\ForceHeight{4.8cm}\BoxedEPSF{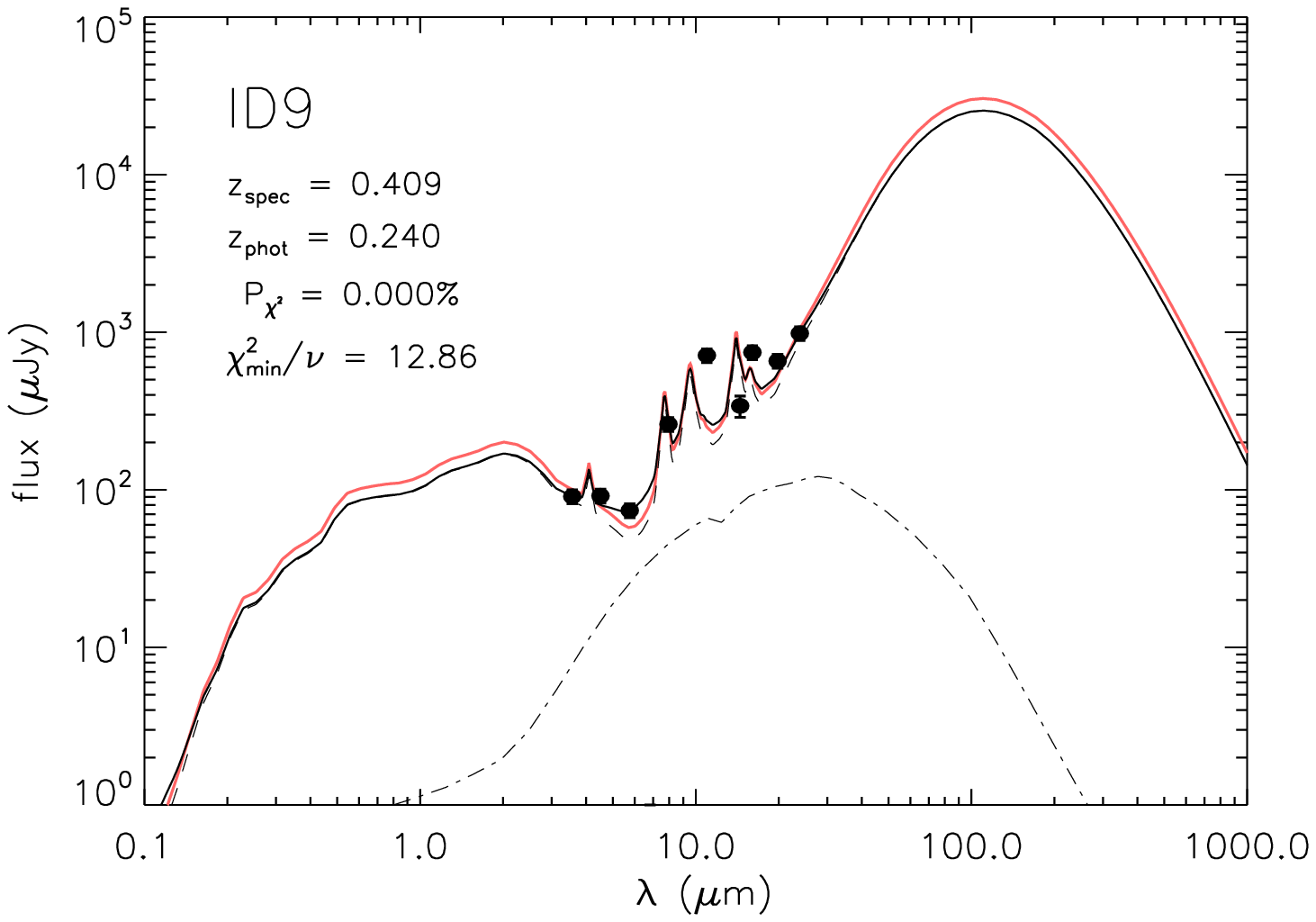}
  \vspace{-5.5cm}\hSlide{-0.5cm}\vSlide{+0.7cm}\ForceHeight{4.8cm}\BoxedEPSF{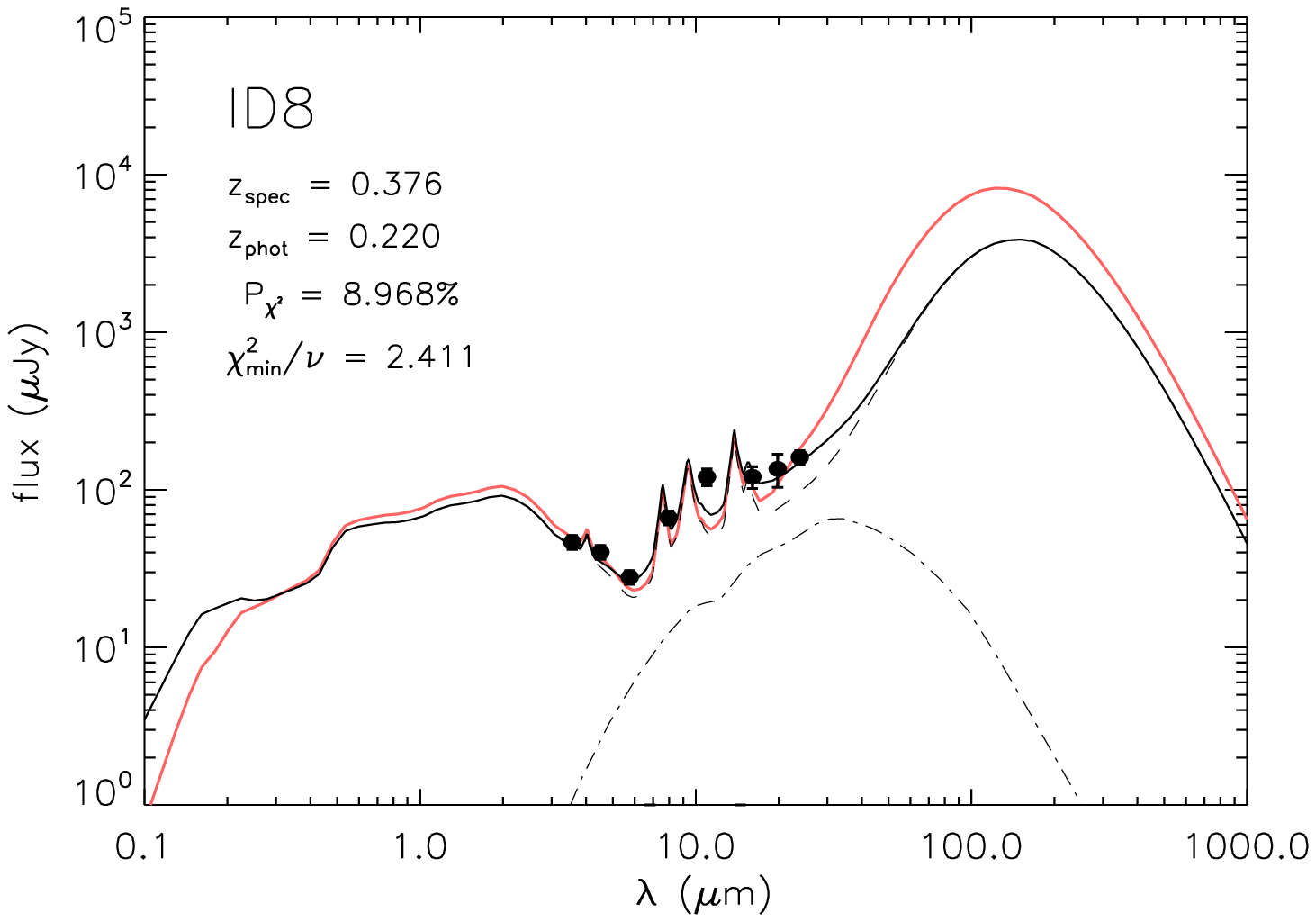} \\
  \hSlide{-3.0cm}\vSlide{+0.0cm}\ForceHeight{4.8cm}\BoxedEPSF{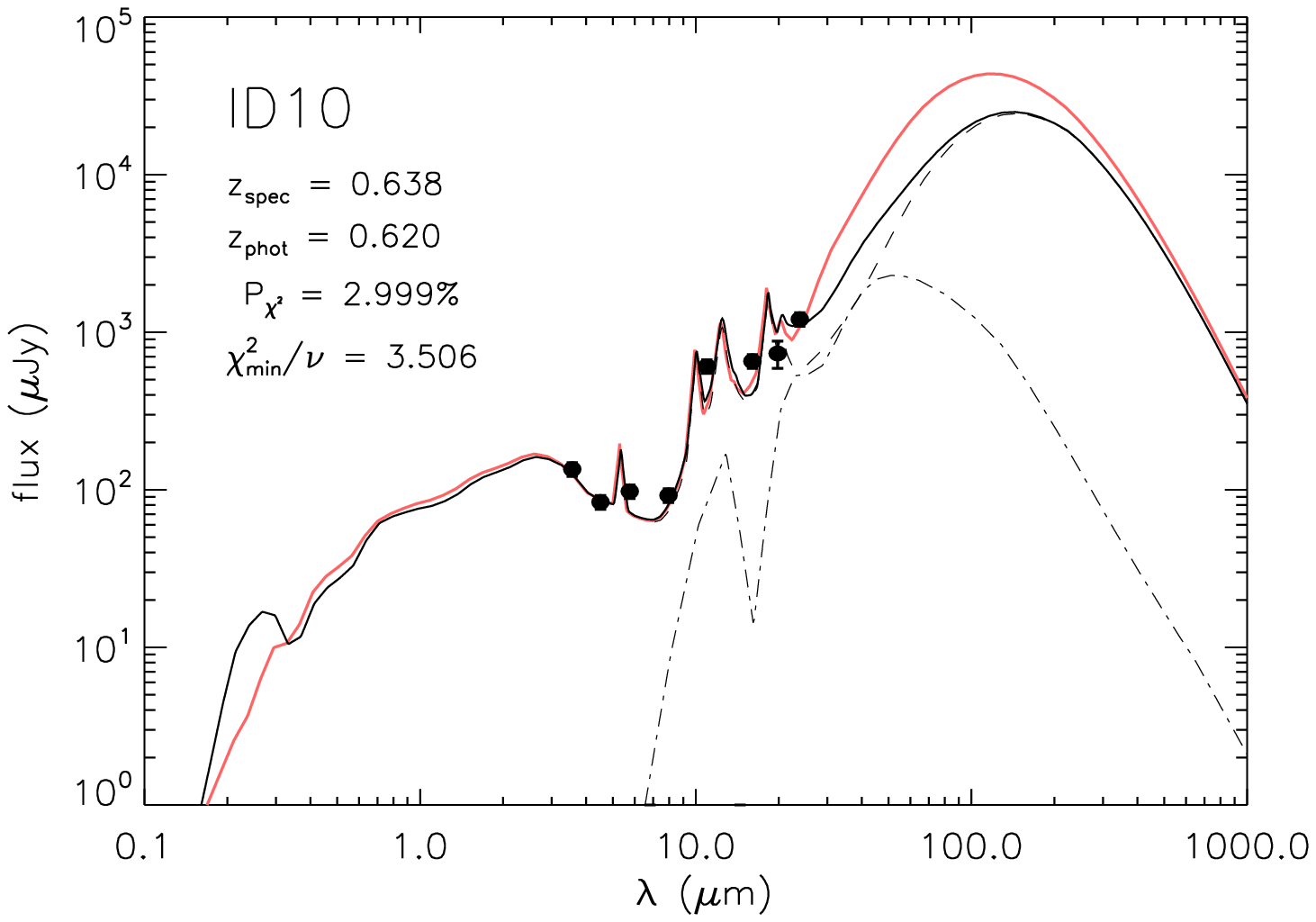}
  \hSlide{+2.0cm}\vSlide{+0.0cm}\ForceHeight{4.8cm}\BoxedEPSF{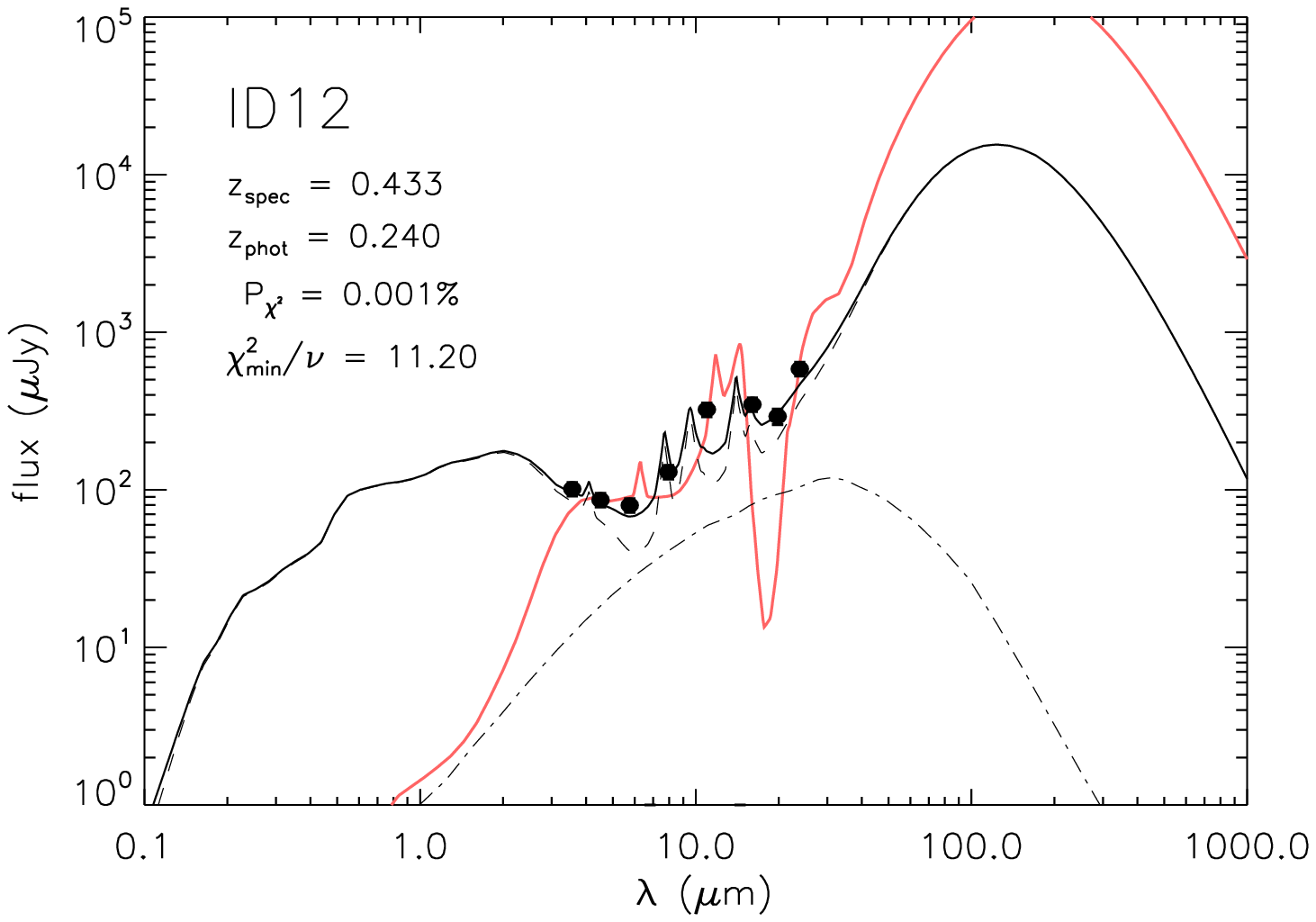}
  \vspace{-5.5cm}\hSlide{-0.5cm}\vSlide{+0.7cm}\ForceHeight{4.8cm}\BoxedEPSF{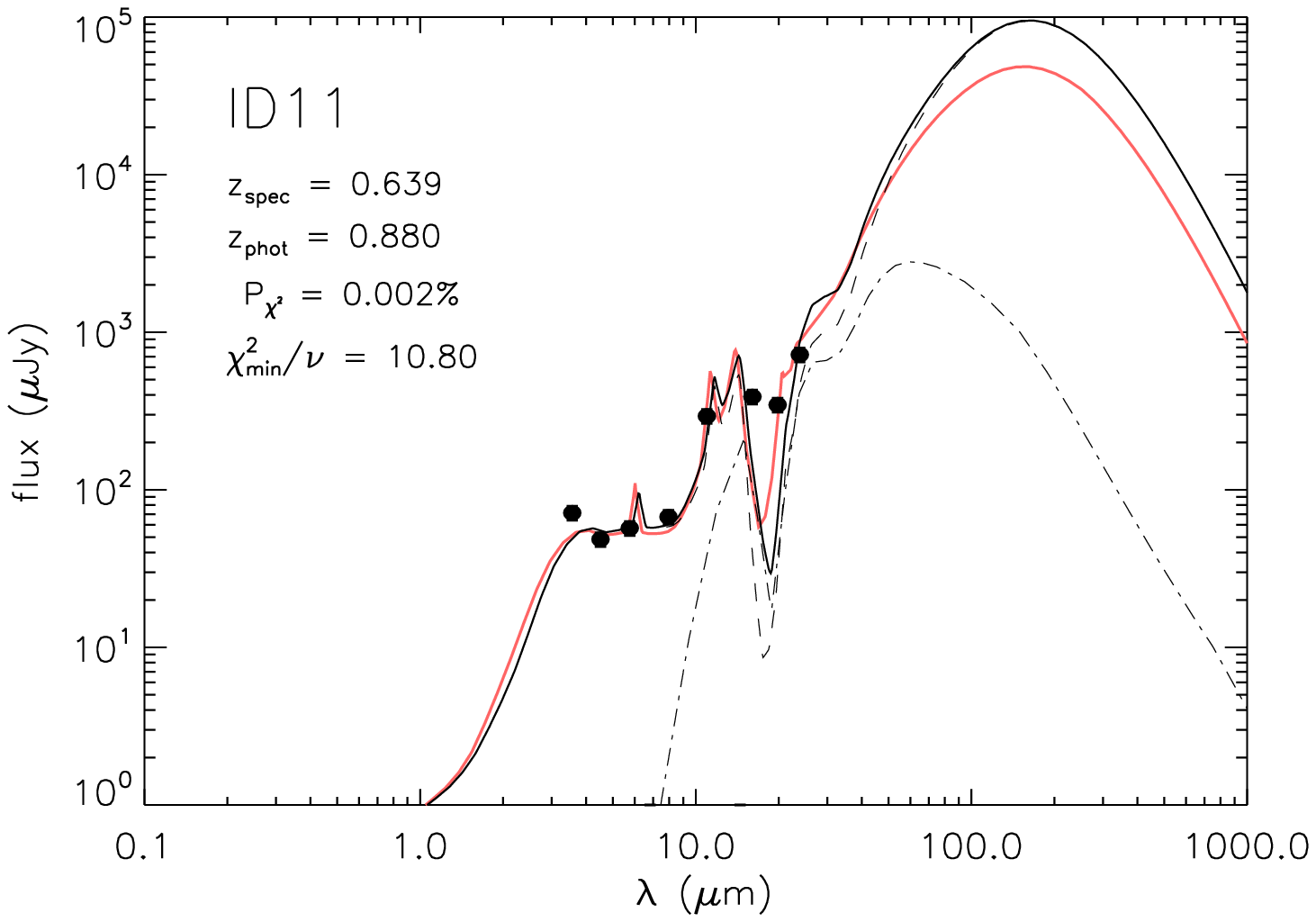} \\
  \hSlide{-3.0cm}\vSlide{+0.0cm}\ForceHeight{4.8cm}\BoxedEPSF{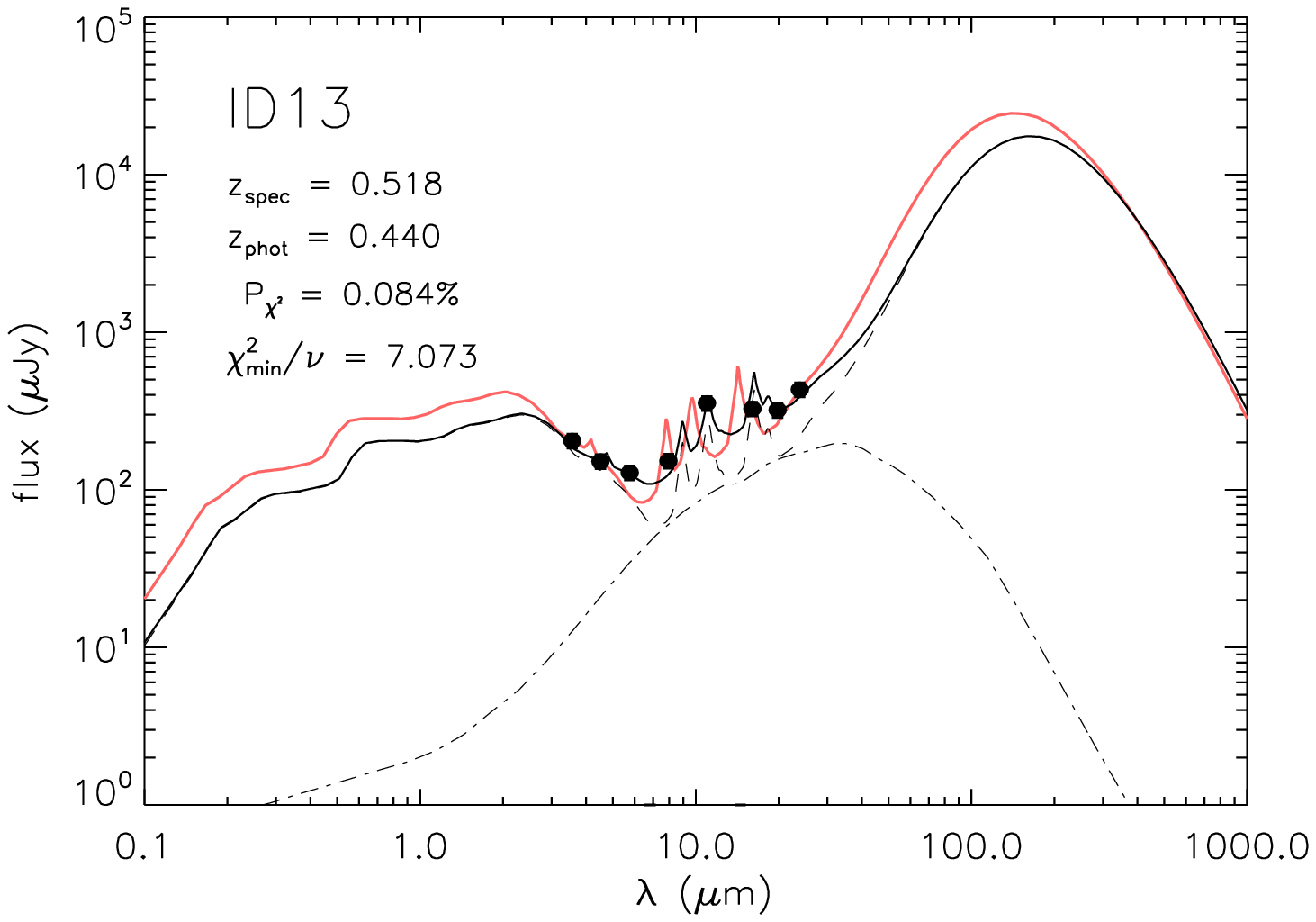}
  \hSlide{+2.0cm}\vSlide{+0.0cm}\ForceHeight{4.8cm}\BoxedEPSF{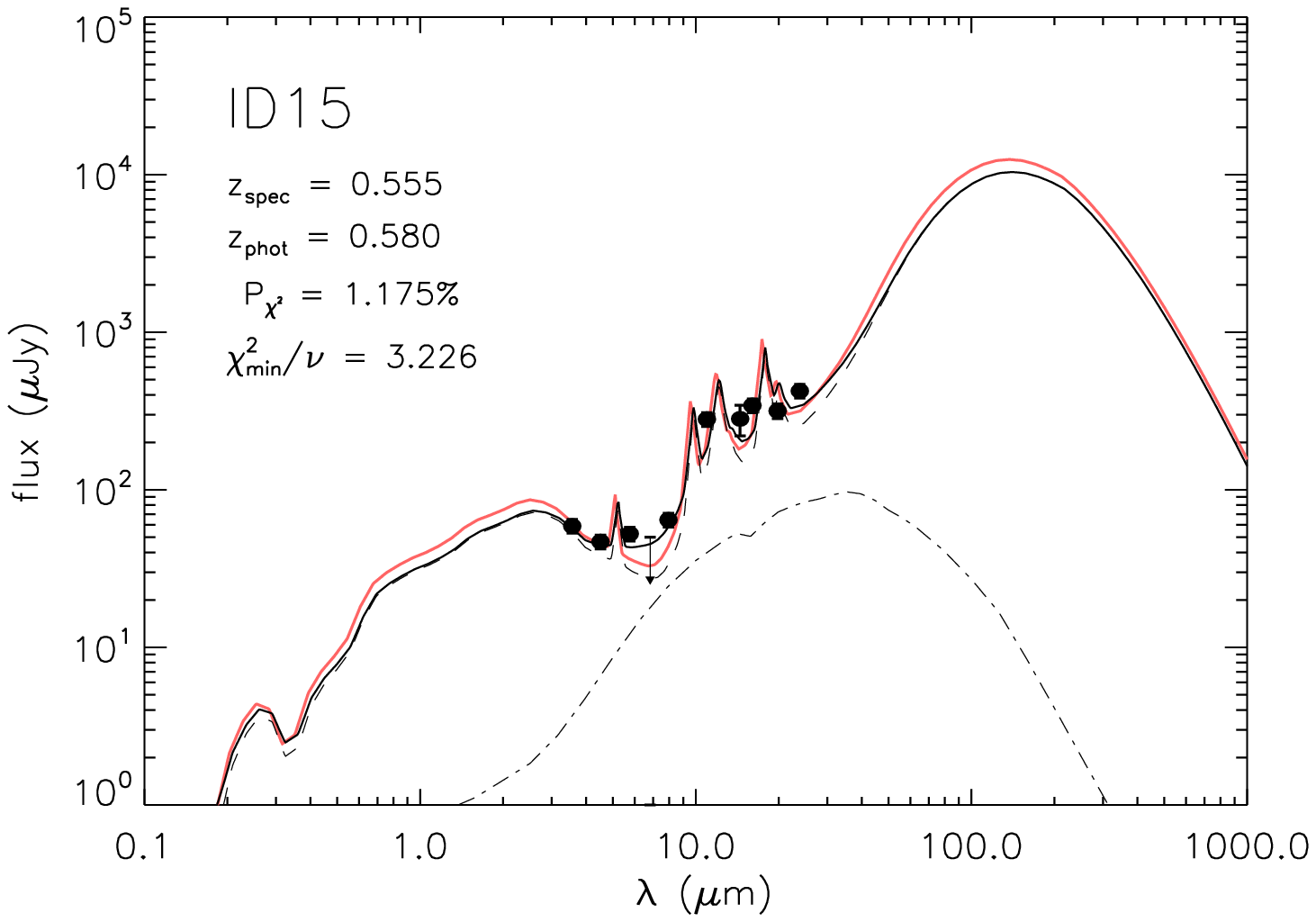}
  \vspace{-5.5cm}\hSlide{-0.5cm}\vSlide{+0.7cm}\ForceHeight{4.8cm}\BoxedEPSF{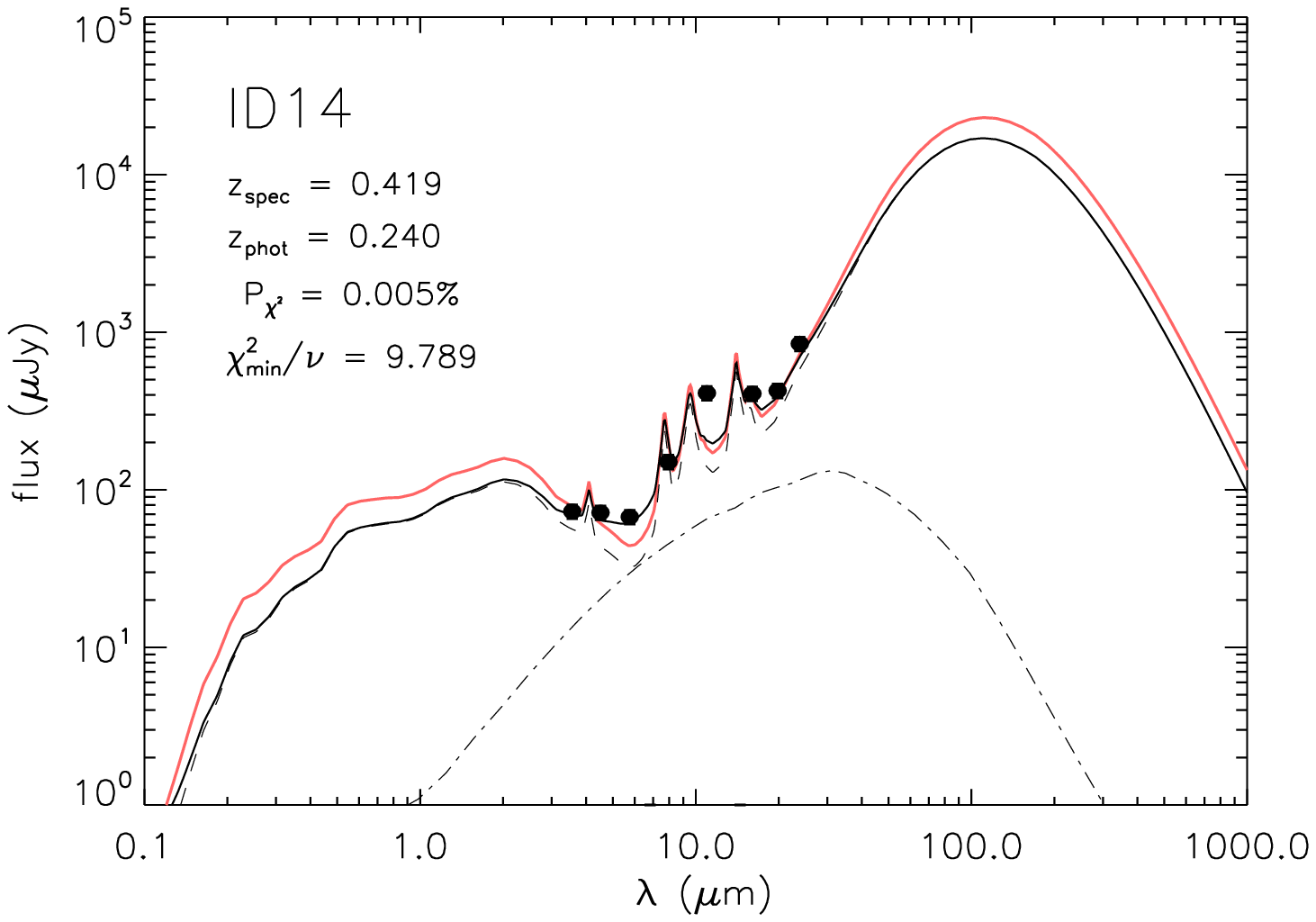}
  \vskip0.2truecm\caption{Comparison between the best-fit SED model
    (black solid curve) and the measured infrared spectrum (dots with
    error bars) for the infrared sources listed in
    Table~\ref{tab:realcat_info}. The inferred contributions from a
    starburst and an AGN torus, are represented by the dashed and the
    dot-dashed curves, respectively. Also shown are the spectroscopic
    redshift of the sources ($z_{\rm spec}$), the derived photometric
    redshifts ($z_{\rm phot}$), the corresponding value of the minimum
    reduced $\chi^{2}$ ($\chi^2_{\rm min}/\nu$) and the associated
    probability ($P_{\chi^2}$). The red solid curve is the best-fit
    SED model obtained by assuming {\it a priori} that the AGN
    component is null.}
  \label{fig:SEDs}
\end{figure*}

\setcounter{figure}{3}
\begin{figure*}
  \HideDisplacementBoxes
  \hSlide{-3.0cm}\vSlide{+0.0cm}\ForceHeight{4.8cm}\BoxedEPSF{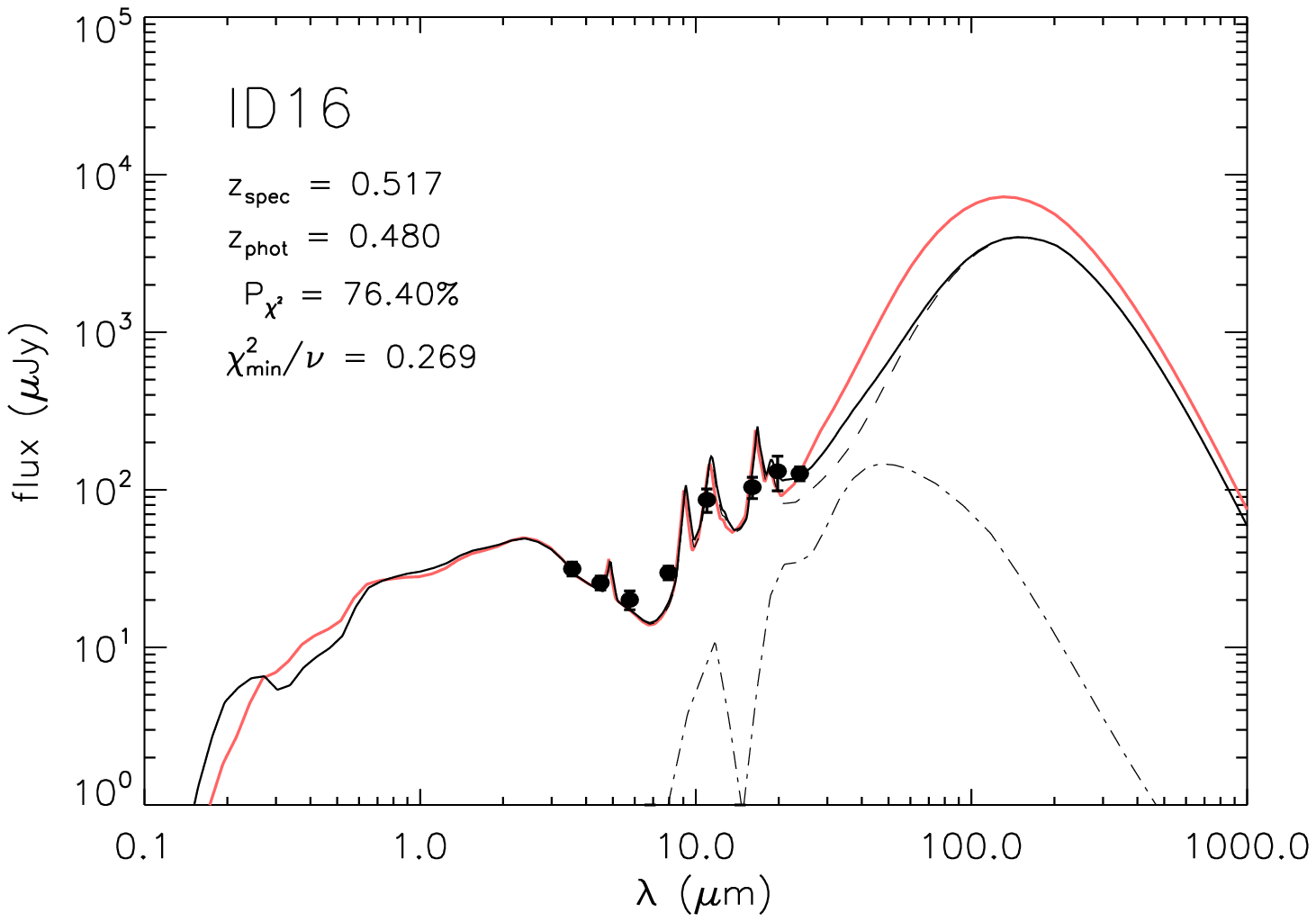}
  \hSlide{+2.0cm}\vSlide{+0.0cm}\ForceHeight{4.8cm}\BoxedEPSF{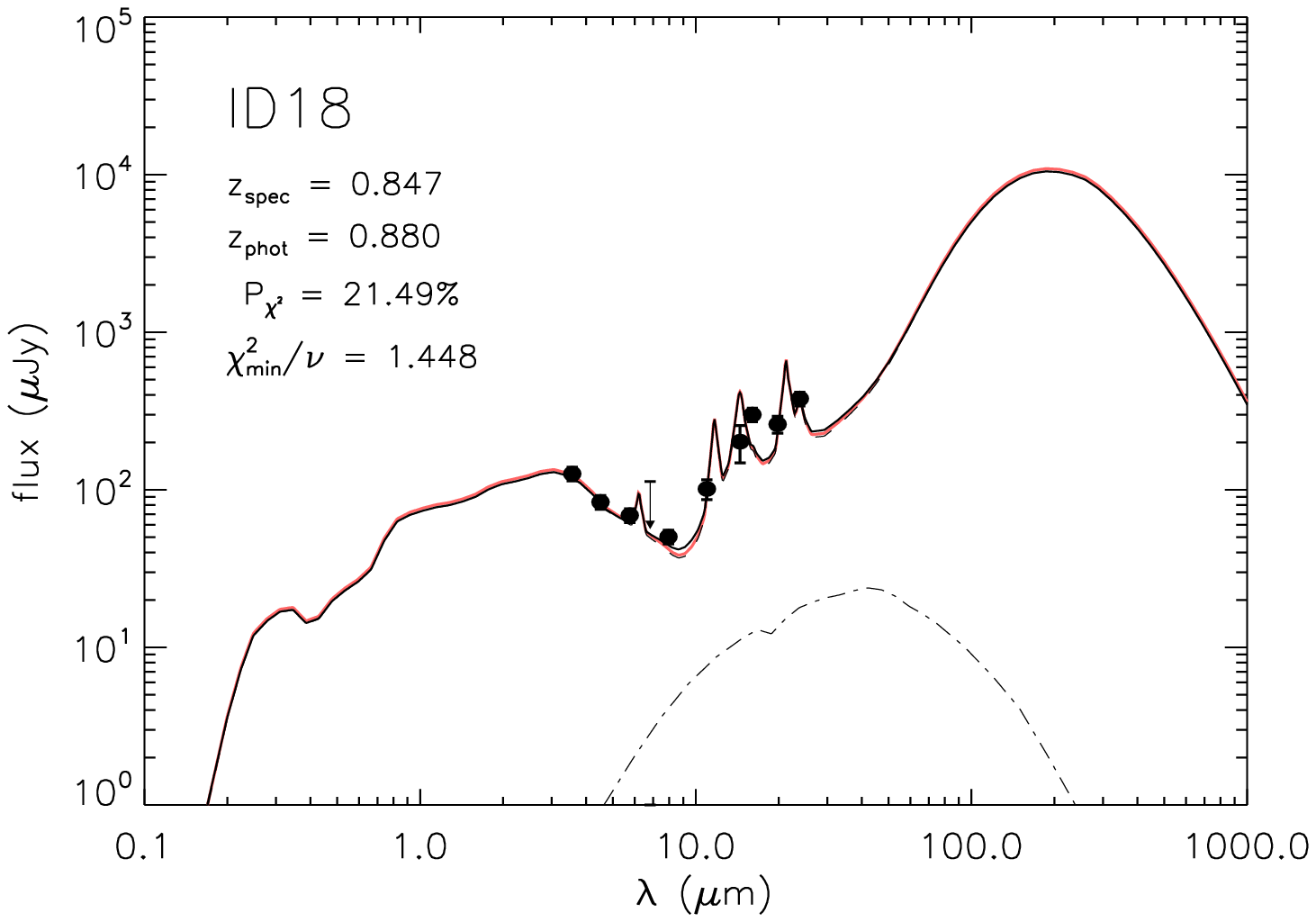}
  \vspace{-5.5cm}\hSlide{-0.5cm}\vSlide{+0.7cm}\ForceHeight{4.8cm}\BoxedEPSF{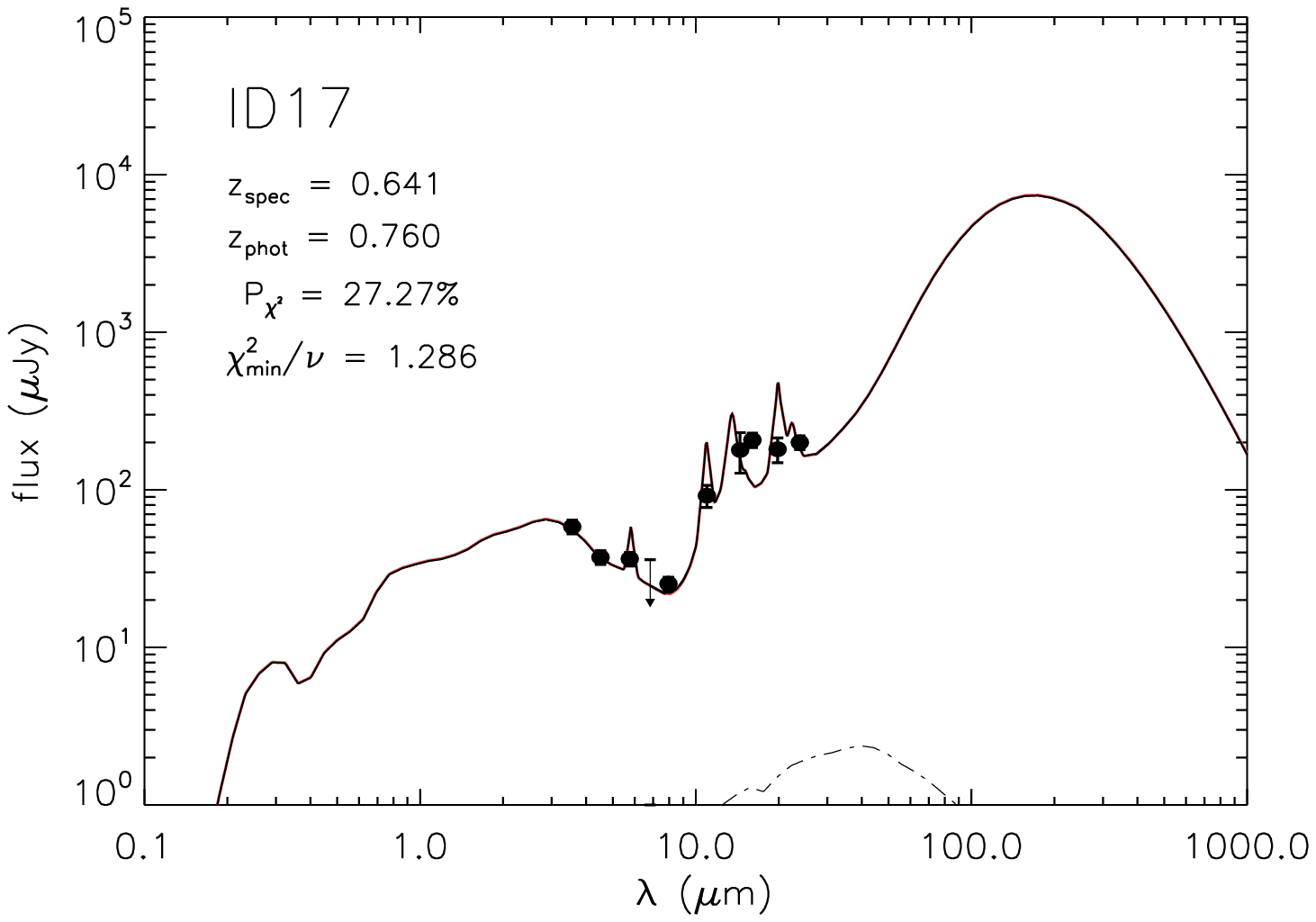} \\
  \hSlide{-3.0cm}\vSlide{+0.0cm}\ForceHeight{4.8cm}\BoxedEPSF{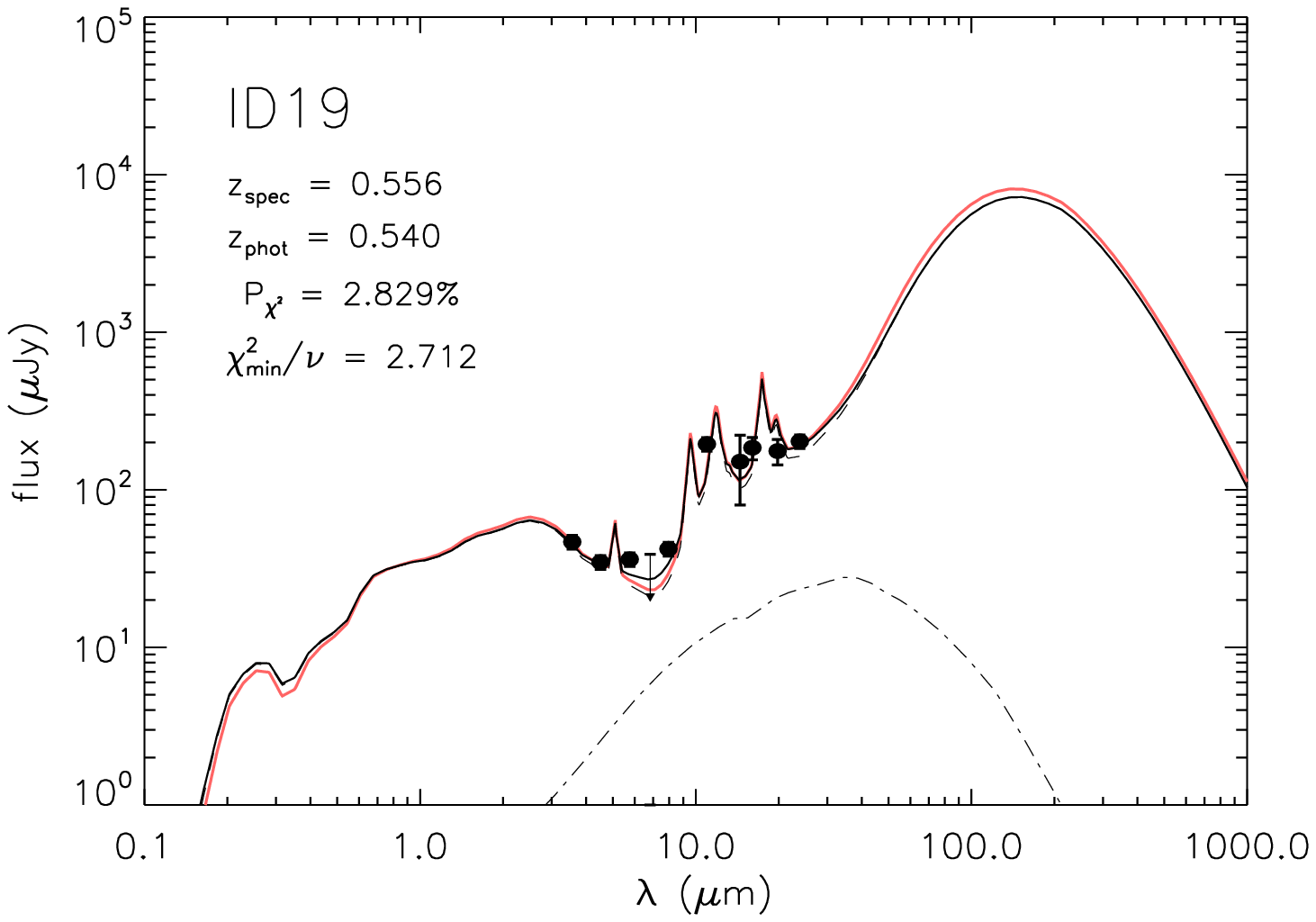}
  \hSlide{+2.0cm}\vSlide{+0.0cm}\ForceHeight{4.8cm}\BoxedEPSF{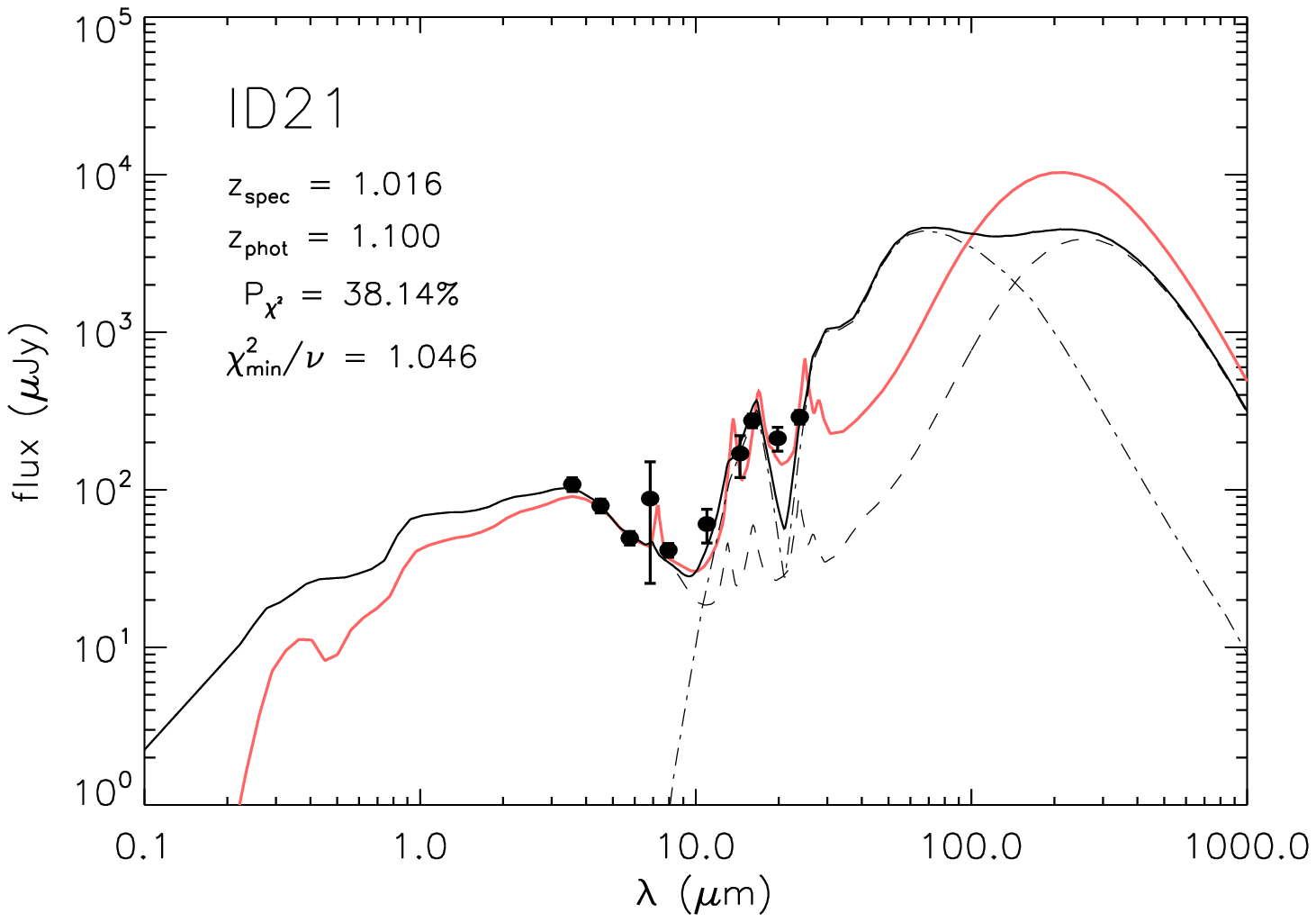}
  \vspace{-5.5cm}\hSlide{-0.5cm}\vSlide{+0.7cm}\ForceHeight{4.8cm}\BoxedEPSF{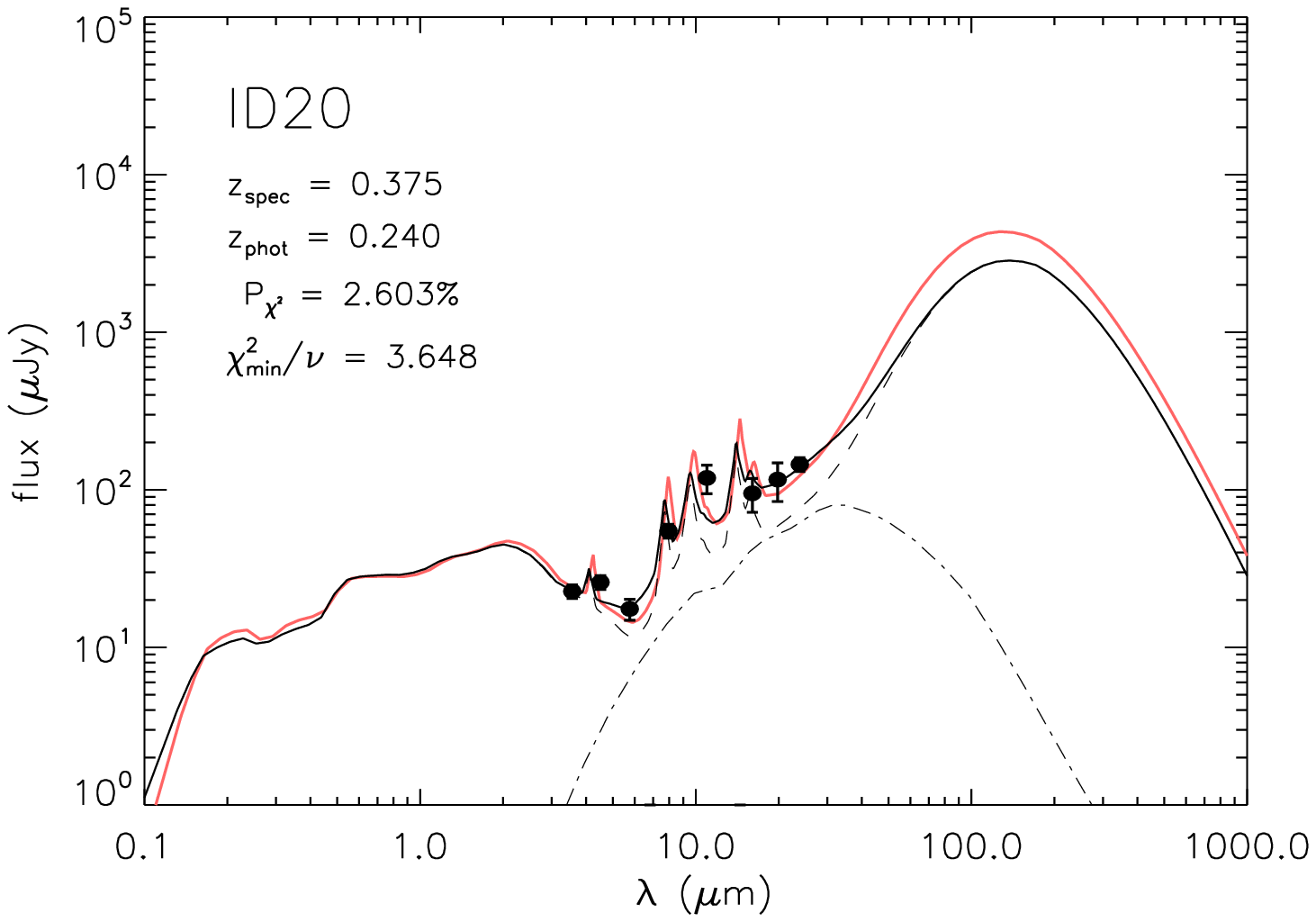} \\
  \hSlide{-3.0cm}\vSlide{+0.0cm}\ForceHeight{4.8cm}\BoxedEPSF{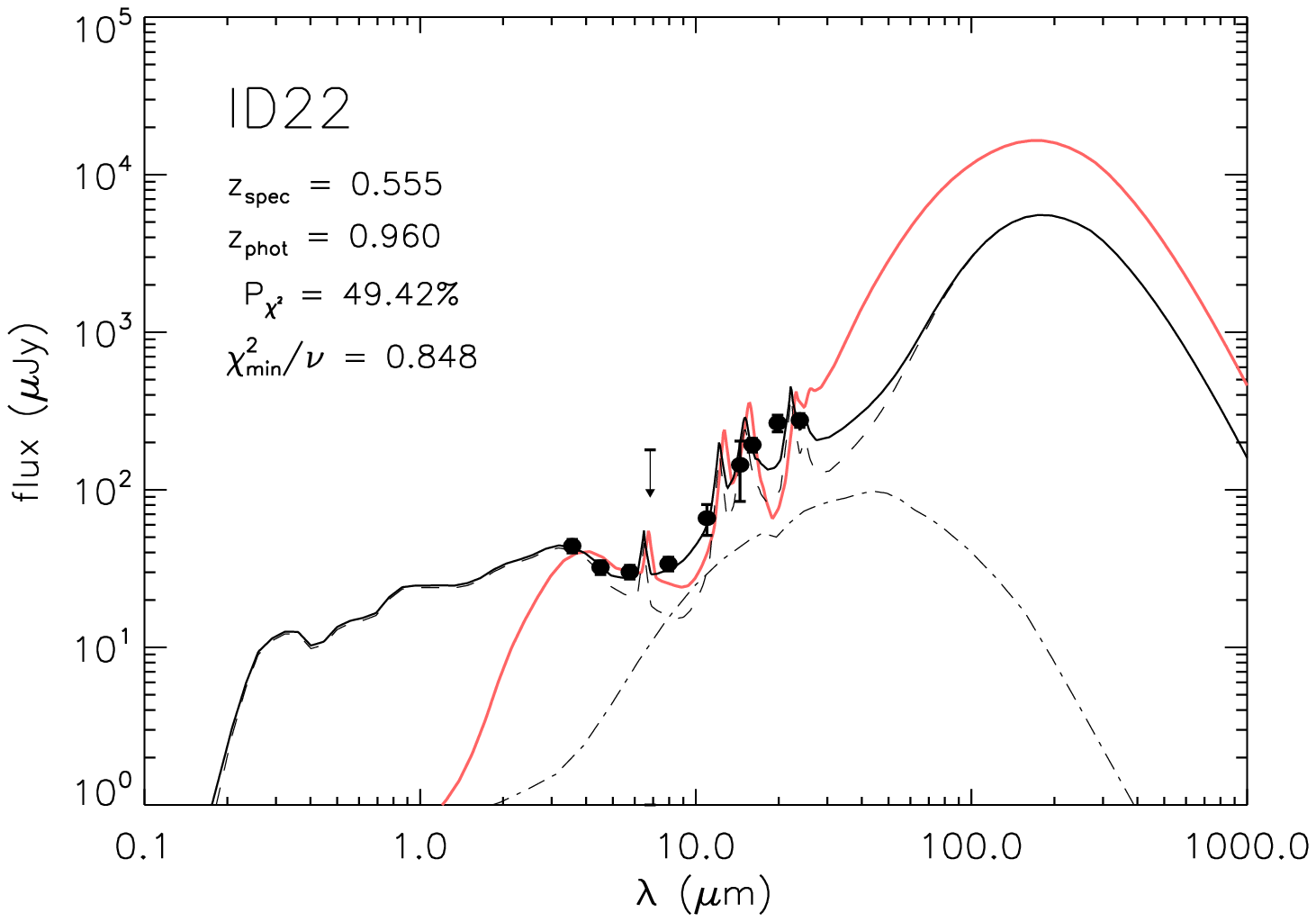}
  \hSlide{+2.0cm}\vSlide{+0.0cm}\ForceHeight{4.8cm}\BoxedEPSF{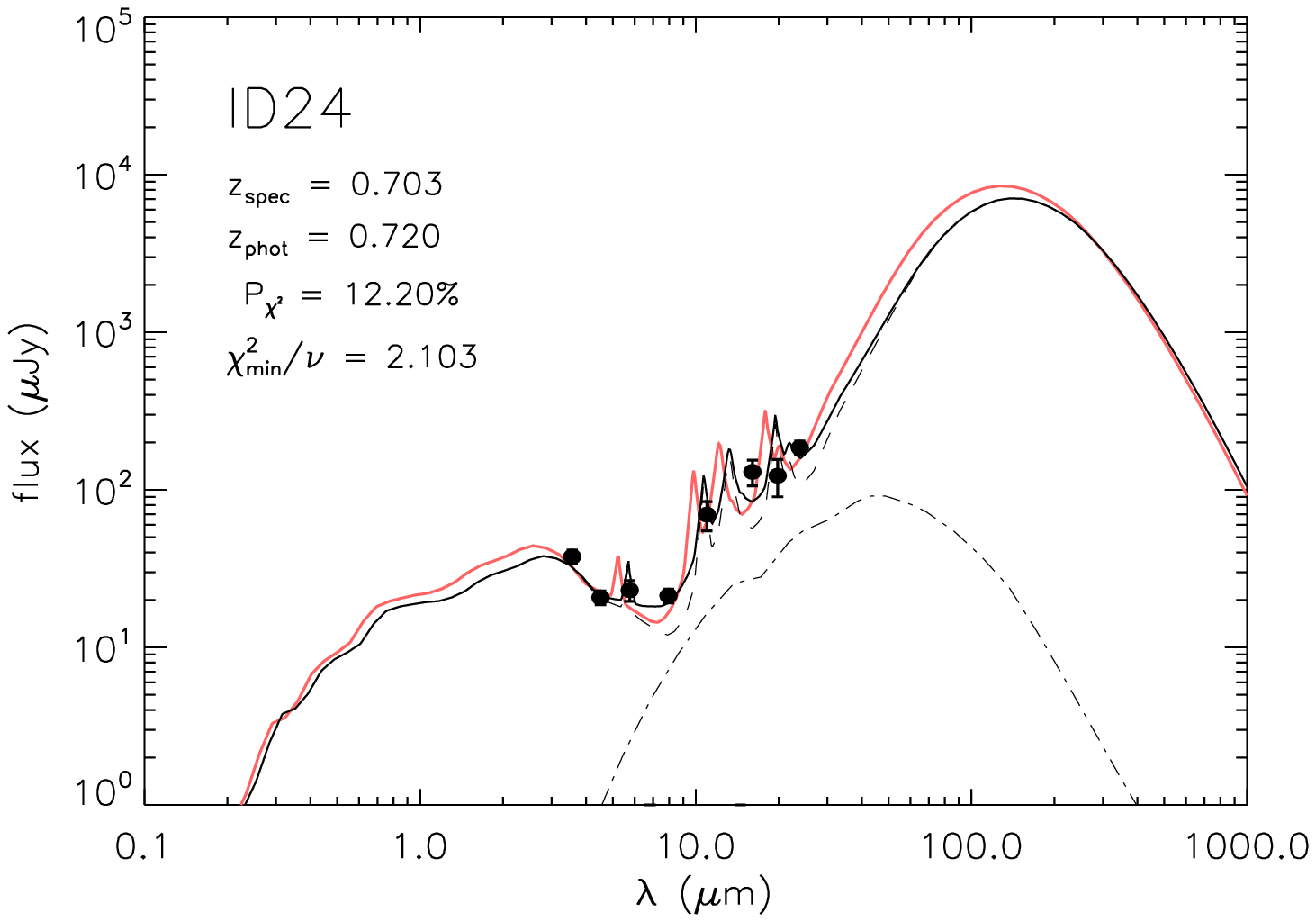}
  \vspace{-5.5cm}\hSlide{-0.5cm}\vSlide{+0.7cm}\ForceHeight{4.8cm}\BoxedEPSF{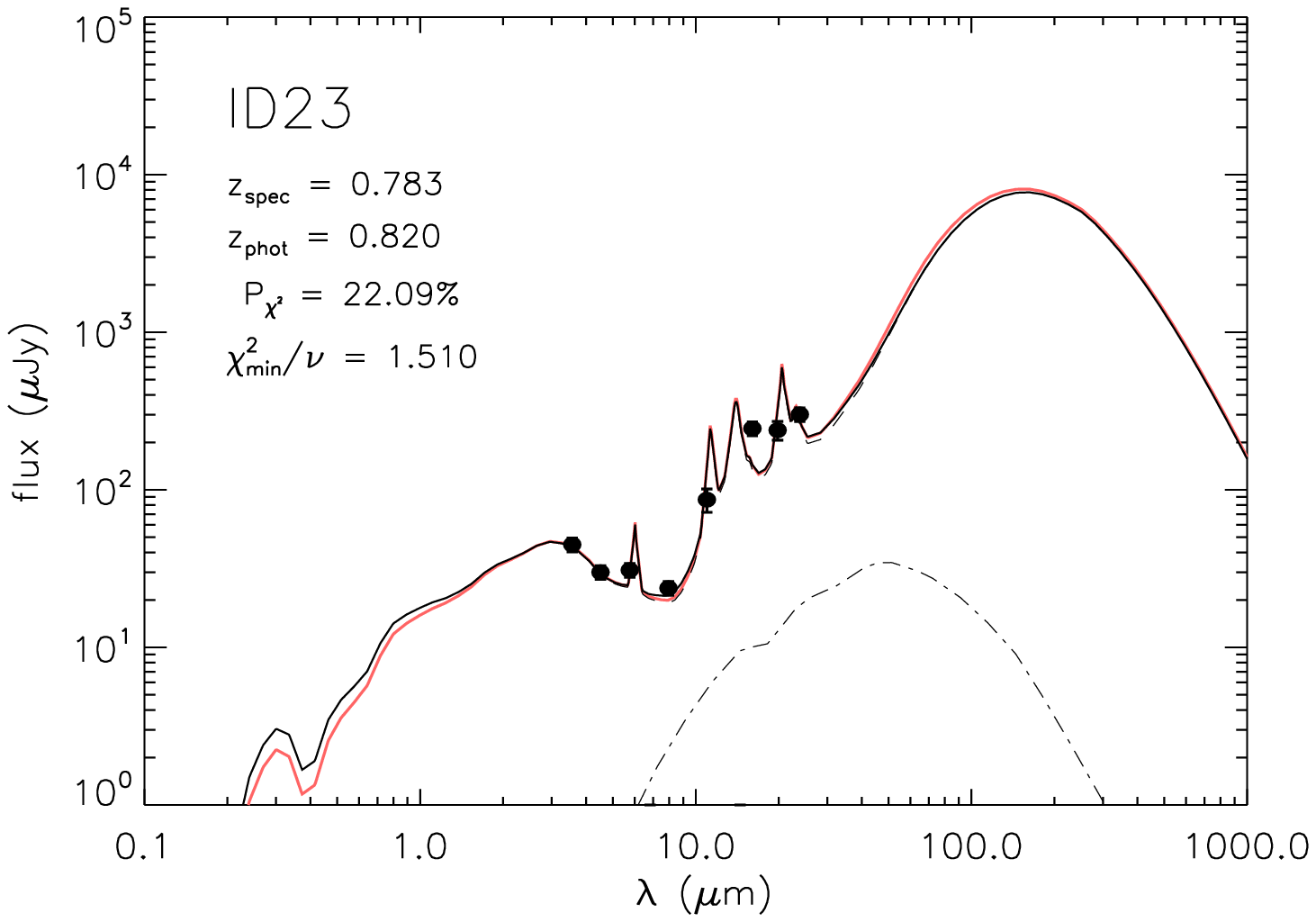} \\
  \hSlide{-3.0cm}\vSlide{+0.0cm}\ForceHeight{4.8cm}\BoxedEPSF{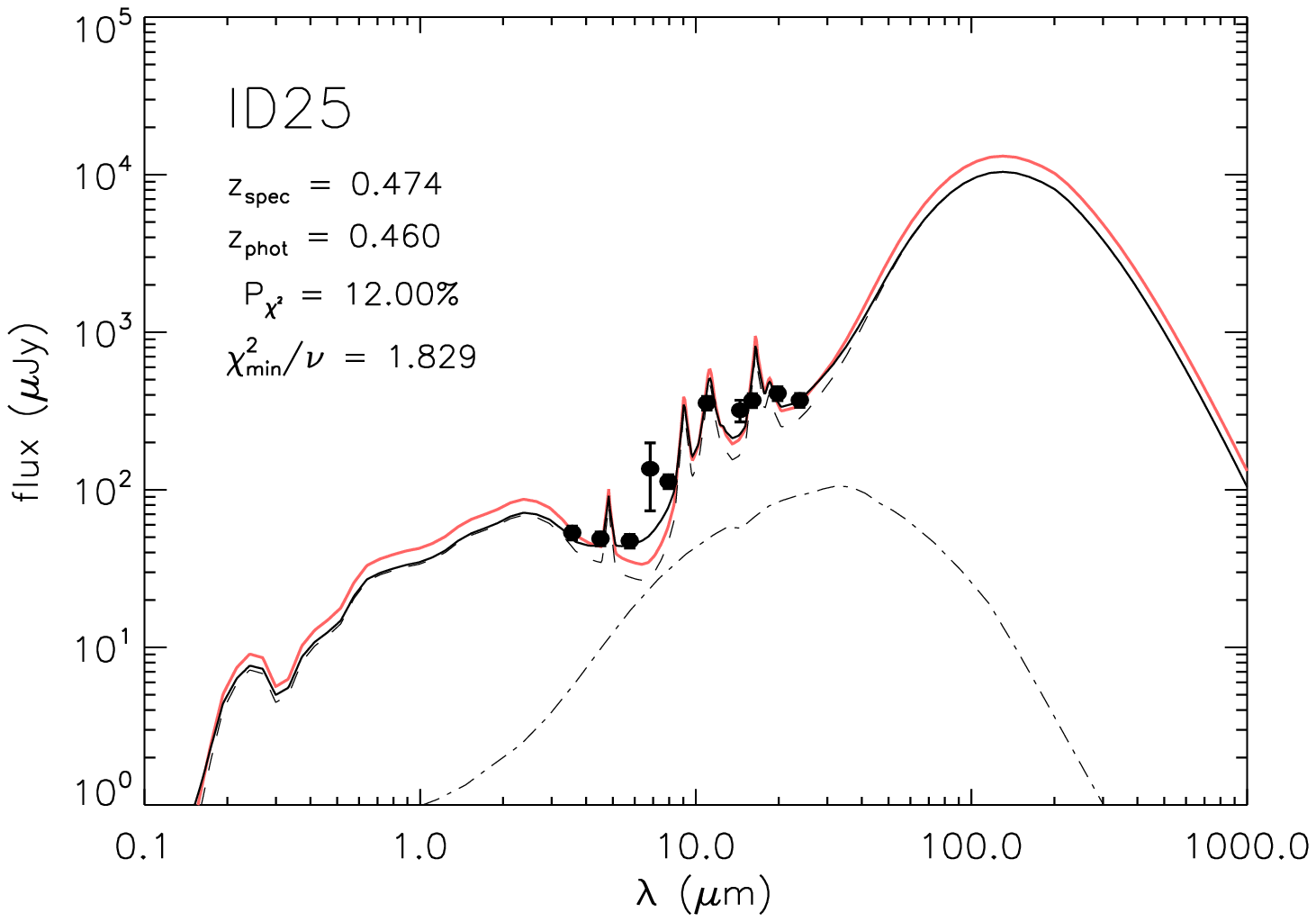}
  \hSlide{+2.0cm}\vSlide{+0.0cm}\ForceHeight{4.8cm}\BoxedEPSF{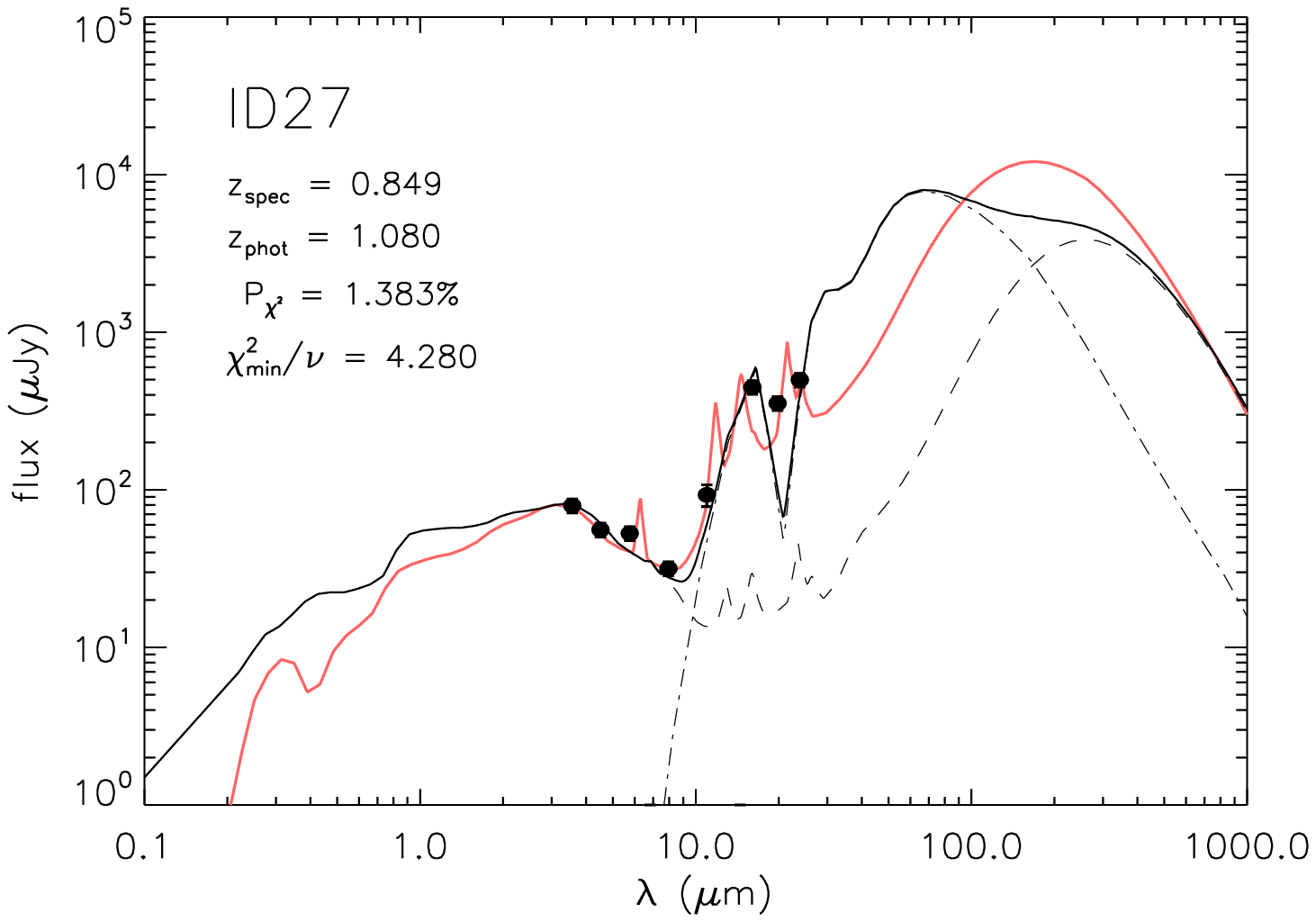}
  \vspace{-5.5cm}\hSlide{-0.5cm}\vSlide{+0.7cm}\ForceHeight{4.8cm}\BoxedEPSF{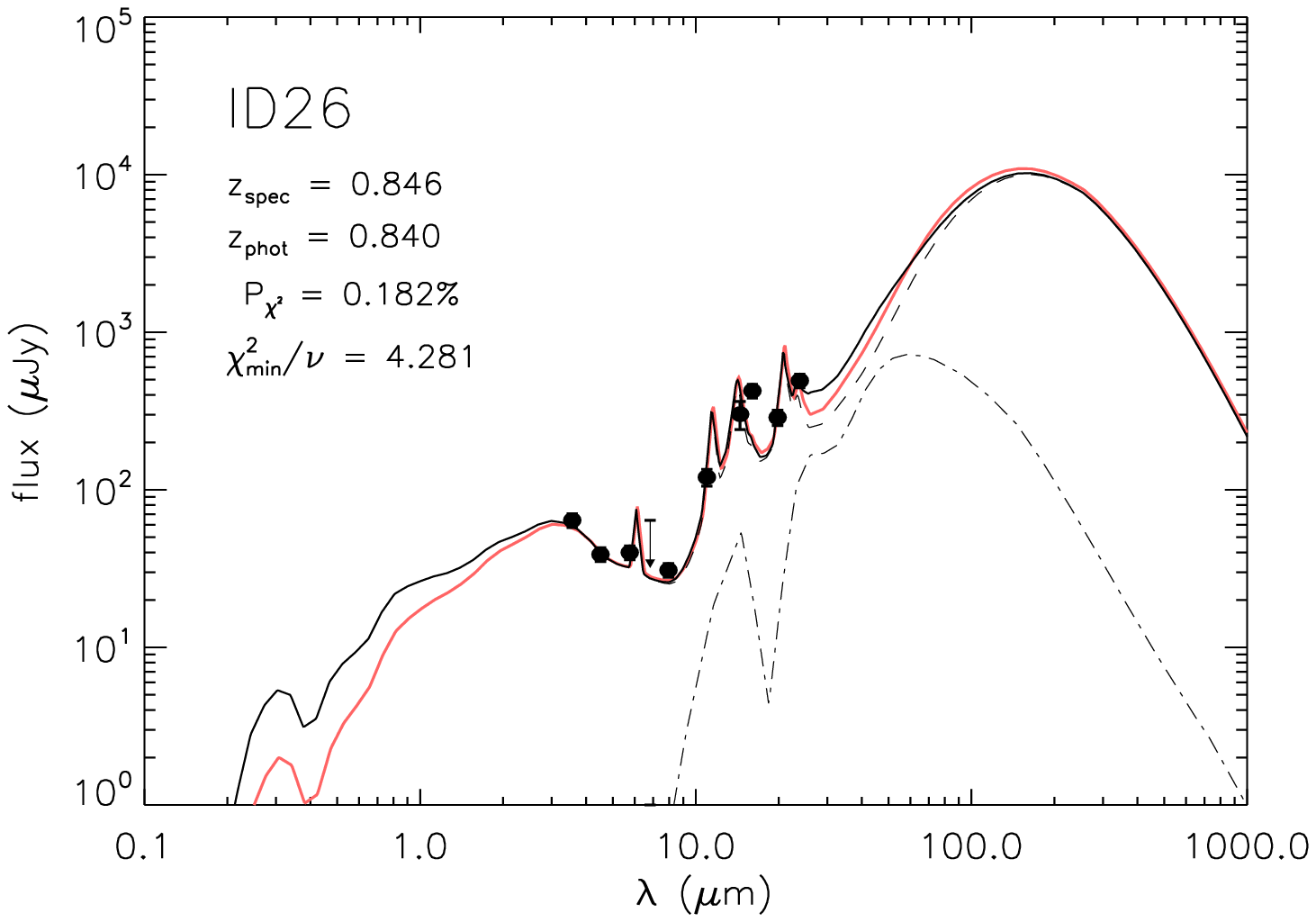} \\
  \hSlide{-3.0cm}\vSlide{+0.0cm}\ForceHeight{4.8cm}\BoxedEPSF{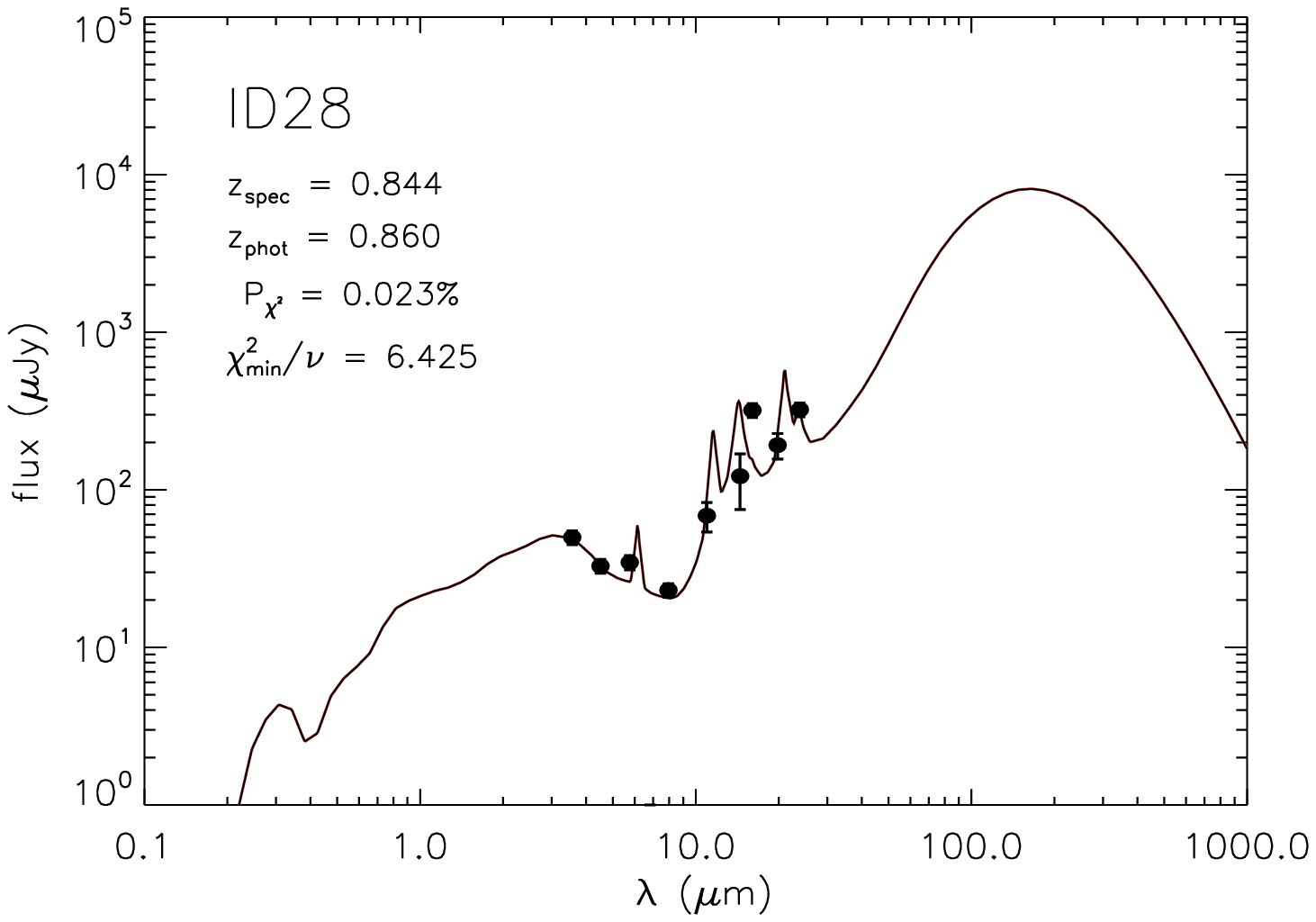}
  \hSlide{+2.0cm}\vSlide{+0.0cm}\ForceHeight{4.8cm}\BoxedEPSF{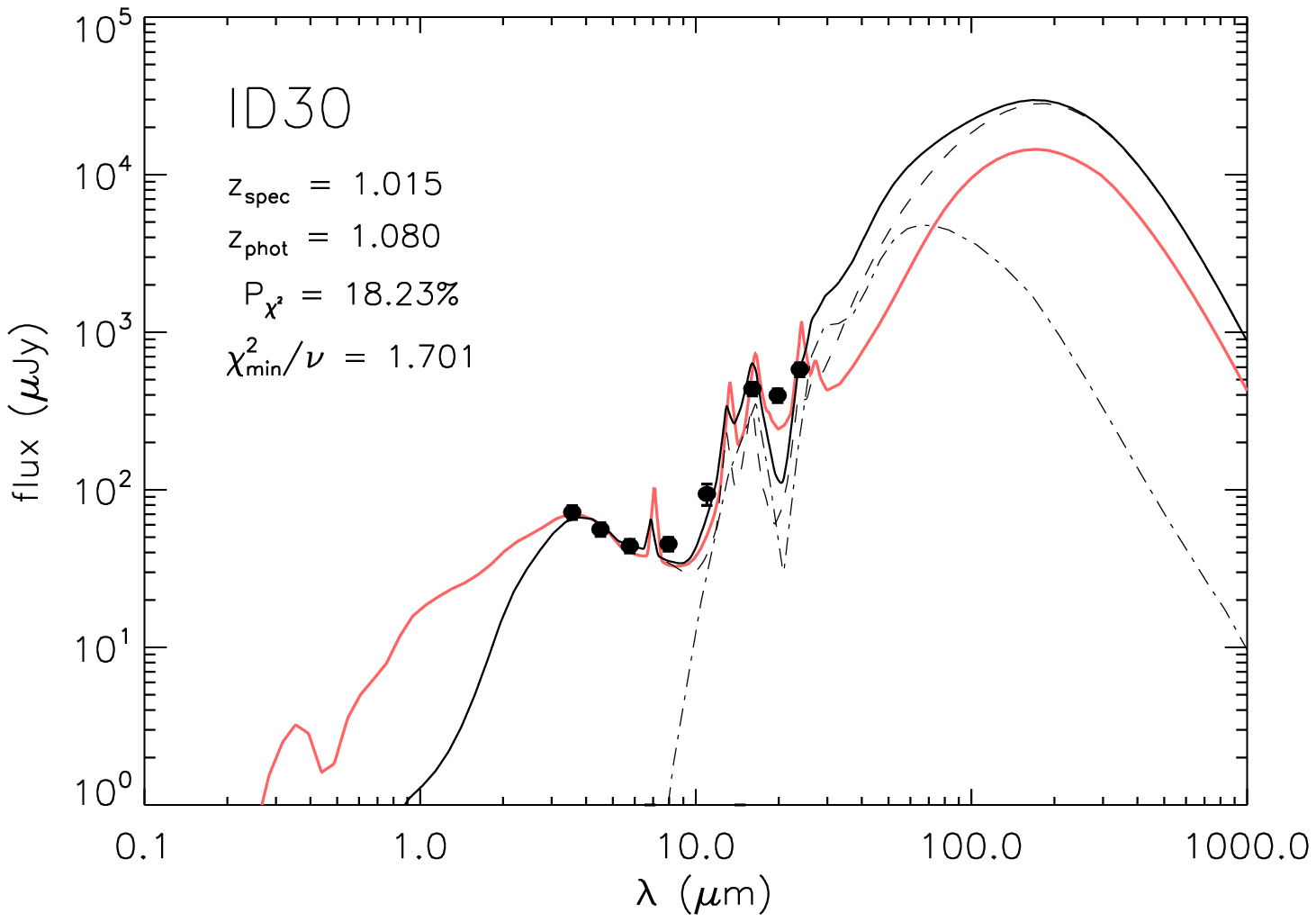}
  \vspace{-5.5cm}\hSlide{-0.5cm}\vSlide{+0.7cm}\ForceHeight{4.8cm}\BoxedEPSF{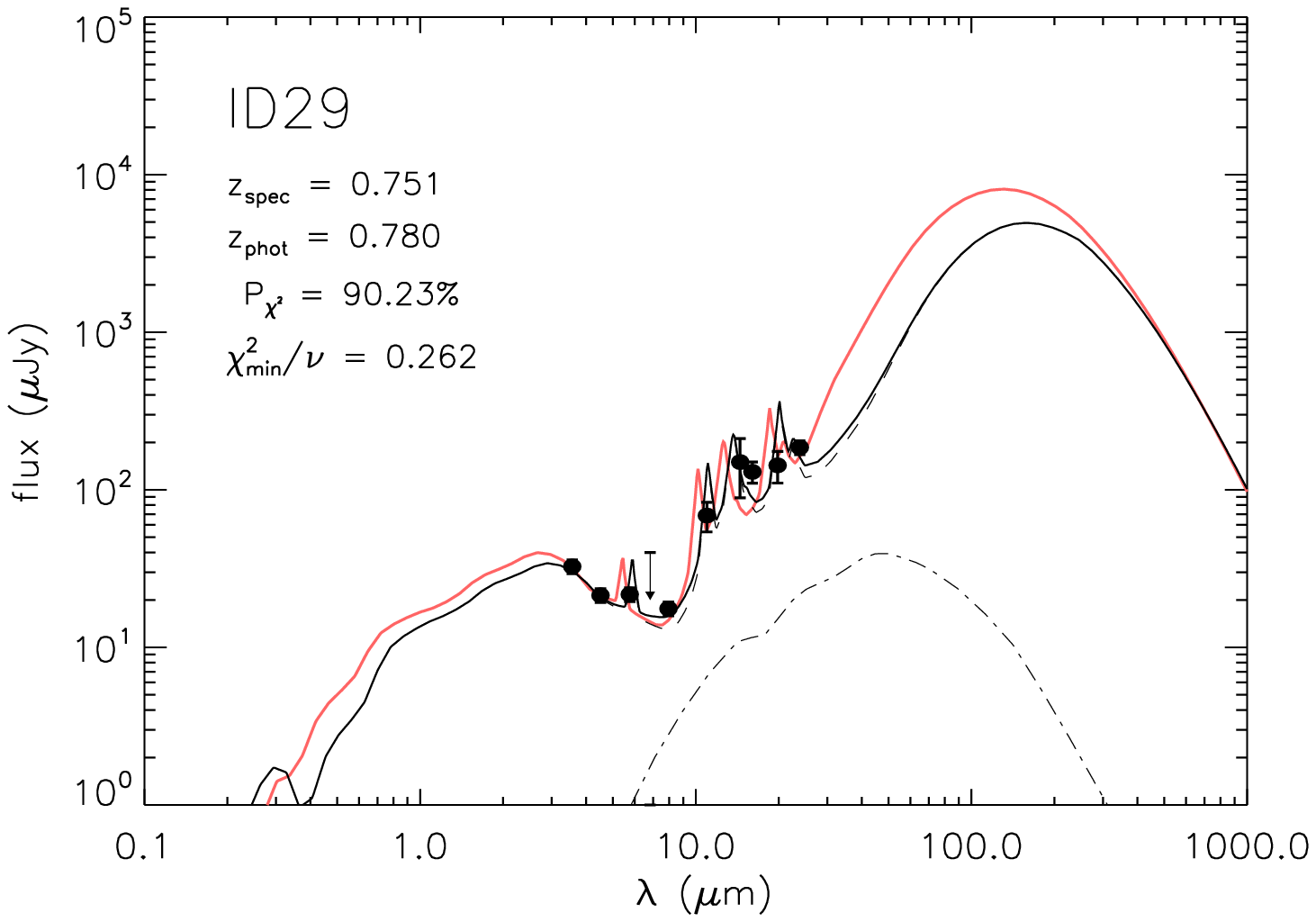}
  \vskip0.2truecm\caption{{\it Continued.}}
\end{figure*}

\setcounter{figure}{3}
\begin{figure*}
  \HideDisplacementBoxes
  \hSlide{-3.0cm}\vSlide{+0.0cm}\ForceHeight{4.8cm}\BoxedEPSF{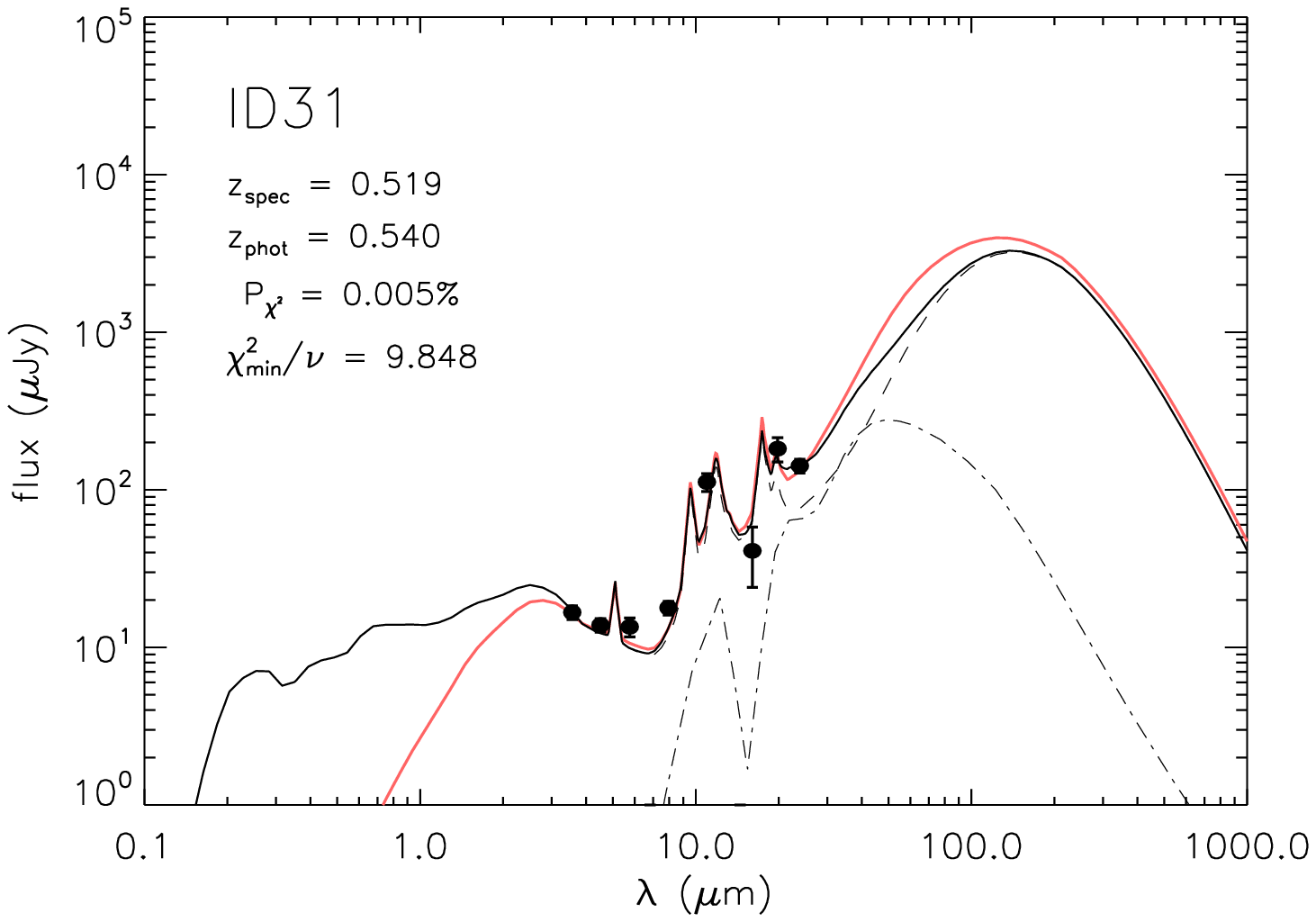}
  \hSlide{+2.0cm}\vSlide{+0.0cm}\ForceHeight{4.8cm}\BoxedEPSF{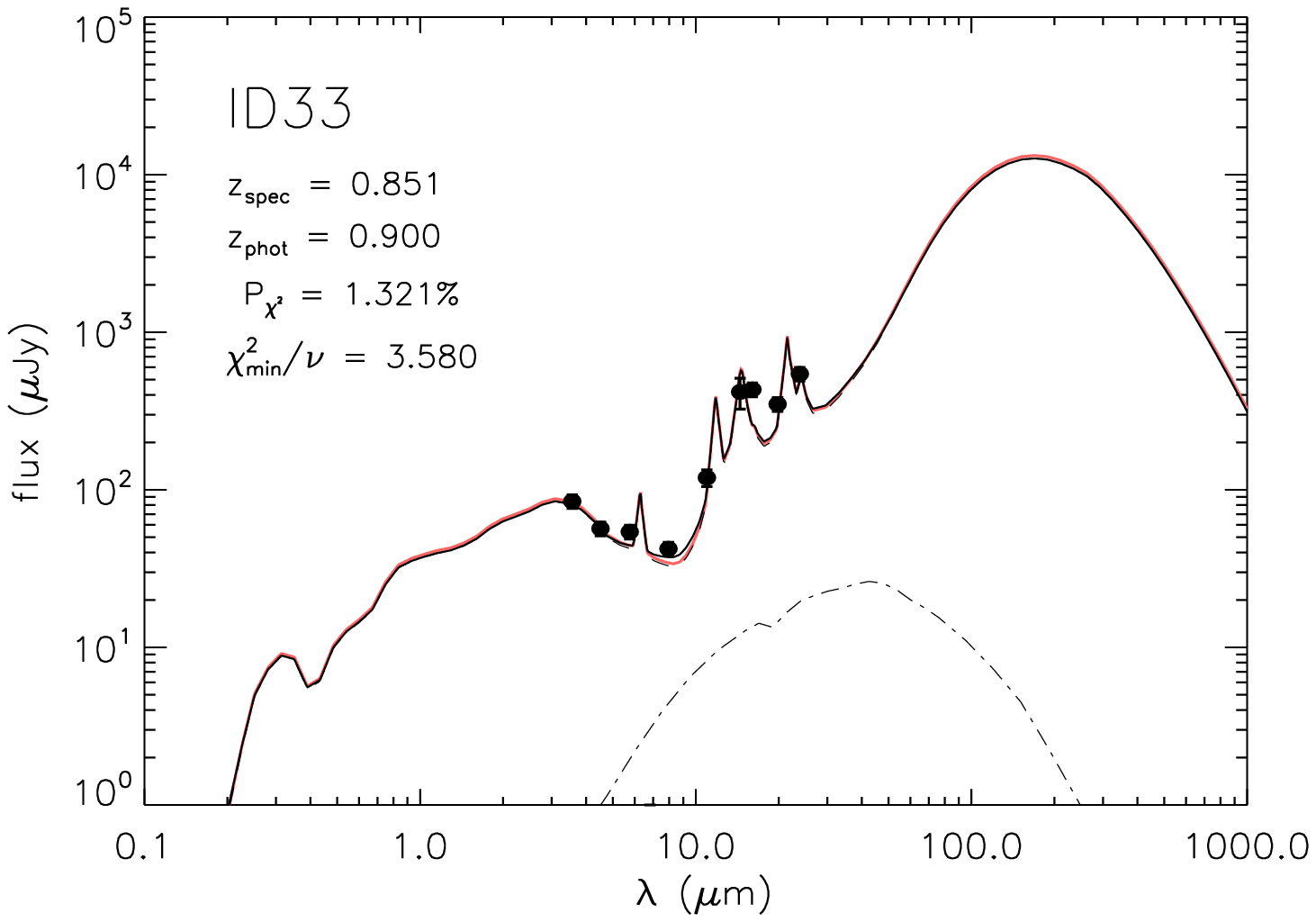}
  \vspace{-5.5cm}\hSlide{-0.5cm}\vSlide{+0.7cm}\ForceHeight{4.8cm}\BoxedEPSF{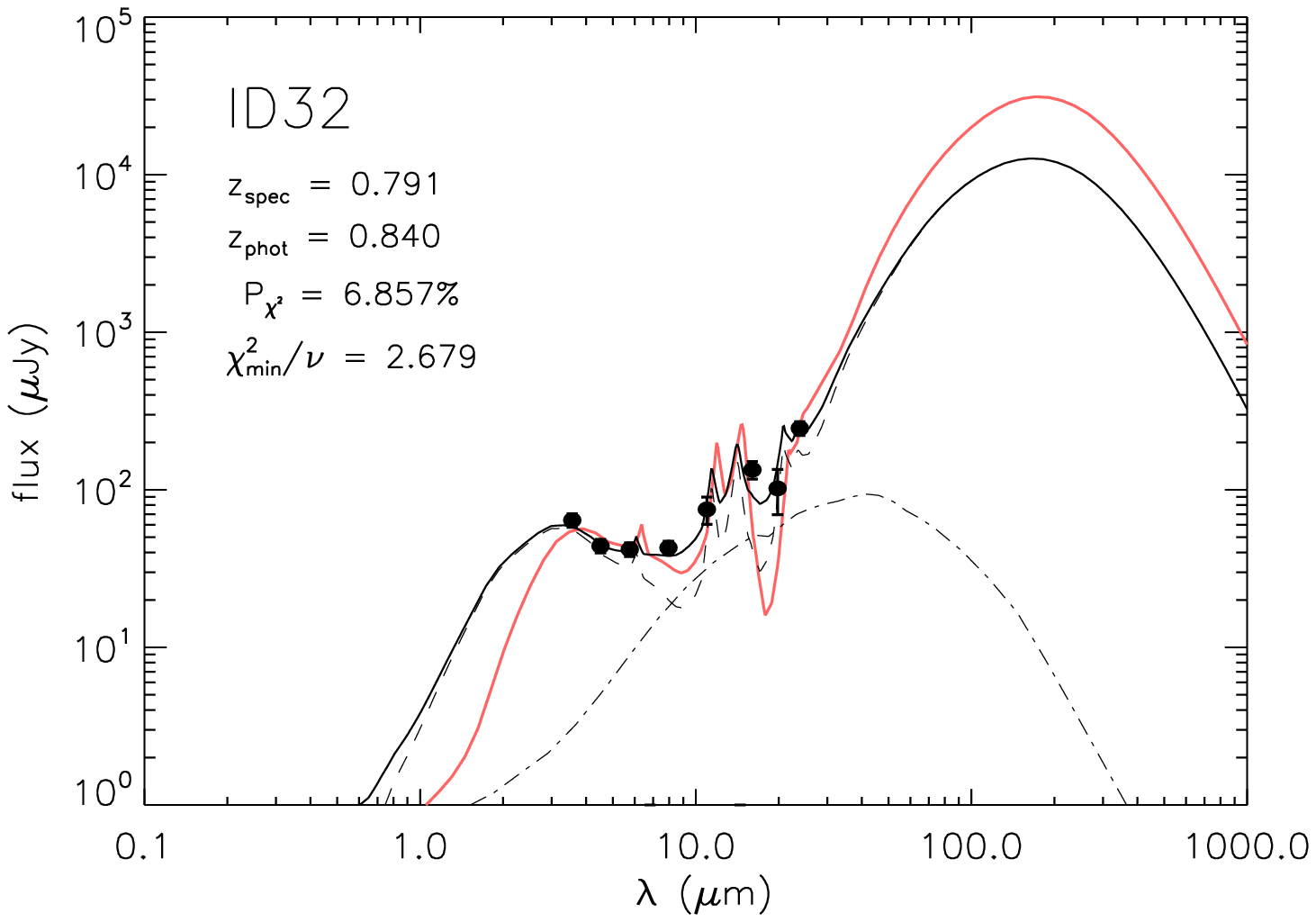} \\
  \hSlide{-3.0cm}\vSlide{+0.0cm}\ForceHeight{4.8cm}\BoxedEPSF{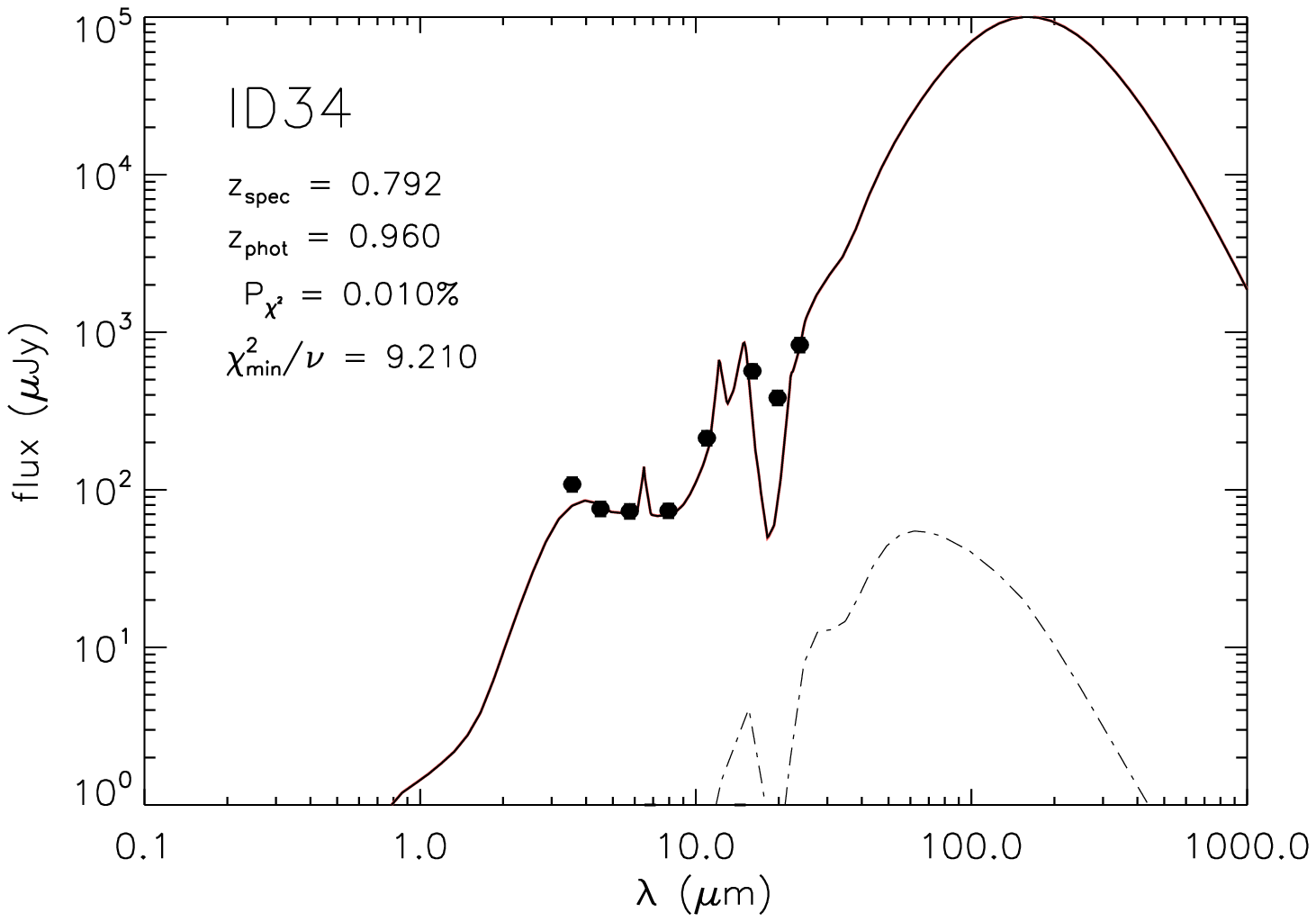}
  \hSlide{+2.0cm}\vSlide{+0.0cm}\ForceHeight{4.8cm}\BoxedEPSF{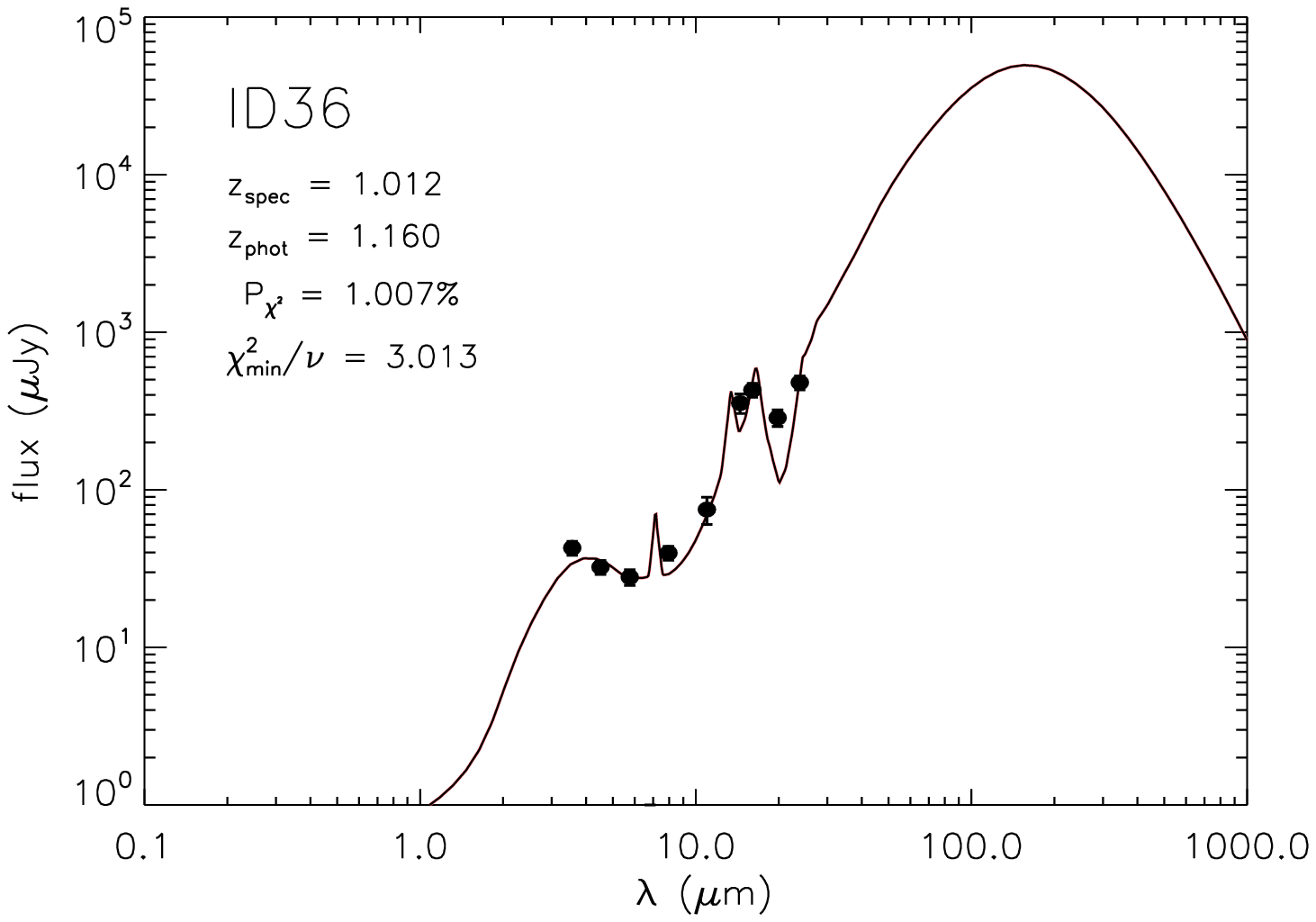}
  \vspace{-5.5cm}\hSlide{-0.5cm}\vSlide{+0.7cm}\ForceHeight{4.8cm}\BoxedEPSF{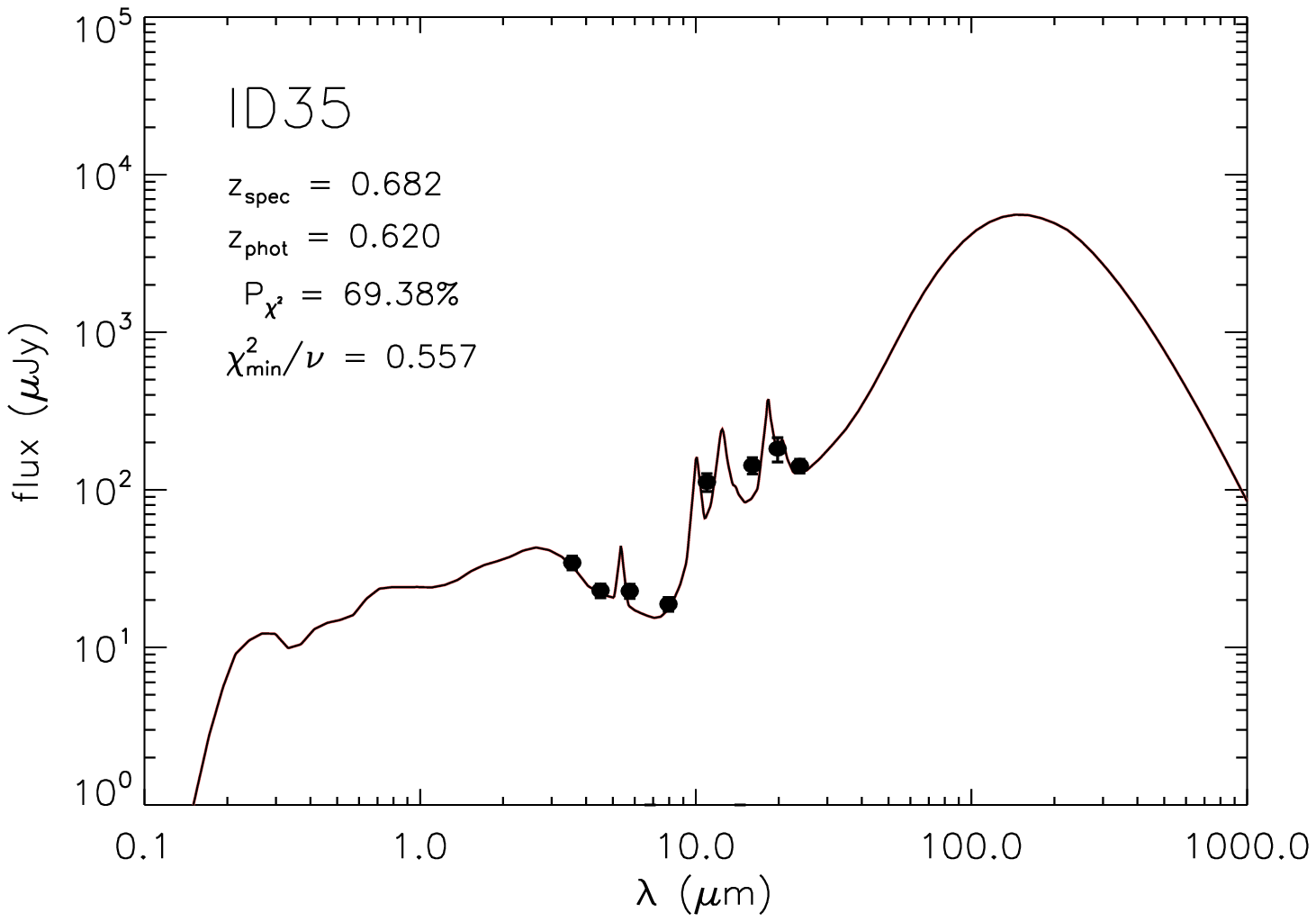} \\
  \hSlide{-3.0cm}\vSlide{+0.0cm}\ForceHeight{4.8cm}\BoxedEPSF{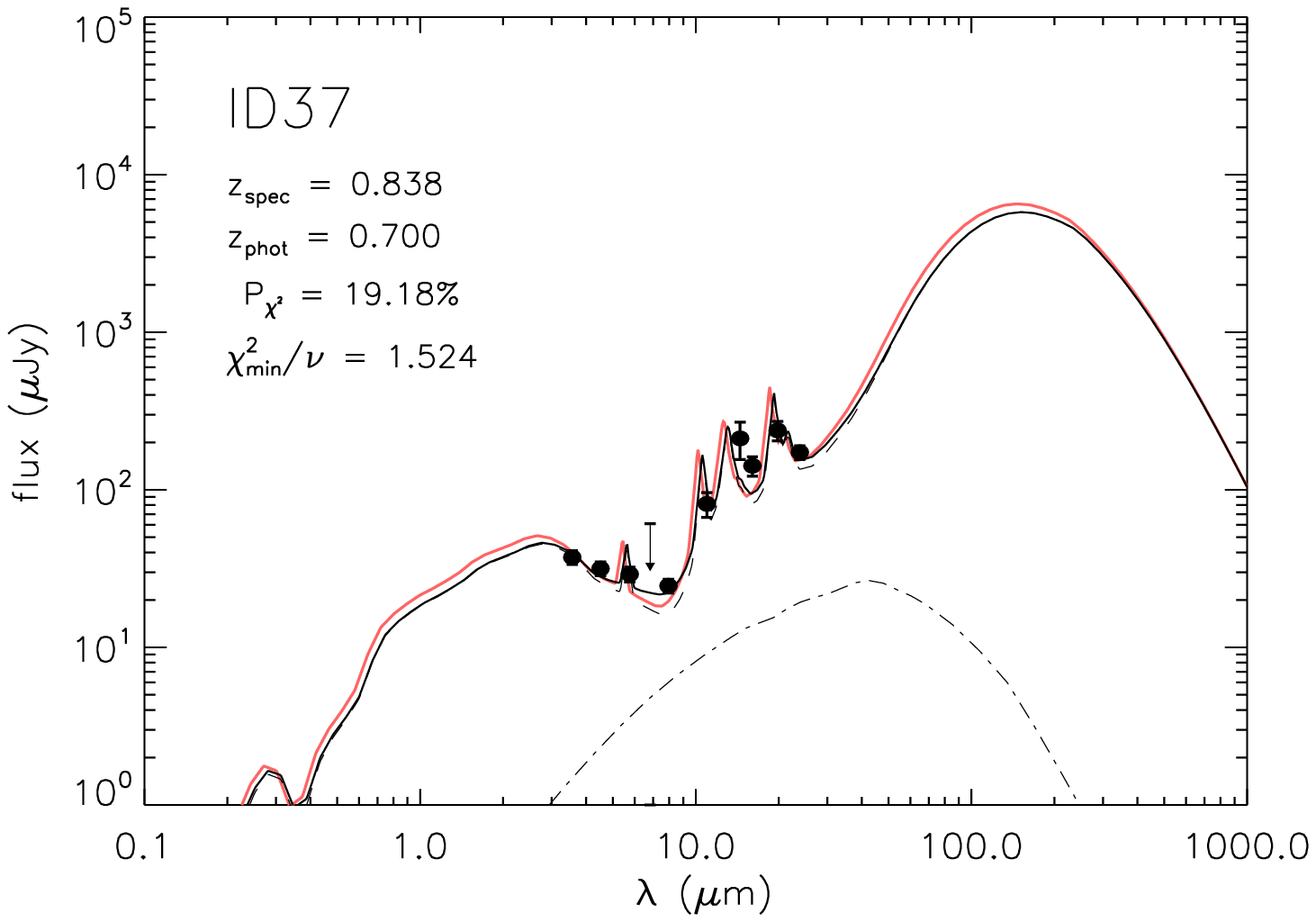}
  \hSlide{+2.0cm}\vSlide{+0.0cm}\ForceHeight{4.8cm}\BoxedEPSF{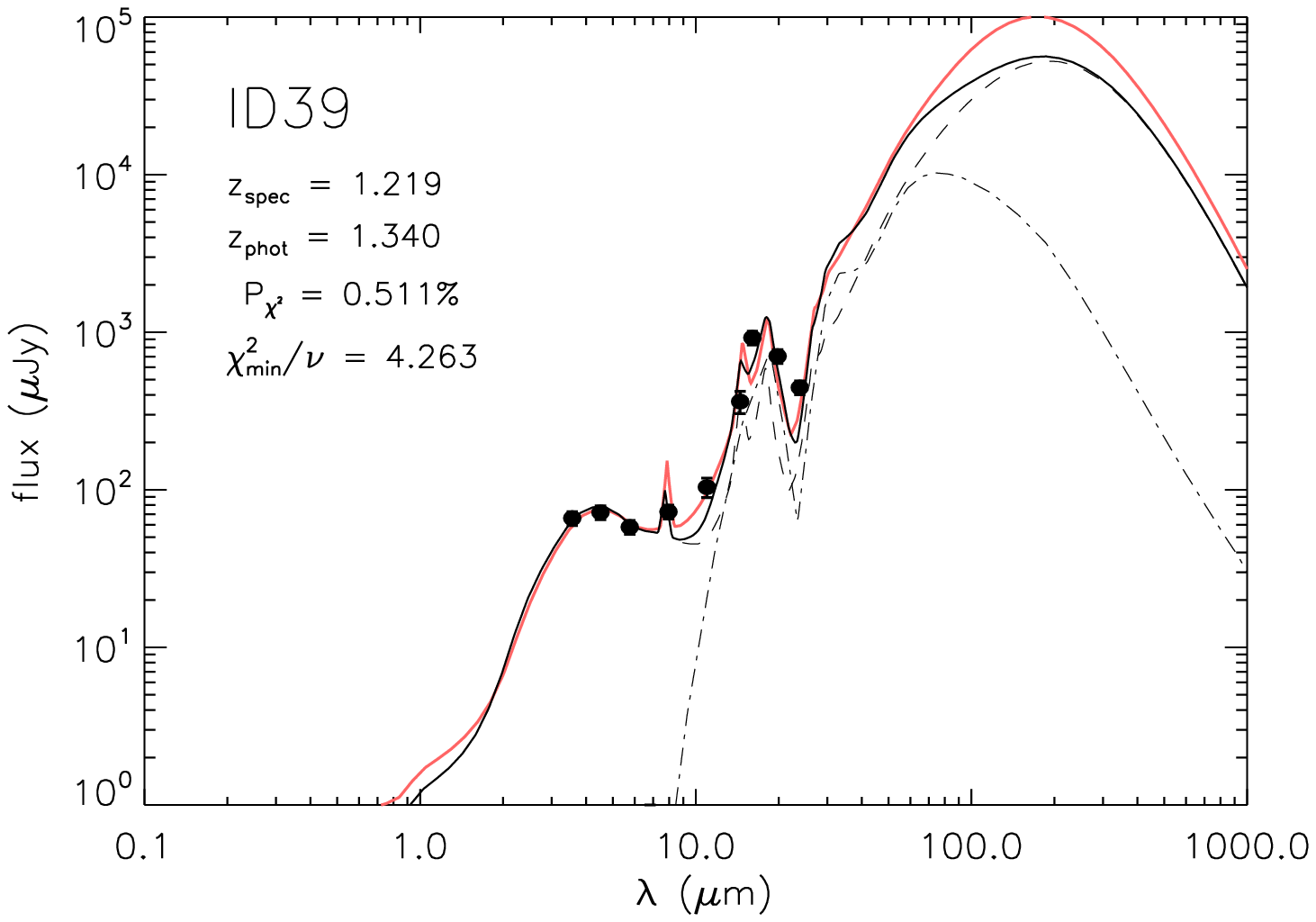}
  \vspace{-5.5cm}\hSlide{-0.5cm}\vSlide{+0.7cm}\ForceHeight{4.8cm}\BoxedEPSF{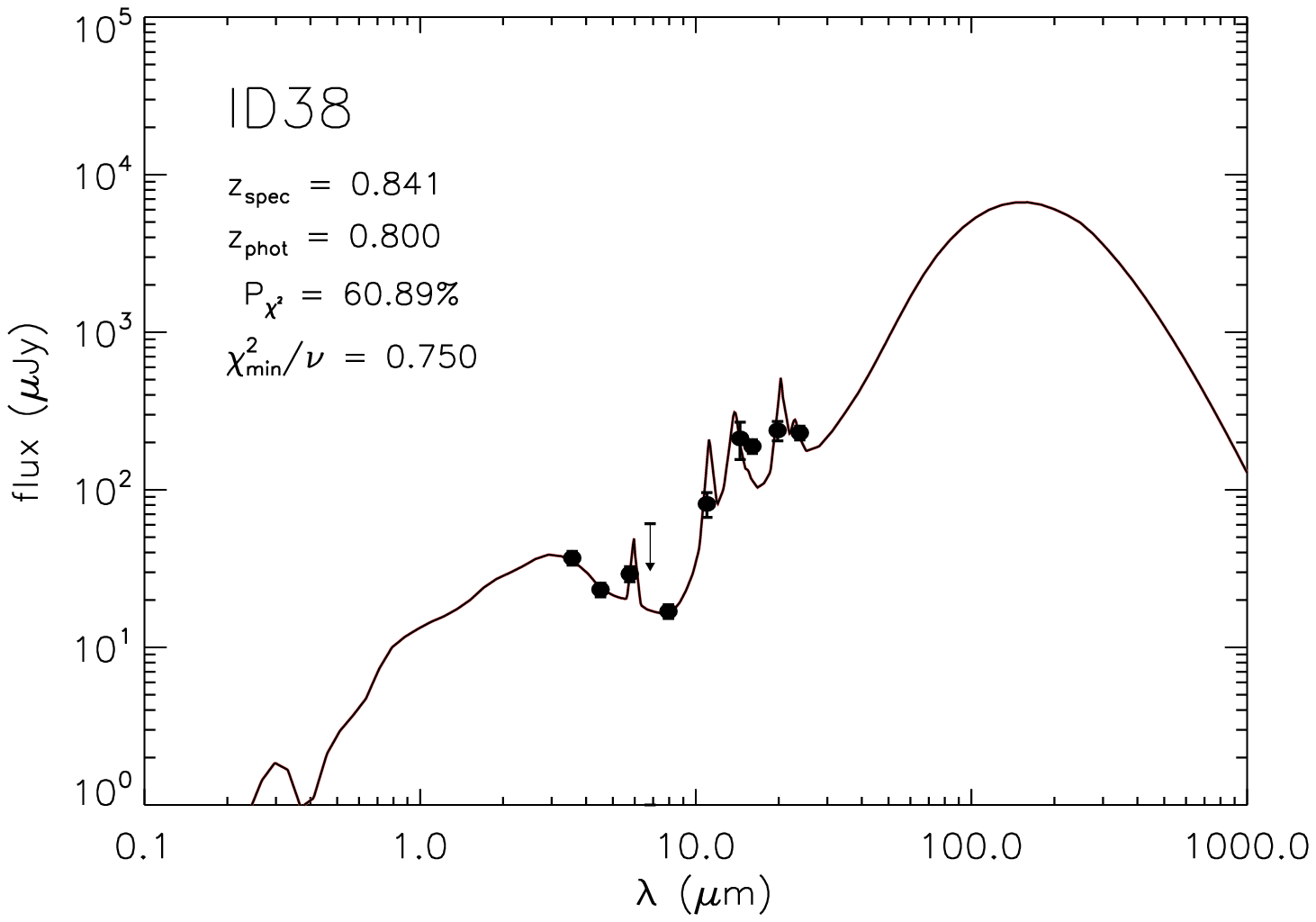} \\
  \hSlide{-3.0cm}\vSlide{+0.0cm}\ForceHeight{4.8cm}\BoxedEPSF{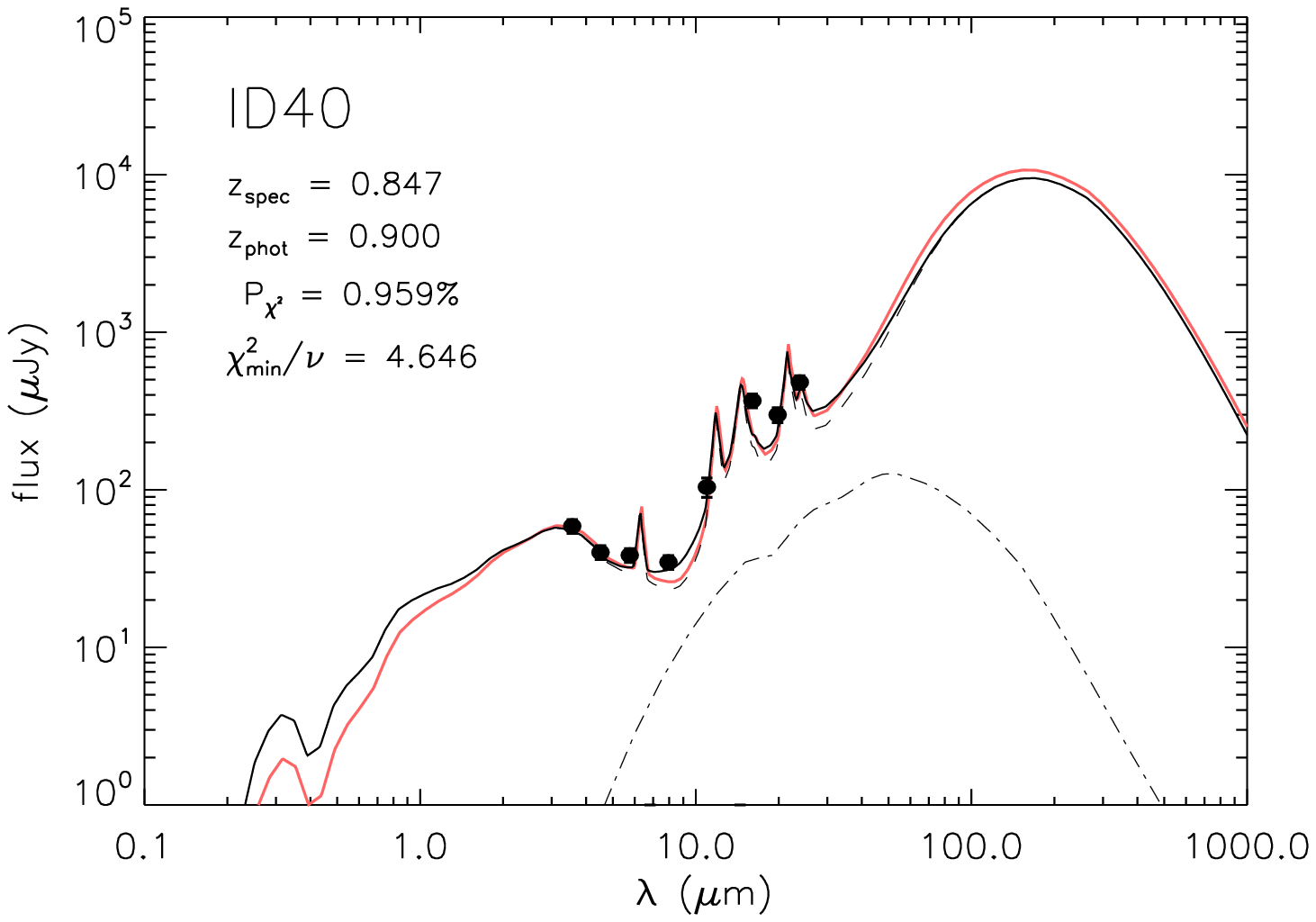}
  \hSlide{+2.0cm}\vSlide{+0.0cm}\ForceHeight{4.8cm}\BoxedEPSF{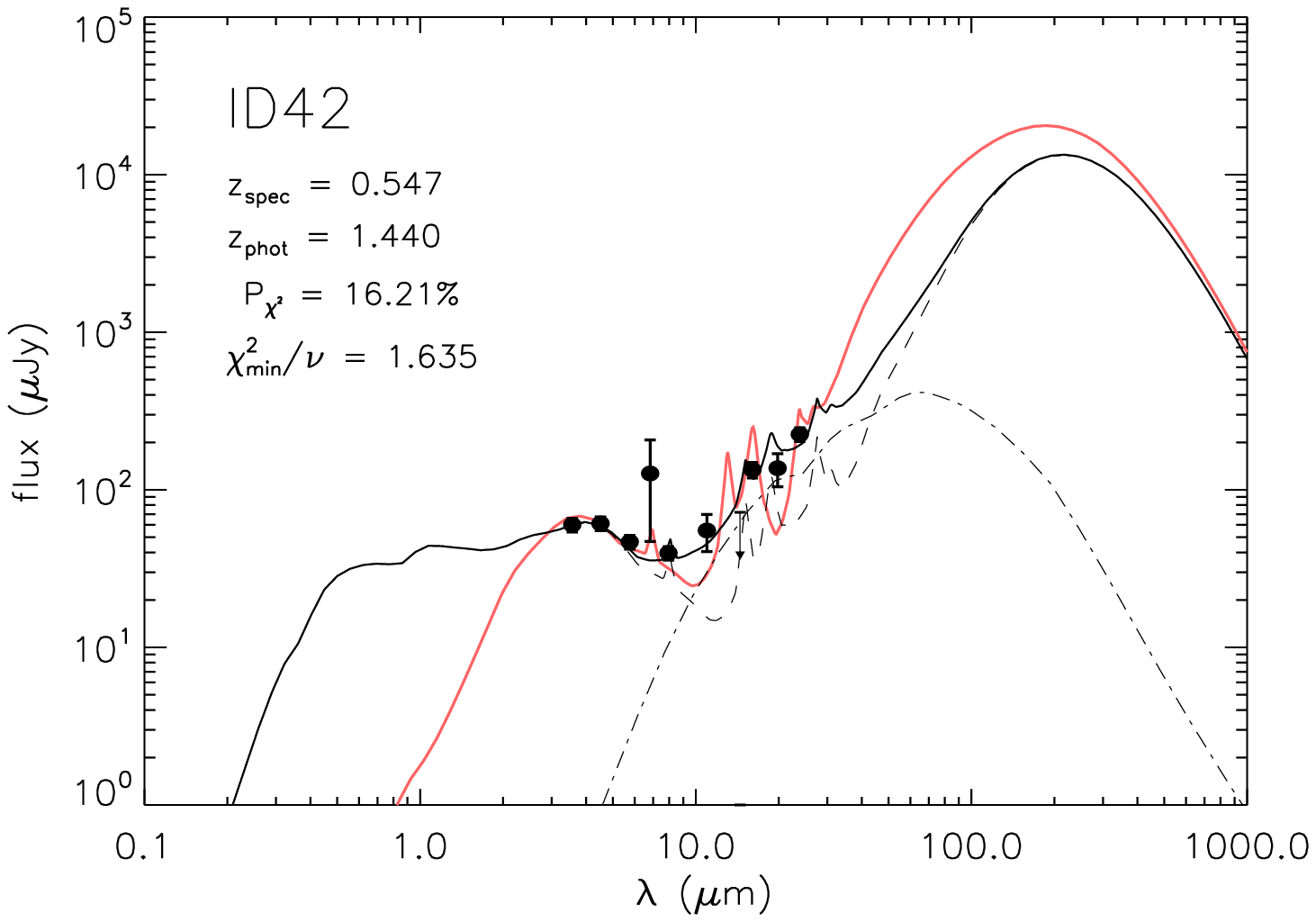}
  \vspace{-5.5cm}\hSlide{-0.5cm}\vSlide{+0.7cm}\ForceHeight{4.8cm}\BoxedEPSF{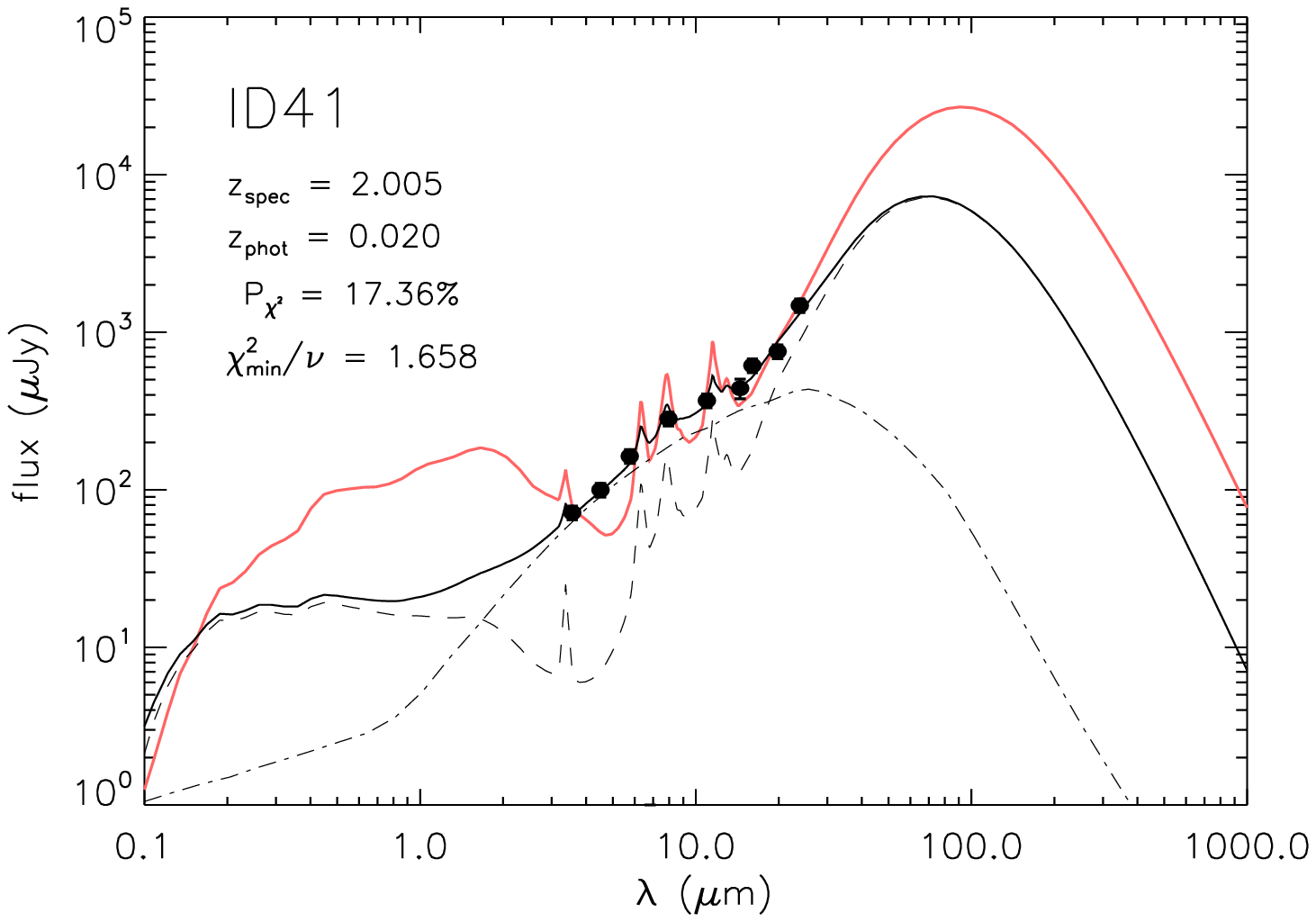} \\
  \hSlide{-3.0cm}\vSlide{+0.0cm}\ForceHeight{4.8cm}\BoxedEPSF{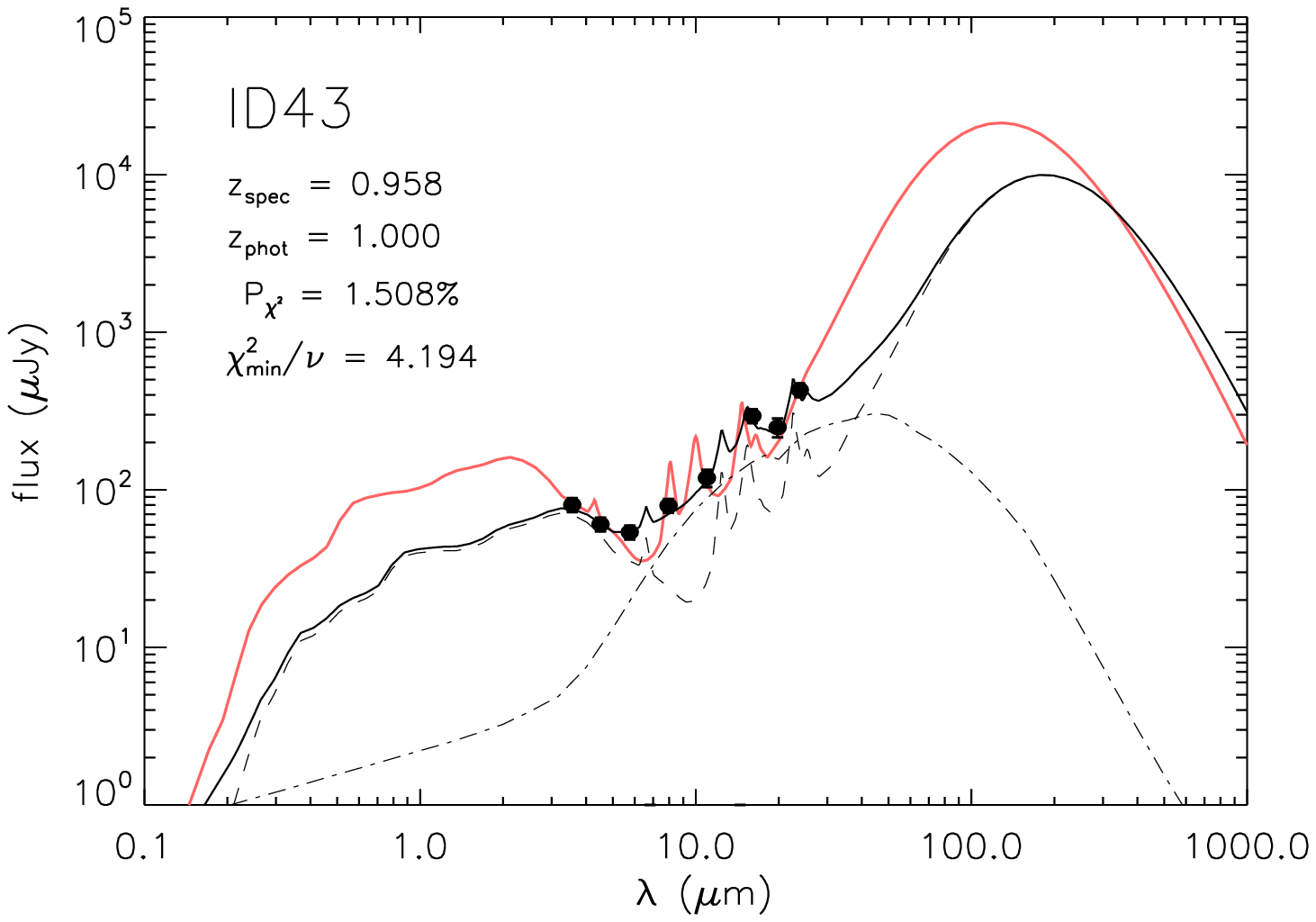}
  \hSlide{+2.0cm}\vSlide{+0.0cm}\ForceHeight{4.8cm}\BoxedEPSF{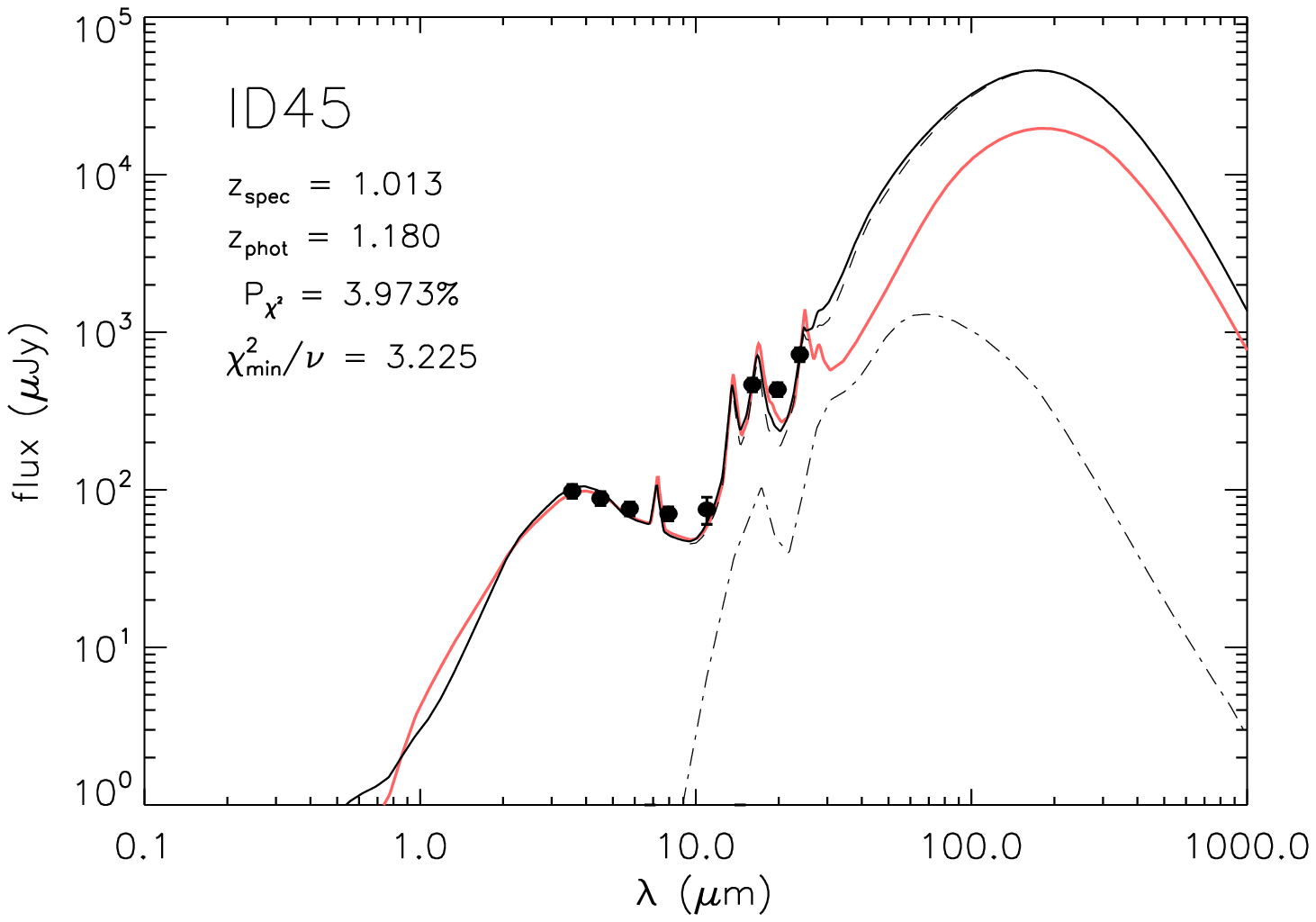}
  \vspace{-5.5cm}\hSlide{-0.5cm}\vSlide{+0.7cm}\ForceHeight{4.8cm}\BoxedEPSF{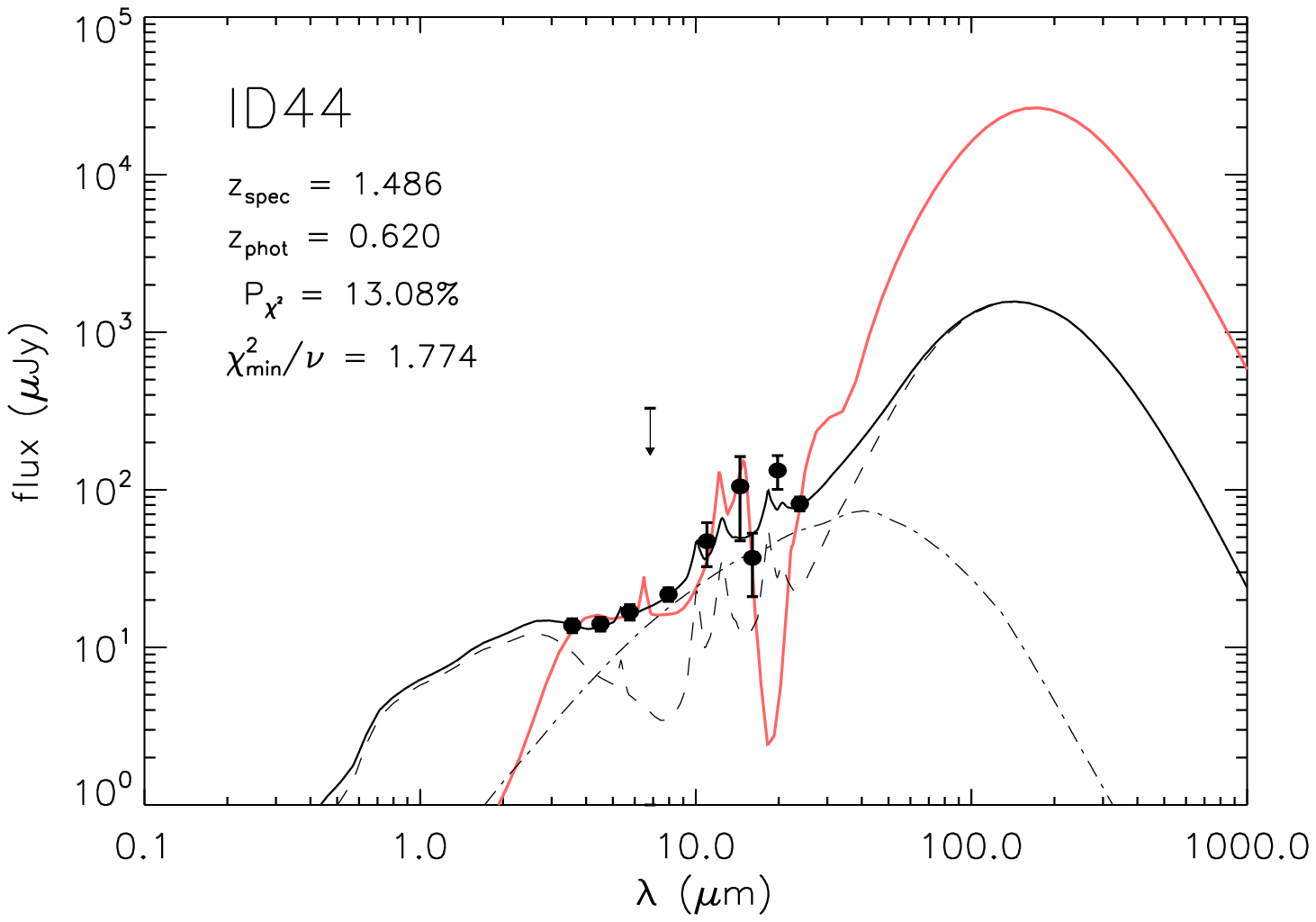}
  \vskip0.2truecm\caption{{\it Continued.}}
\end{figure*}

\setcounter{figure}{3}
\begin{figure*}
  \HideDisplacementBoxes
  \hSlide{-3.0cm}\vSlide{+0.0cm}\ForceHeight{4.8cm}\BoxedEPSF{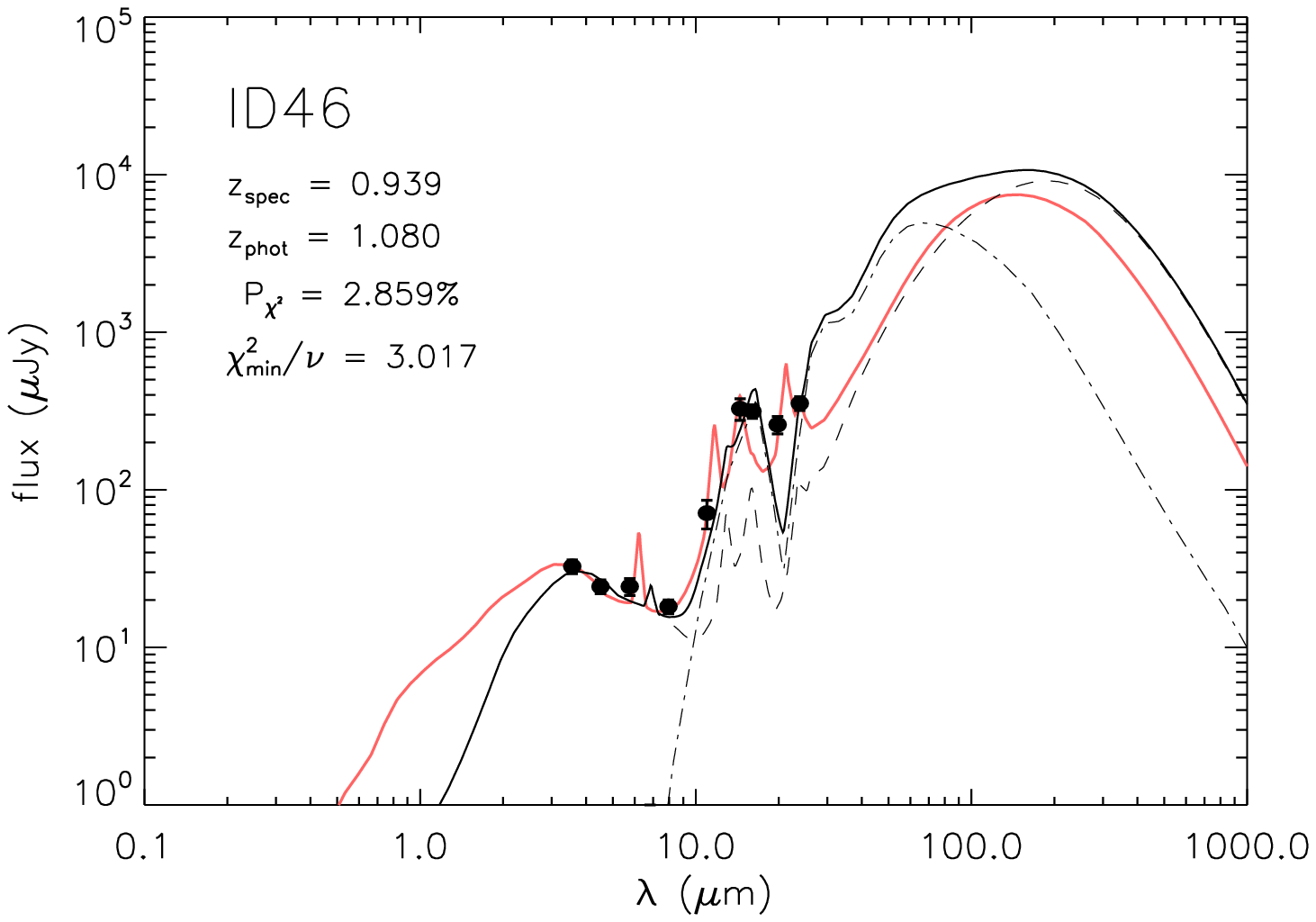}
  \hSlide{+2.0cm}\vSlide{+0.0cm}\ForceHeight{4.8cm}\BoxedEPSF{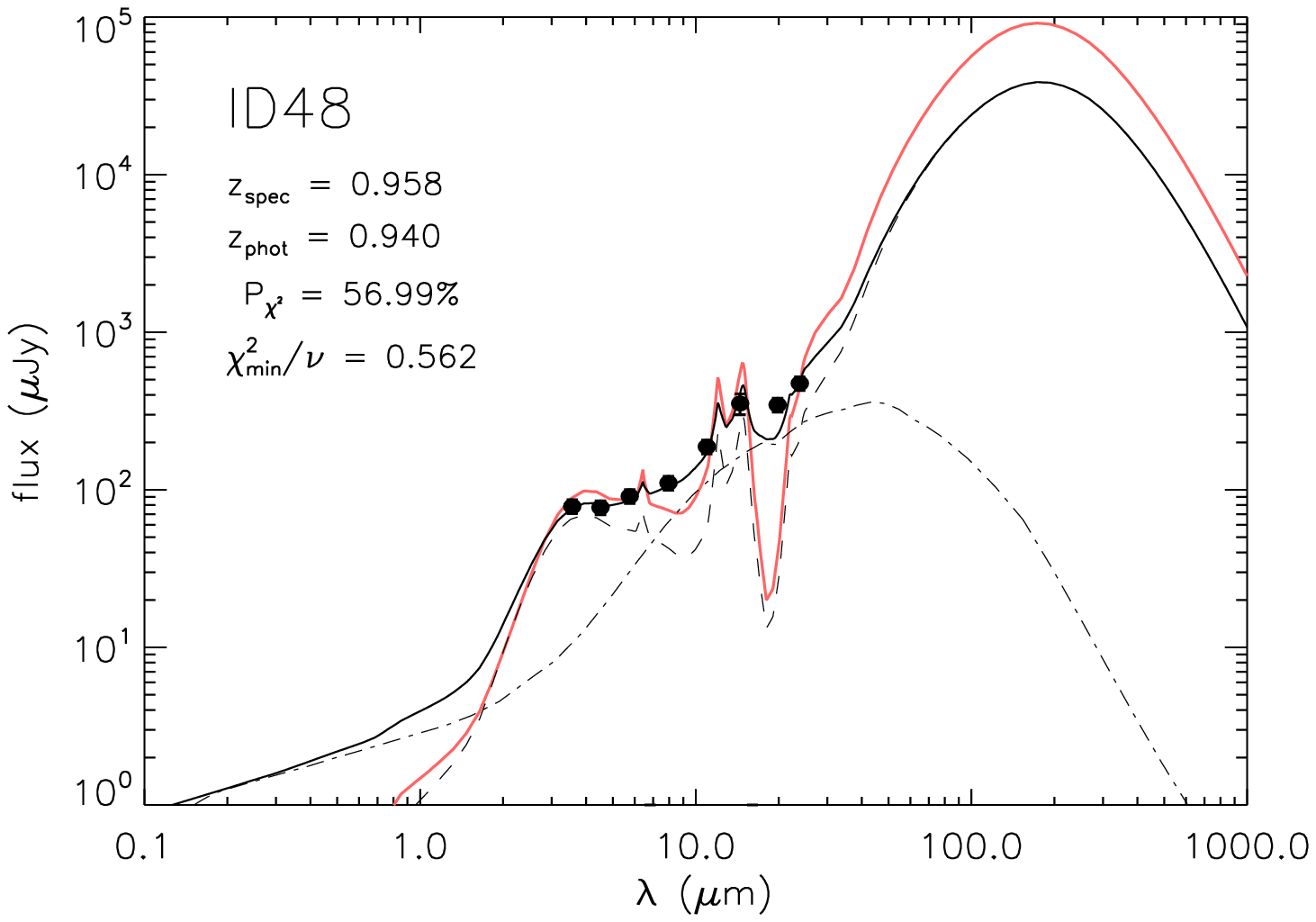}
  \vspace{-5.5cm}\hSlide{-0.5cm}\vSlide{+0.7cm}\ForceHeight{4.8cm}\BoxedEPSF{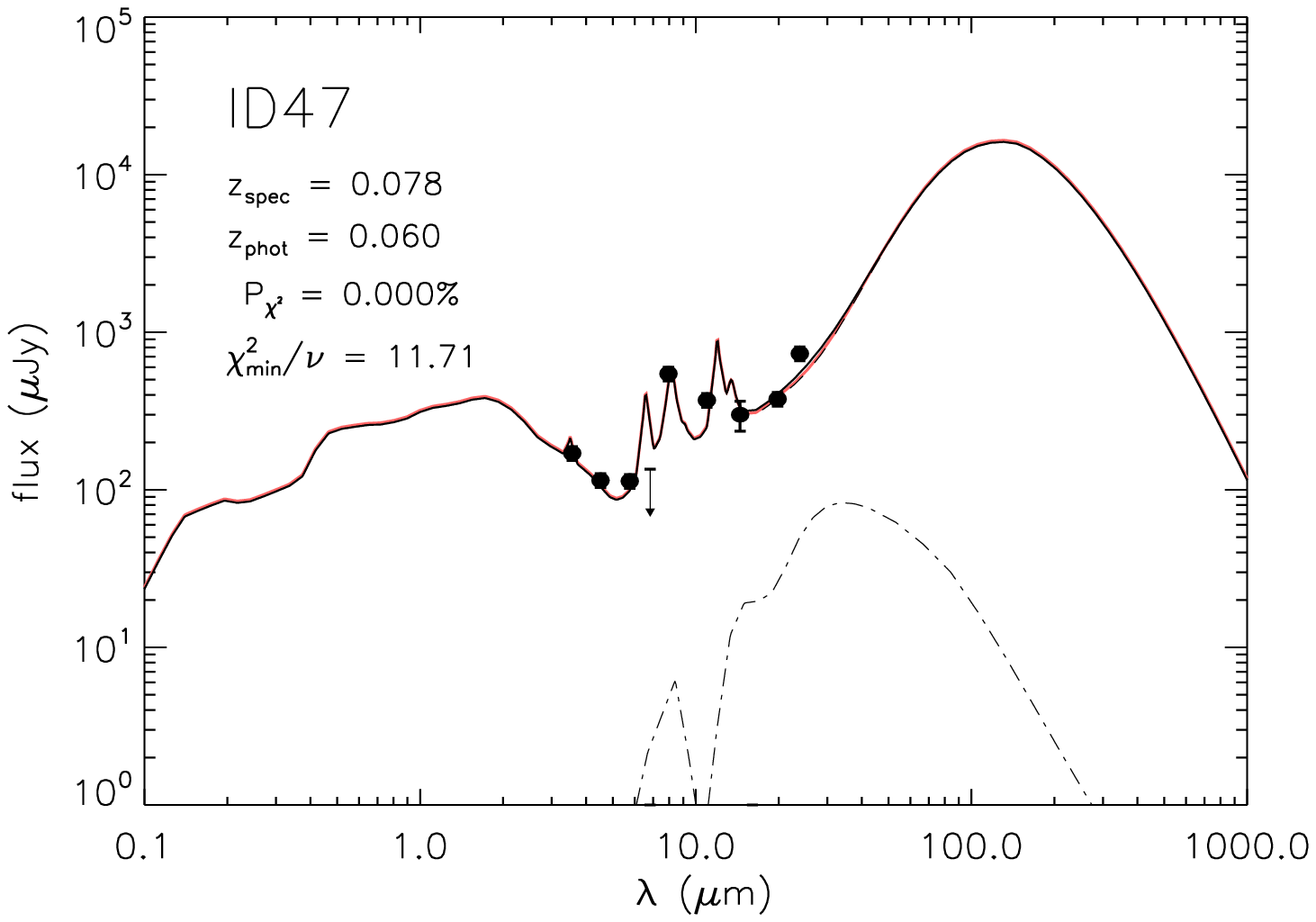} \\
  \hSlide{-3.0cm}\vSlide{+0.0cm}\ForceHeight{4.8cm}\BoxedEPSF{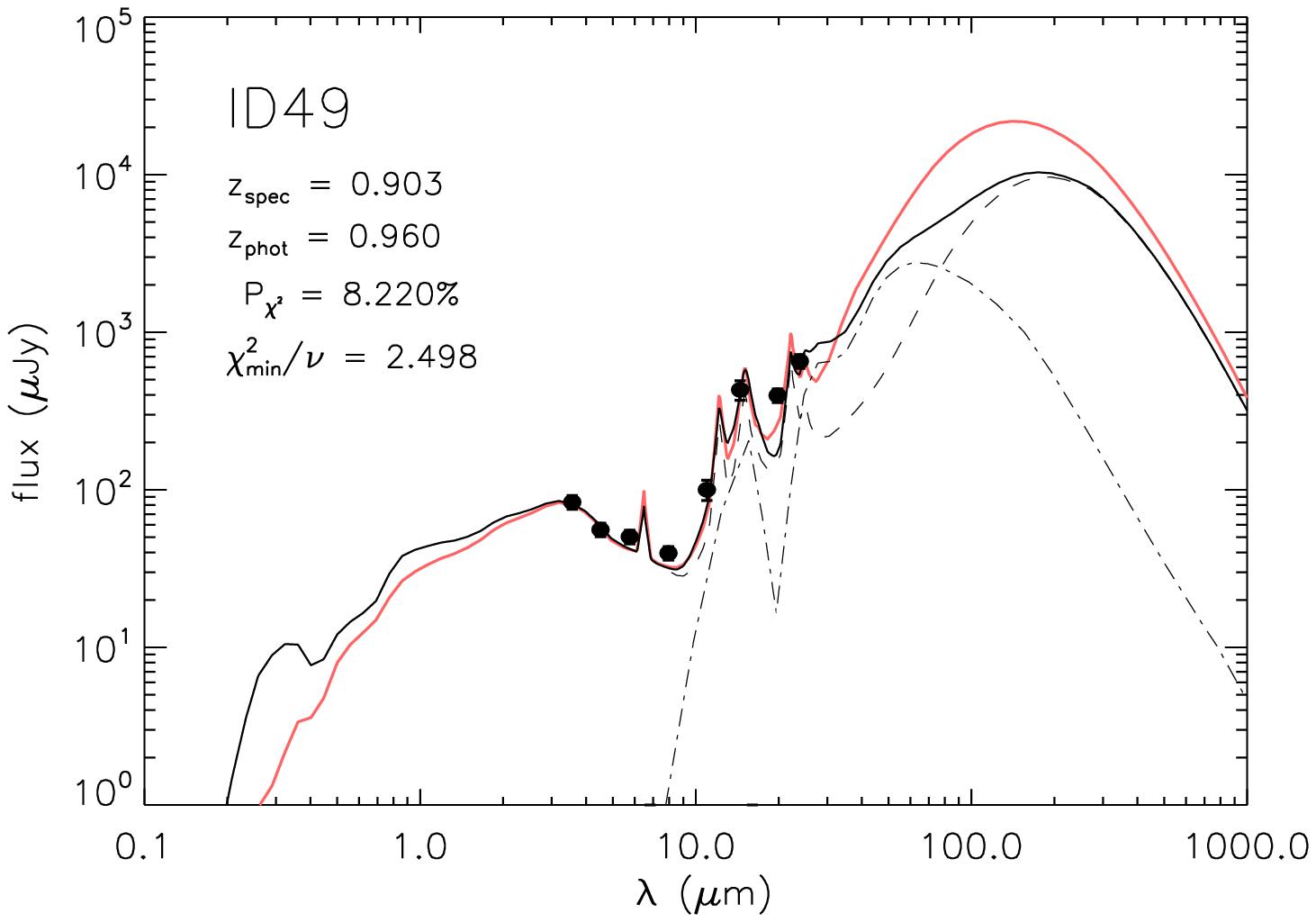}
  \hSlide{+2.0cm}\vSlide{+0.0cm}\ForceHeight{4.8cm}\BoxedEPSF{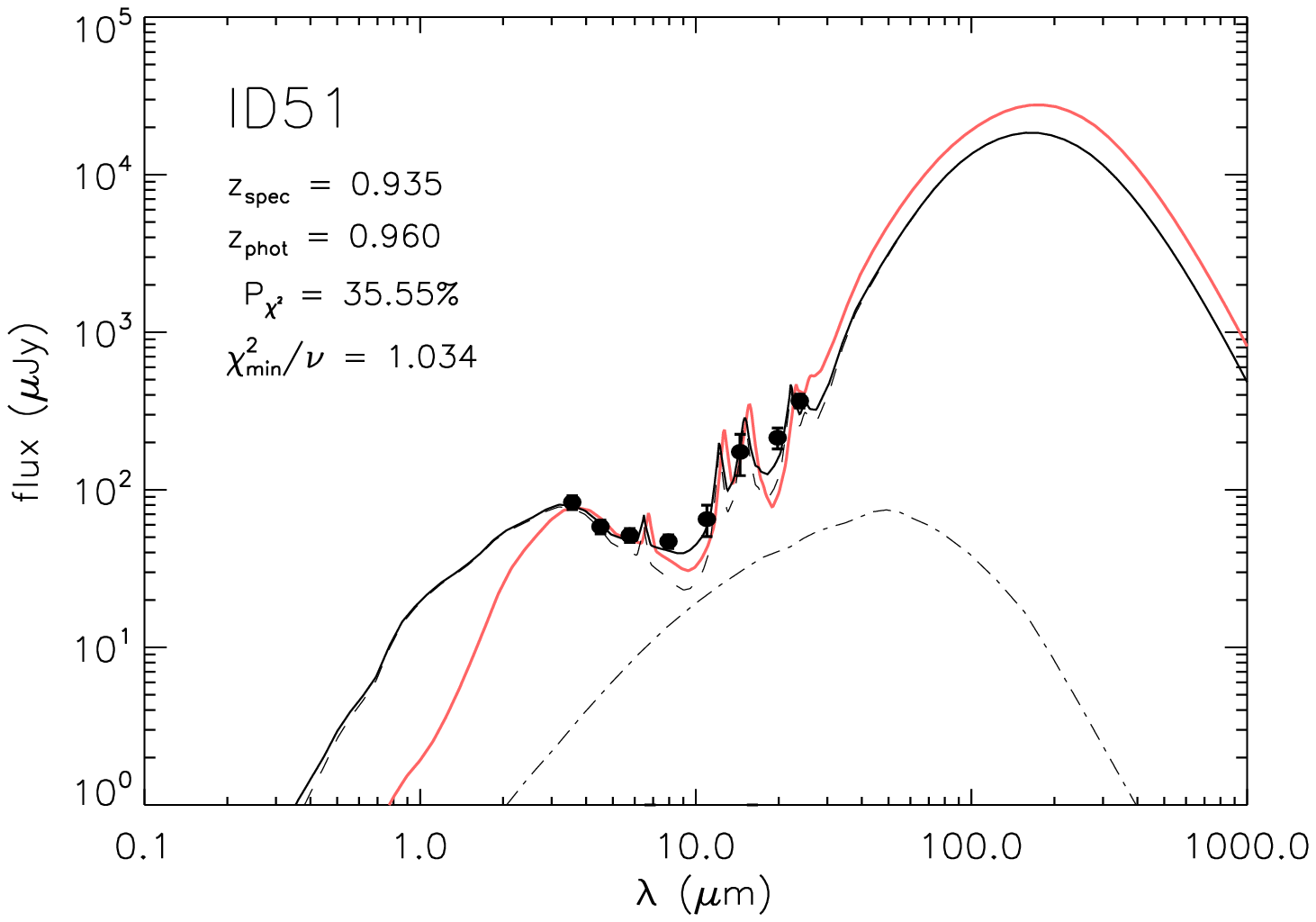}
  \vspace{-5.5cm}\hSlide{-0.5cm}\vSlide{+0.7cm}\ForceHeight{4.8cm}\BoxedEPSF{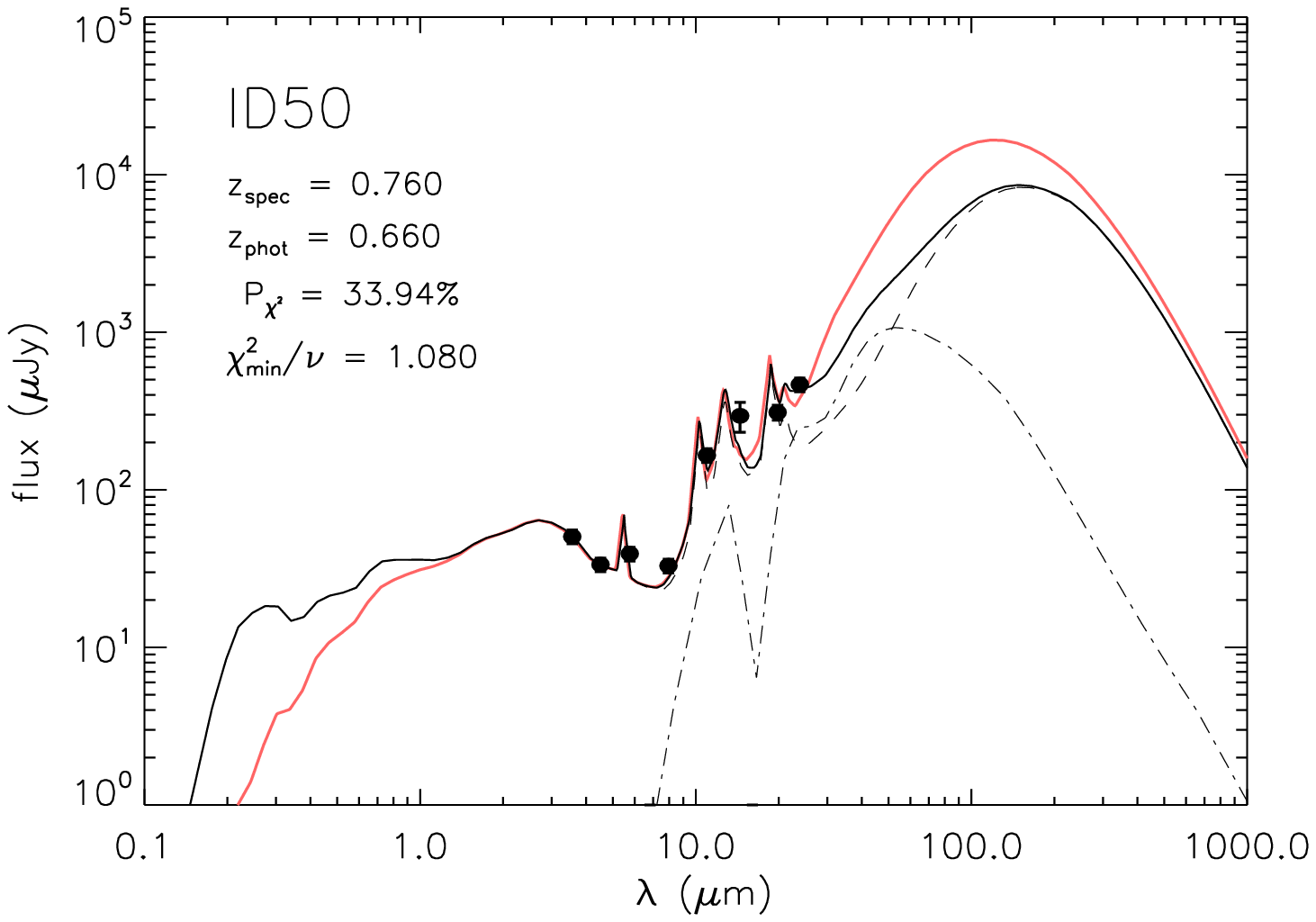} \\
  \hSlide{-3.0cm}\vSlide{+0.0cm}\ForceHeight{4.8cm}\BoxedEPSF{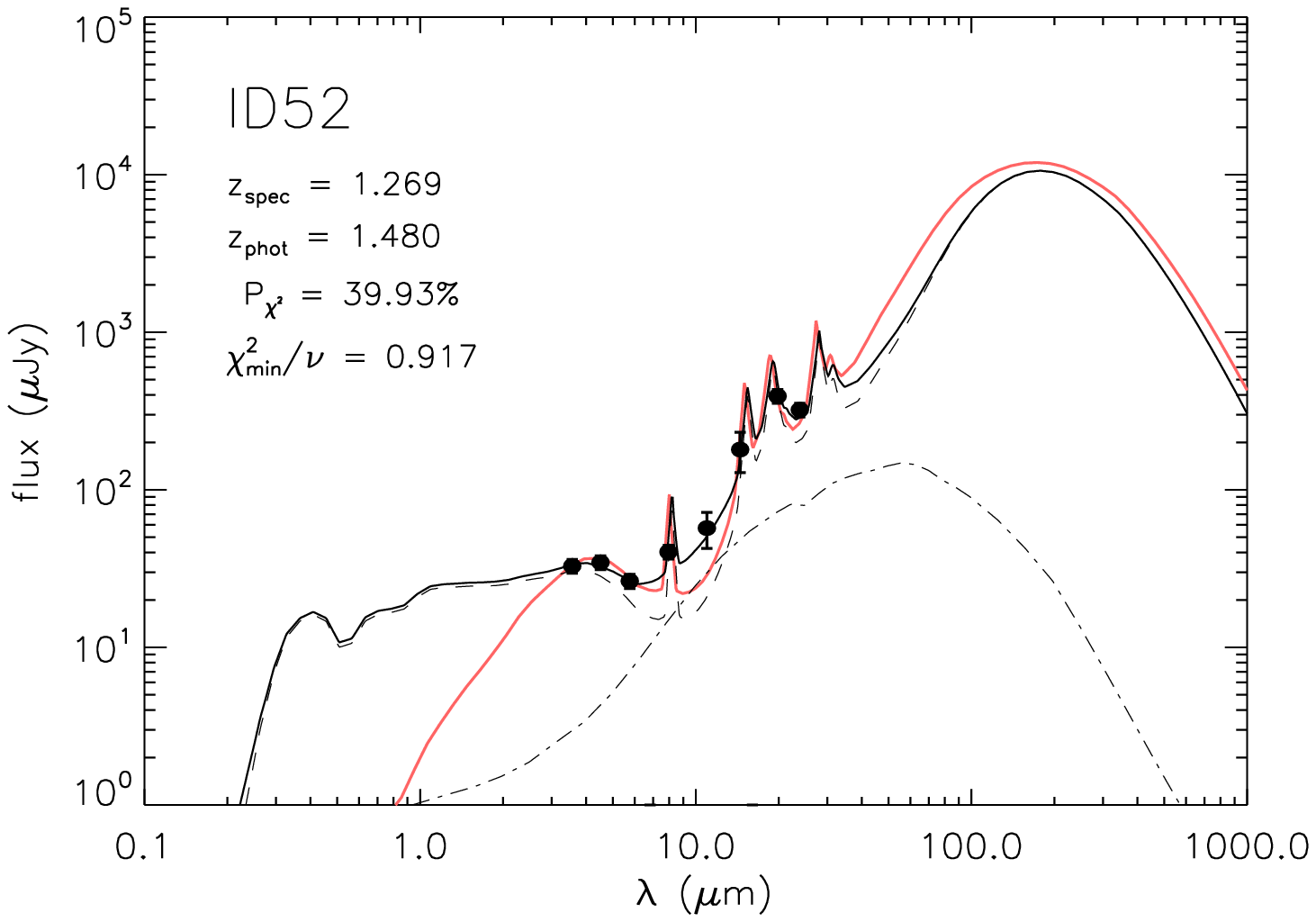}
  \hSlide{+2.0cm}\vSlide{+0.0cm}\ForceHeight{4.8cm}\BoxedEPSF{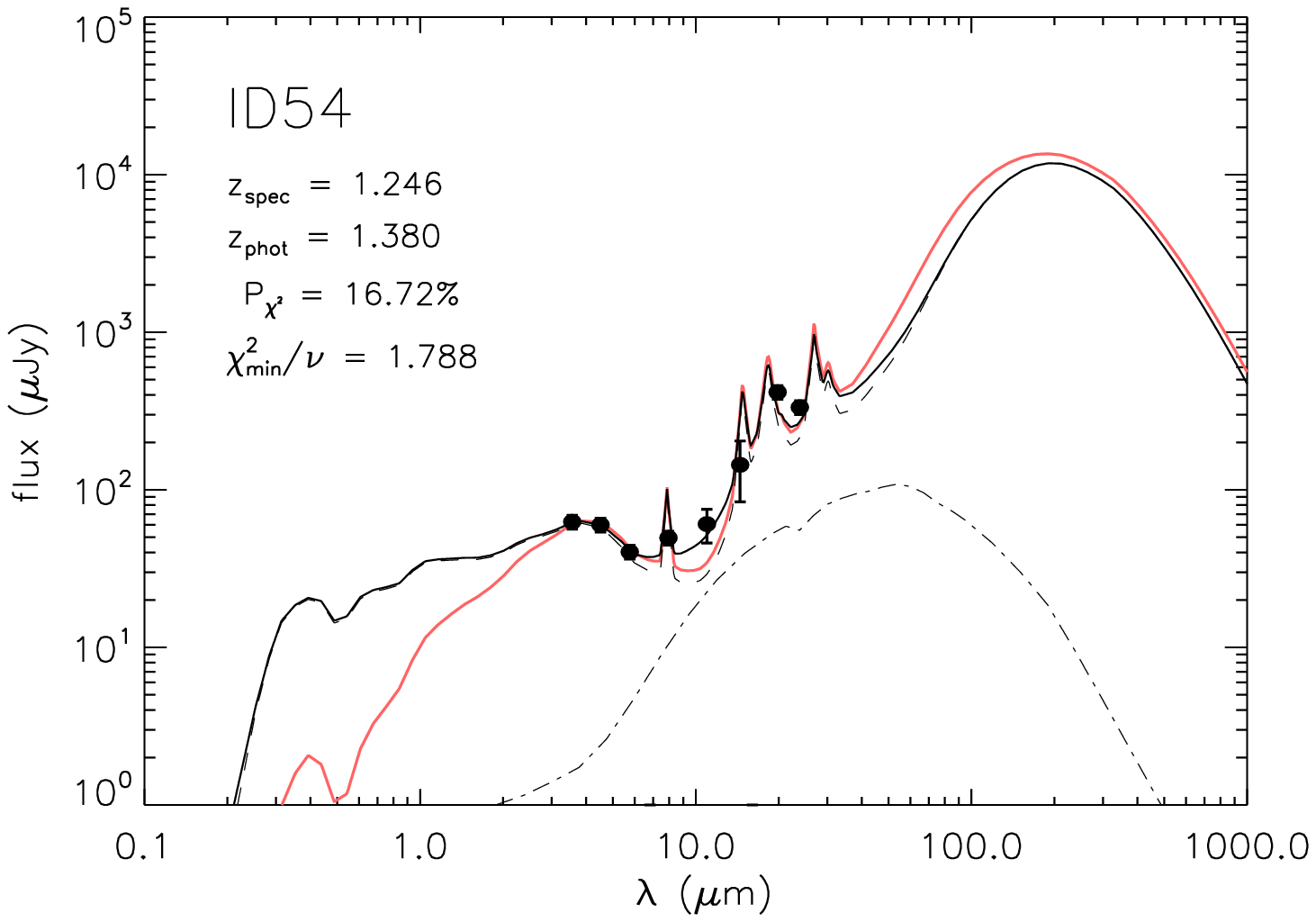}
  \vspace{-5.5cm}\hSlide{-0.5cm}\vSlide{+0.7cm}\ForceHeight{4.8cm}\BoxedEPSF{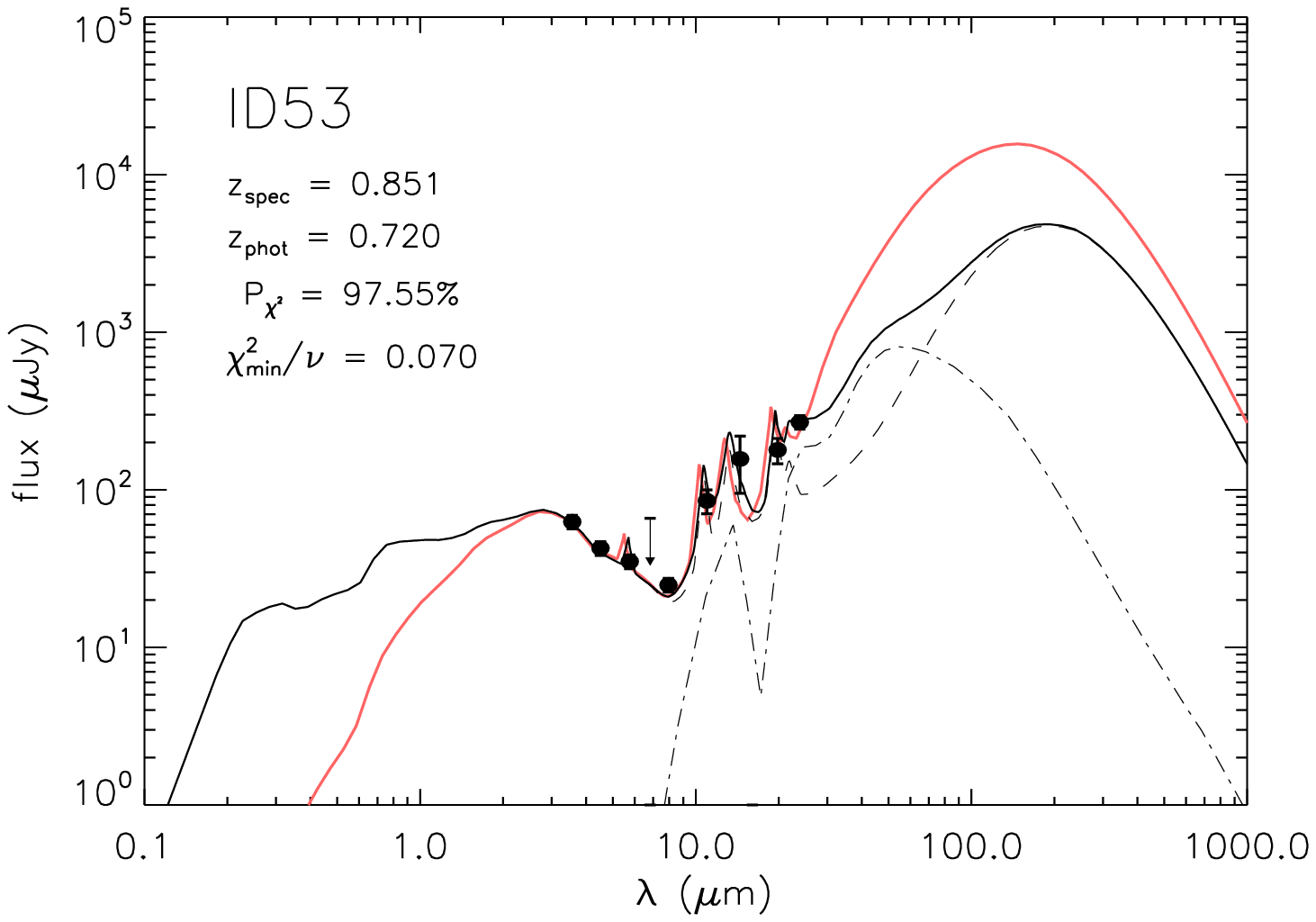} \\
  \hSlide{-3.0cm}\vSlide{+0.0cm}\ForceHeight{4.8cm}\BoxedEPSF{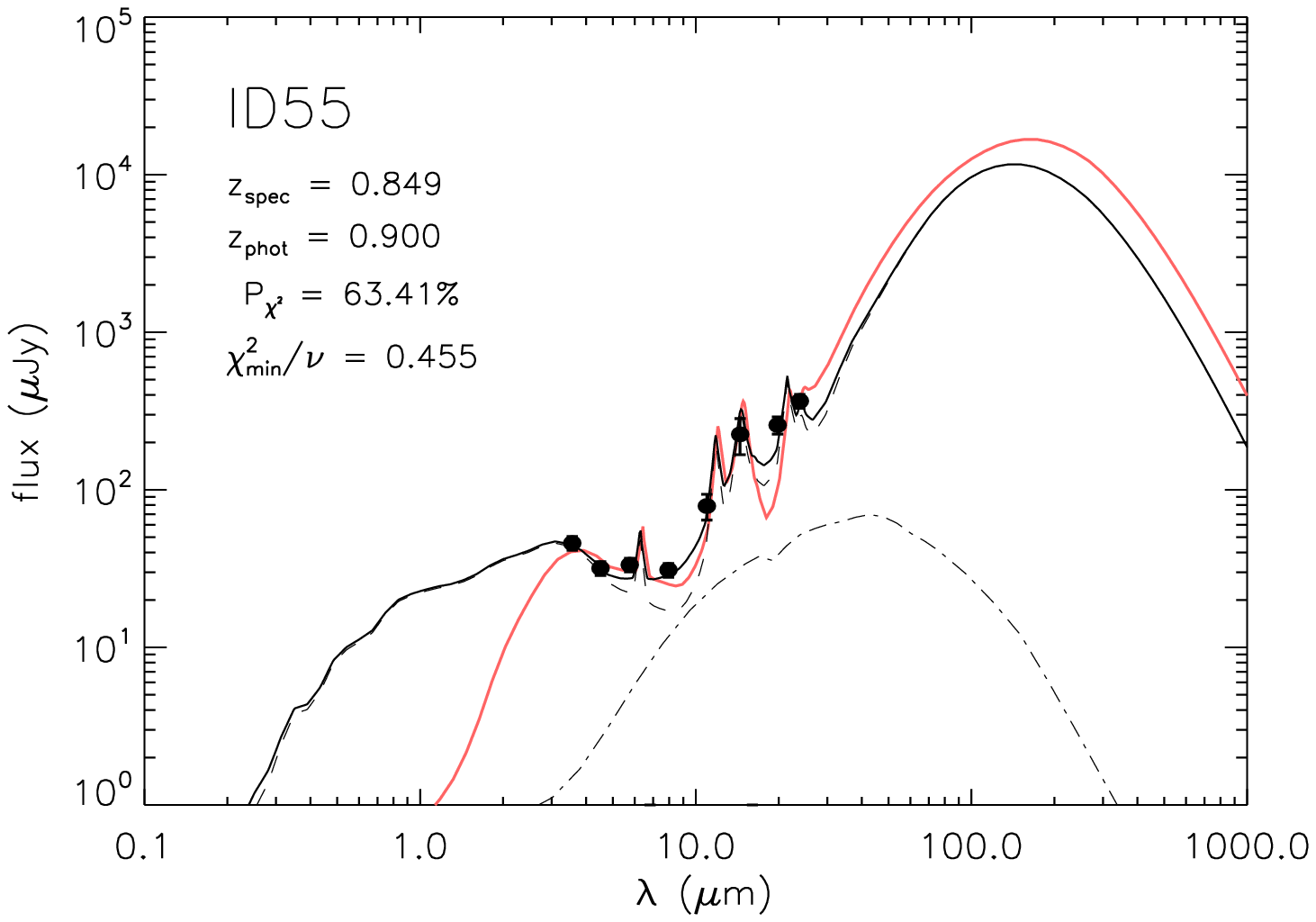}
  \hSlide{+2.0cm}\vSlide{+0.0cm}\ForceHeight{4.8cm}\BoxedEPSF{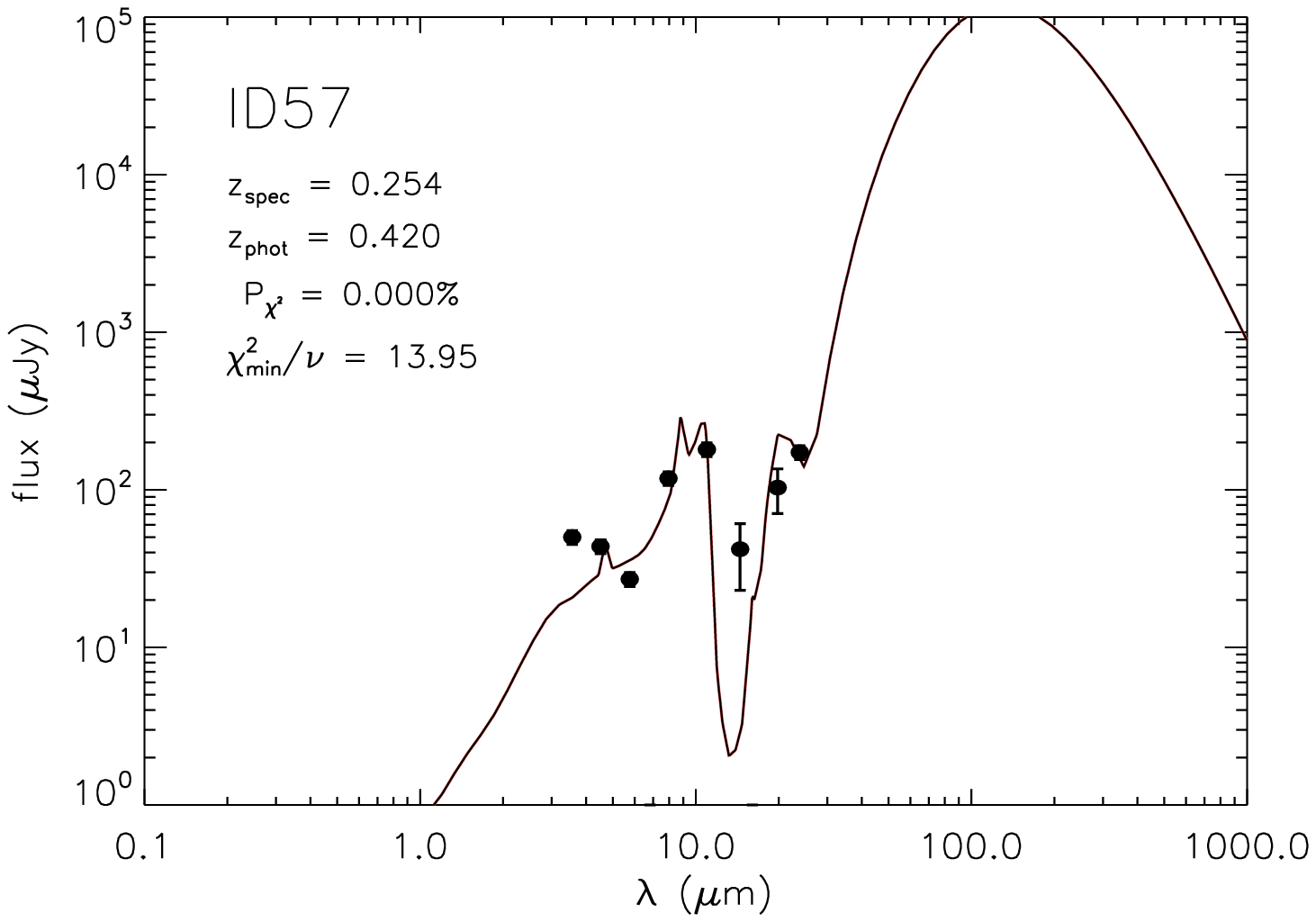}
  \vspace{-5.5cm}\hSlide{-0.5cm}\vSlide{+0.7cm}\ForceHeight{4.8cm}\BoxedEPSF{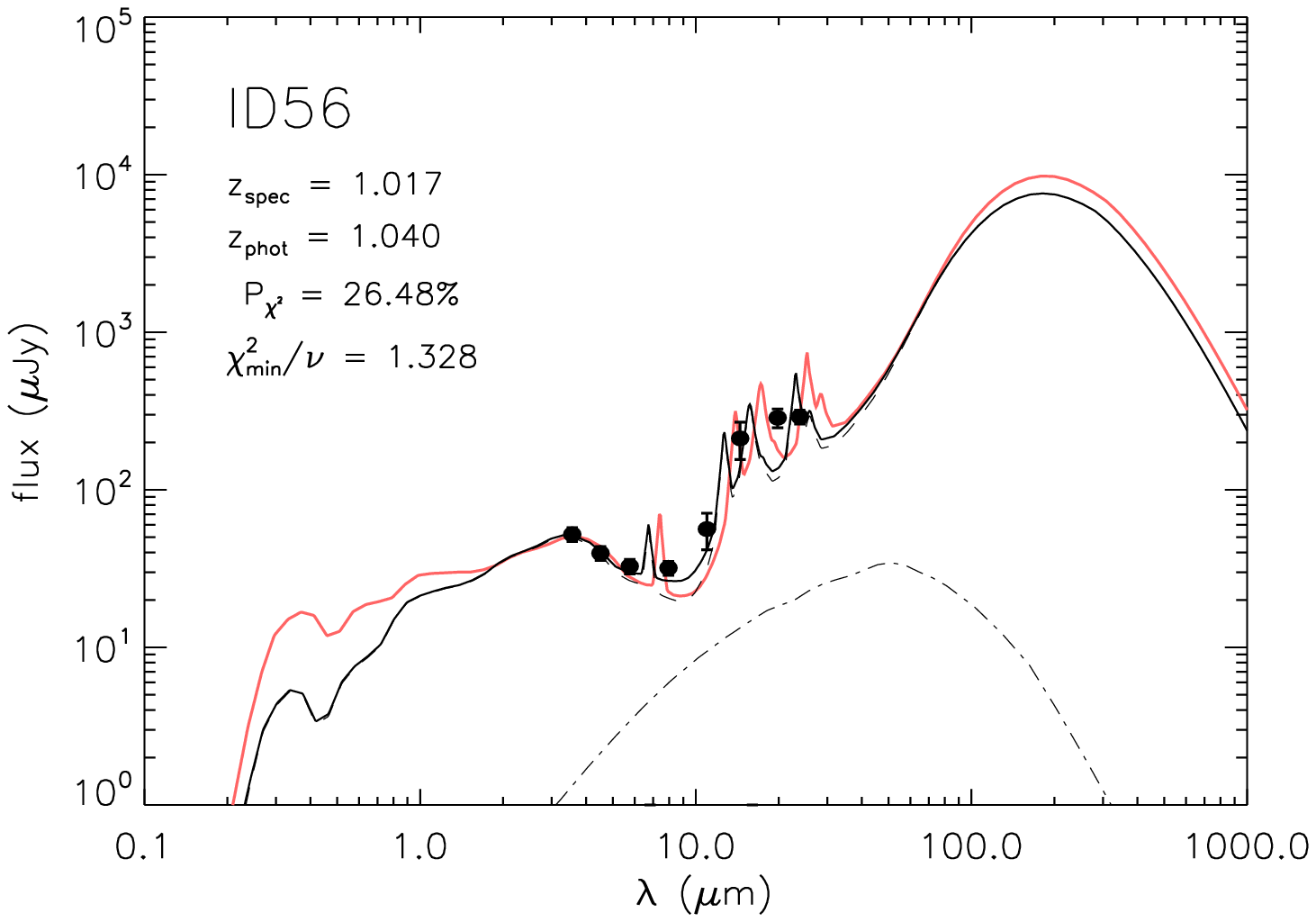} \\
  \hSlide{-3.0cm}\vSlide{+0.0cm}\ForceHeight{4.8cm}\BoxedEPSF{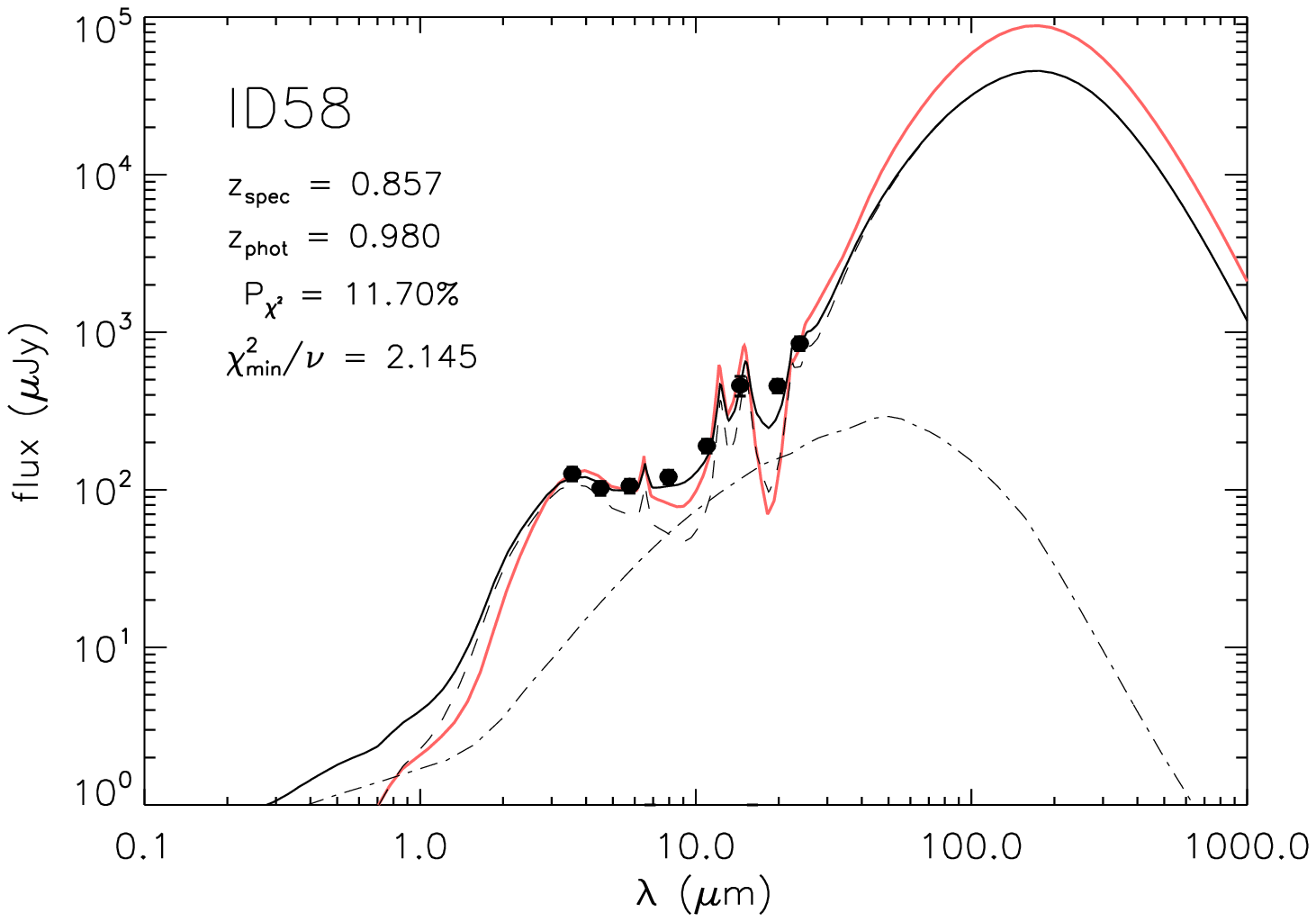}
  \hSlide{+2.0cm}\vSlide{+0.0cm}\ForceHeight{4.8cm}\BoxedEPSF{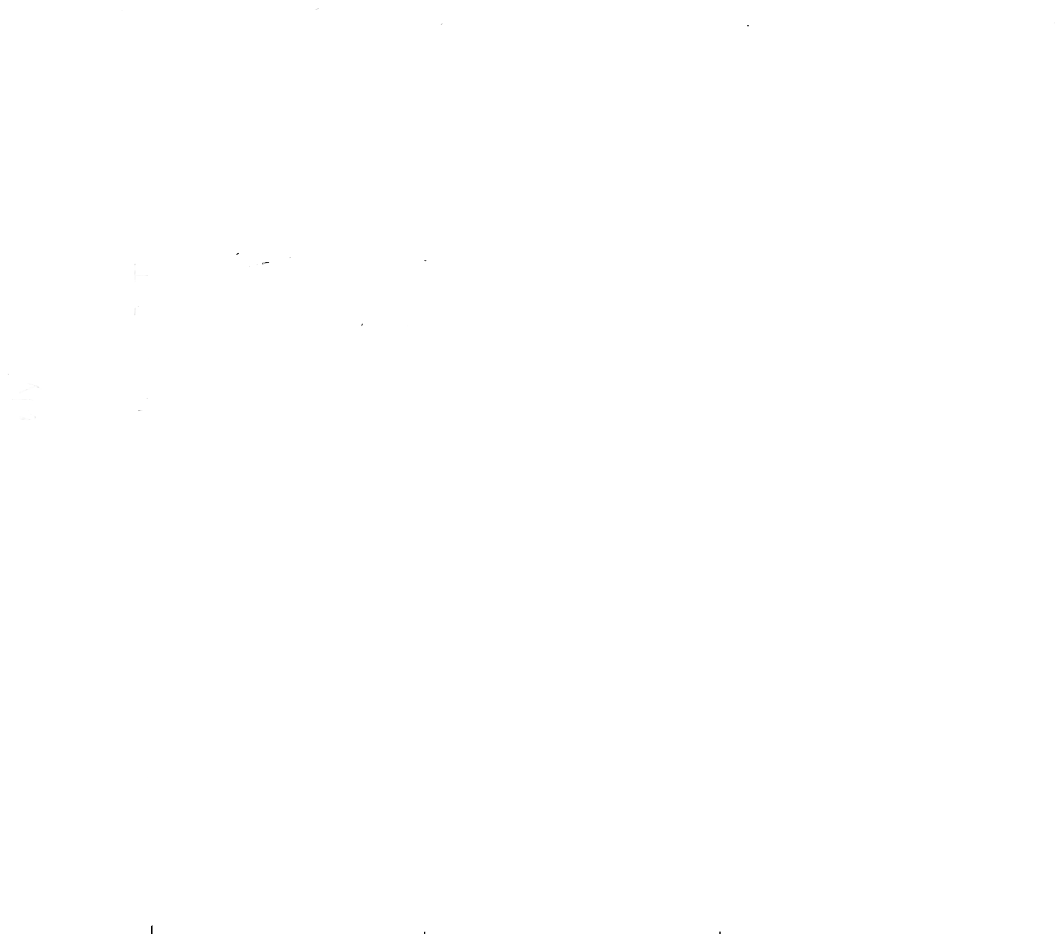}
  \vspace{-5.5cm}\hSlide{-0.5cm}\vSlide{+0.7cm}\ForceHeight{4.8cm}\BoxedEPSF{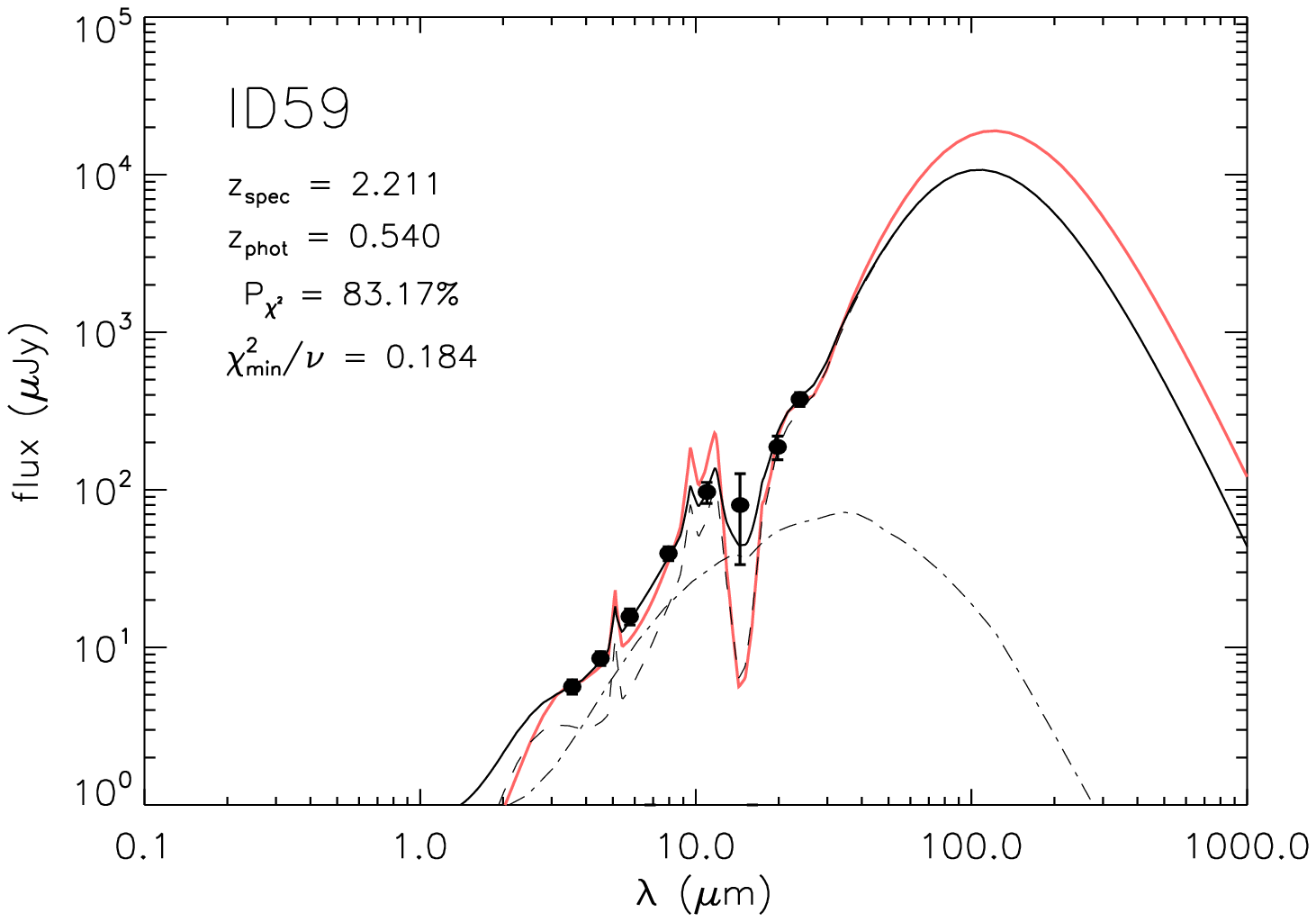}
  \vskip0.2truecm\caption{{\it Continued.}}
\end{figure*}

\setcounter{table}{2}
\begin{table*}
  \vspace{-0.0cm}
  \begin{center}
    \scriptsize
    \caption{\label{tab:SEDfit_results} Best SED-fit parameters
      obtained for the sample of infrared sources listed in
      Table~\ref{tab:realcat_info}: photometric redshif ($z_{\rm
        phot}$, second column) with errors corresponding to a 99$\%$
      confidence limit, minimum reduced $\chi^{2}$ ($\chi_{\rm
        min}^{2}/\nu$, third column), number of degrees of freedom
      ($\nu$, fourth column), probability associated to the minimum
      $\chi^{2}$ ($P_{\chi^2}$, fifth column), age of the starburst
      (Age, sixth column), compactness factor ($\Theta$, seventh
      column), extinction curve (Ext., eighth column), viewing angle
      of the AGN dust torus ($\theta_{\rm view}$, ninth column),
      luminosity contributed by the starburst component ($\log L_{\rm
        SB}$, tenth column), luminosity contributed by the AGN
      component ($\log L_{\rm AGN}$, eleventh column).  The best-fit
      luminosities have been derived integrating the rest-frame SED
      from 0.1 to 1000 $\mu$m.  The spectroscopic redshifts of the
      sources ($z_{\rm spec}$) are listed for comparison.  The other
      columns show the results obtained by setting {\it a priori} the
      AGN fraction, $f_{2}$, to zero and fitting the observed SED with
      the starburst SED templates alone.  An asterisk marks the
      sources with uncertain counterpart at {\it AKARI} and {\it ISO}
      wavebands.}
    \vspace{-0.2cm}
    \begin{tabular}{lrrrrrrrrrrrrrrr}
      \hline
      \hline
      ID & $z_{\rm spec}$ & $z_{\rm phot}$ & $\chi_{\rm min}^{2}/\nu$ & $\nu$ & $P_{\chi^2}$ & Age   &  $\Theta$  & Ext. & $\theta_{\rm view}$ & $\log L_{\rm SB}$  & $\log L_{\rm AGN}$ &
      $z_{\rm phot}$ & $\chi_{\rm min}^{2}/\nu$ & $\nu$ & $P_{\chi^2}$ \\
      &              &               &                        &        &             & (Gyr)  &           &       & ($^{\circ}$)        & ($L_{\odot}$)      & ($L_{\odot}$) &
      &   &     &        \\
      \hline
      \hline
      ID1         & 0.1389 & 0.10$^{+0.04}_{-0.02}$ &  6.67 & 4 & $\ll0.01$ &  0.5 &  5.0 &  MW &  0 & 10.17 &  6.60 & 0.10$^{+0.04}_{-0.02}$ &  4.45 & 6 & $\ll0.01$ \\ 
      ID2         & 0.1133 & 0.10$^{+0.04}_{-0.04}$ &  6.19 & 4 & $\ll0.01$ &  0.6 &  5.0 &  MW &  - & 10.11 &     - & 0.10$^{+0.04}_{-0.04}$ &  6.19 & 4 & $\ll0.01$ \\
      ID3$^{\ast}$ & 0.2759 & 0.22$^{+0.22}_{-0.04}$ &  3.53 & 2 &     0.029 &  0.4 &  5.0 &  MW & 27 & 11.09 &  9.52 & 0.24$^{+0.14}_{-0.04}$ &  2.05 & 4 & 0.085   \\
      ID4         & 0.4573 & 0.42$^{+0.08}_{-0.20}$ &  8.37 & 3 & $\ll0.01$ &  0.6 &  5.0 &  MW &  0 & 11.87 & 10.59 & 0.44$^{+0.04}_{-0.12}$ &  6.61 & 5 & $\ll0.01$ \\
      ID5         & 2.5920 & 0.04$^{+4.96}_{-0.02}$ &  2.68 & 2 &     0.069 &  0.6 &  5.0 & SMC & 36 &  8.93 &  8.02 & 0.98$^{+0.06}_{-0.06}$ &  7.19 & 4 & $\ll0.01$  \\
      ID6         & 0.4225 & 0.24$^{+0.24}_{-0.06}$ & 13.31 & 2 & $\ll0.01$ &  0.6 &  1.8 &  MW & 45 & 10.66 &  9.05 & 0.24$^{+0.22}_{-0.04}$ & 7.37 & 4 & $\ll0.01$  \\
      ID7         & 0.5560 & 0.54$^{+0.08}_{-0.06}$ &  1.71 & 2 &     0.181 &  0.6 &  3.0 &  MW & 45 & 11.36 &  9.39 & 0.52$^{+0.06}_{-0.04}$ & 0.97 & 4 & 0.423  \\
      ID8         & 0.3765 & 0.22$^{+0.22}_{-0.10}$ &  2.41 & 2 &     0.090 &  0.6 &  5.0 &  MW & 36 & 10.50 &  8.71 & 0.22$^{+0.18}_{-0.08}$ & 2.14 & 4 & 0.073 \\
      ID9         & 0.4099 & 0.24$^{+0.22}_{-0.06}$ & 12.87 & 3 & $\ll0.01$ &  0.4 &  2.6 & LMC & 90 & 10.98 &  9.38 & 0.24$^{+0.22}_{-0.04}$ & 8.85 & 5 & $\ll0.01$ \\
      ID10        & 0.6382 & 0.62$^{+0.08}_{-0.06}$ &  3.51 & 2 &     0.030 &  0.4 &  3.0 &  MW &  0 & 11.81 & 10.86 & 0.60$^{+0.12}_{-0.06}$ & 2.45 & 4 & 0.044  \\
      ID11        & 0.6393 & 0.88$^{+0.08}_{-0.36}$ & 10.80 & 2 & $\ll0.01$ &  0.3 &  0.5 & SMC &  0 & 12.45 & 11.25 & 0.82$^{+0.12}_{-0.04}$ & 6.05 & 4 & $\ll0.01$ \\
      ID12        & 0.4335 & 0.24$^{+0.38}_{-0.10}$ & 11.21 & 2 & $\ll0.01$ &  0.6 &  2.4 & LMC & 45 & 10.87 &  9.38 & 0.90$^{+0.06}_{-0.06}$ & 5.75 & 4 & $\ll0.01$  \\
      ID13        & 0.5186 & 0.44$^{+0.36}_{-0.28}$ &  7.07 & 2 & $\ll0.01$ &  0.5 &  5.0 & LMC & 63 & 11.70 & 10.18 & 0.26$^{+0.22}_{-0.06}$ & 4.66 & 4 & $\ll0.01$  \\
      ID14        & 0.4191 & 0.24$^{+0.28}_{-0.06}$ &  9.79 & 2 & $\ll0.01$ &  0.4 &  2.6 & LMC & 45 & 10.80 &  9.42 & 0.24$^{+0.08}_{-0.04}$ & 6.87 & 4 & $\ll0.01$  \\
      ID15        & 0.5560 & 0.58$^{+0.16}_{-0.08}$ &  3.23 & 4 &     0.012 &  0.5 &  2.2 &  MW & 81 & 11.36 & 10.09 & 0.54$^{+0.06}_{-0.06}$ & 2.83 & 6 & $\ll0.01$  \\
      ID16        & 0.5174 & 0.48$^{+0.24}_{-0.18}$ &  0.27 & 2 &     0.764 &  0.6 &  3.0 &  MW &  0 & 10.94 &  9.43 & 0.46$^{+0.08}_{-0.16}$ & 0.32 & 4 & 0.866  \\
      ID17        & 0.6419 & 0.76$^{+0.08}_{-0.18}$ &  1.29 & 4 &     0.273 &  0.5 &  3.0 &  MW & 90 & 11.53 &  8.71 & 0.76$^{+0.08}_{-0.18}$ & 0.86 & 6 & 0.052  \\
      ID18        & 0.8477 & 0.88$^{+0.28}_{-0.18}$ &  1.45 & 4 &     0.215 &  0.6 &  3.0 &  MW & 90 & 11.90 &  9.84 & 0.88$^{+0.22}_{-0.14}$ & 1.06 & 6 & 0.390  \\
      ID19        & 0.5561 & 0.54$^{+0.10}_{-0.04}$ &  2.71 & 4 &     0.028 &  0.5 &  3.0 &  MW & 63 & 11.22 &  9.52 & 0.54$^{+0.04}_{-0.04}$ & 2.05 & 6 & 0.055  \\
      ID20$^{\ast}$ & 0.3758 & 0.24$^{+0.22}_{-0.14}$ & 3.65 & 2 &     0.026 &  0.5 &  5.0 &  MW & 36 & 10.32 &  8.87 & 0.28$^{+0.16}_{-0.08}$ & 2.73 & 4 & 0.028  \\
      ID21        & 1.0164 & 1.10$^{+0.22}_{-0.08}$ &  1.05 & 4 &     0.400 &  0.6 &  5.0 & LMC &  0 & 11.98 & 11.64 & 1.20$^{+0.14}_{-0.28}$ & 1.56 & 6 & 0.154  \\
      ID22 & 0.5560 & 0.96$^{+0.80}_{-0.30}$ & 0.85 & 4 &     0.494 &  0.3 &  5.0 &  MW & 90 & 11.67 & 10.53 & 1.04$^{+0.54}_{-0.30}$ & 1.97 & 6 & 0.067  \\
      ID23        & 0.7840 & 0.82$^{+0.12}_{-0.18}$ &  1.51 & 2 &     0.221 &  0.4 &  2.6 &  MW & 36 & 11.53 &  9.61 & 0.82$^{+0.10}_{-0.12}$ & 0.82 & 4 & 0.512  \\
      ID24$^{\ast}$ & 0.7038 & 0.72$^{+0.26}_{-0.22}$ &  2.10 & 2 &      0.122 &  0.3 &  2.8 & LMC & 36 & 11.39 &  9.92 & 0.58$^{+0.26}_{-0.10}$ & 1.69 & 4 & 0.150  \\
      ID25$^{\ast}$  & 0.4745 & 0.46$^{+0.06}_{-0.14}$ &  1.83 & 4 &    0.120 &  0.4 &  3.0 &  MW & 72 & 11.18 &  9.93 & 0.46$^{+0.04}_{-0.16}$ & 2.31 & 6 & 0.031  \\
      ID26        & 0.8462 & 0.84$^{+0.20}_{-0.10}$ &  4.28 & 4 &   $<0.01$ &  0.4 &   2.8 &  MW &  0 & 11.68 & 10.62 & 0.86$^{+0.10}_{-0.12}$ & 3.01 & 6 & $<0.01$   \\
      ID27        & 0.8497 & 1.08$^{+0.10}_{-0.10}$ &  4.28 & 2 &     0.014 &  0.6 &  5.0 & SMC &  0 & 11.87 & 11.87 & 0.90$^{+0.06}_{-0.12}$ & 3.64 & 4 & $<0.01$  \\
      ID28        & 0.8446 & 0.86$^{+0.36}_{-0.12}$ &  6.43 & 3 & $\ll0.01$ &  0.4 &   2.8 &  MW & 45 & 11.61 &  7.37 & 0.86$^{+0.10}_{-0.10}$ & 3.86 & 5 & $<0.01$  \\
      ID29        & 0.7517 & 0.78$^{+0.18}_{-0.24}$ &  0.26 & 4 &     0.902 &  0.5 &  2.2 &  MW & 36 & 11.30 &  9.62 & 0.64$^{+0.28}_{-0.10}$ & 0.291 & 6 & 0.941  \\
      ID30        & 1.0156 & 1.08$^{+0.24}_{-0.14}$ &  1.70 & 2 &      0.182 &  0.3 &  0.9 & SMC &  0 & 12.22 & 11.66 & 1.14$^{+0.14}_{-0.20}$ & 1.29 & 4 & 0.272  \\
      ID31$^{\ast}$ & 0.5196 & 0.54$^{+0.14}_{-0.06}$ &  9.85 & 2 & $\ll0.01$ &   0.3 &  5.0 &  MW &  0 & 10.93 &  9.82 & 0.54$^{+0.08}_{-0.06}$ & 5.69 & 4 & $\ll0.01$  \\
      ID32        & 0.7913 & 0.84$^{+0.42}_{-0.54}$ &  2.68 & 2 &      0.069 &  0.6 &  0.7 & SMC & 72 & 11.67 & 10.42 & 0.92$^{+0.14}_{-0.10}$ & 2.69 & 4 & 0.029  \\
      ID33        & 0.8510 & 0.90$^{+0.22}_{-0.12}$ &  3.58 & 3 &      0.013 &  0.4 &  3.0 &  MW & 90 & 11.86 &  9.90 & 0.90$^{+0.10}_{-0.10}$ & 2.28 & 5 & 0.044  \\
      ID34        & 0.7920 & 0.96$^{+0.08}_{-0.18}$ &  9.21 & 2 & $\ll0.01$ &  0.2 &  0.7 & SMC &  0 & 12.61 &  9.62 & 0.96$^{+0.06}_{-0.10}$ & 4.61 & 4 & $\ll0.01$  \\
      ID35$^{\ast}$ & 0.6827 & 0.62$^{+0.08}_{-0.06}$ &  0.56 & 4 &     0.694 &  0.3 &  5.0 &  MW &  - & 11.29 &    - & 0.62$^{+0.08}_{-0.06}$ & 0.56 & 4 & 0.694  \\
      ID36        & 1.0124 & 1.16$^{+0.08}_{-0.14}$ &  3.01 & 5 &       0.010 &  0.1 &  0.9 & SMC &  - & 12.53 &    -  & 1.16$^{+0.08}_{-0.08}$ & 3.01 & 5 & 0.010 \\
      ID37$^{\ast}$  & 0.8380 & 0.70$^{+0.14}_{-0.16}$ &  1.52 & 4 &    0.192 &  0.6 &  1.8 &  MW & 45 & 11.28 &  9.69 & 0.64$^{+0.14}_{-0.10}$ & 1.34 & 6 & 0.234  \\
      ID38$^{\ast}$  & 0.8410 & 0.80$^{+0.06}_{-0.10}$ &  0.75 & 6 &    0.609 &  0.4 &  2.4 &  MW &  - & 11.44 &    - & 0.80$^{+0.06}_{-0.10}$ & 0.75 & 6 & 0.609   \\
      ID39        & 1.2190 & 1.34$^{+0.14}_{-0.12}$ &  4.26 & 3 &  $<0.01$ &  0.2 &  0.9 & SMC &  0 & 12.67 & 12.17 & 1.38$^{+0.08}_{-0.06}$ & 2.86 & 5 & 0.014  \\
      ID40        & 0.8470 & 0.90$^{+0.22}_{-0.16}$ &  4.65 & 2 & $<0.01$ &  0.4 &  2.6 &  MW & 36 & 11.70 & 10.25 & 0.92$^{+0.14}_{-0.14}$ & 2.88 & 4 & 0.021  \\
      ID41        & 2.0050 & 0.02$^{+0.52}_{-0.02}$ &  1.66 & 3 &    0.174 &  0.07 &  5.0 & LMC & 45 &  8.31 &  7.75 & 0.02$^{+0.02}_{-0.02}$ & 7.57 & 5 & $\ll0.01$ \\
      ID42        & 0.5480 & 1.44$^{+0.56}_{-0.64}$ &  1.64 & 4 &    0.162 &  0.2 &  5.0 &  SMC & 36 &  12.30 & 11.17 & 1.10$^{+0.66}_{-0.24}$ & 1.72 & 6 & 0.112  \\
      ID43        & 0.9585 & 1.00$^{+0.40}_{-0.82}$ &  4.19 & 2 &    0.015 &  0.4 &  3.0 &  LMC & 90 &  11.87 & 11.05 & 0.30$^{+0.16}_{-0.10}$ & 3.86 & 4 & $<0.01$  \\
      ID44$^{\ast}$ & 1.4865 & 0.62$^{+4.38}_{-0.58}$ &  1.77 & 4 &   0.131 &  0.6 &  1.4 &  LMC & 45 &  10.60 & 10.02 & 0.96$^{+0.06}_{-0.24}$ & 2.79 & 6 & 0.010  \\
      ID45        & 1.0130 & 1.18$^{+0.28}_{-0.14}$ &  3.23 & 2 &    0.040 &  0.2 &  1.2 &  SMC & 18 &  12.55 & 11.22 & 1.20$^{+0.16}_{-0.14}$ & 1.66 & 4 & 0.156  \\
      ID46        & 0.9399 & 1.08$^{+0.10}_{-0.08}$ &  3.02 & 3 &    0.029 &  0.5 &  0.7 &  SMC &  0 &  11.73 & 11.68 & 0.88$^{+0.10}_{-0.08}$ & 3.28 & 5 & $<0.01$ \\
      ID47        & 0.0786 & 0.06$^{+0.04}_{-0.02}$ & 11.71 & 3 & $\ll0.01$ &  0.6 &  5.0 &  MW &  0 &  9.96 &  7.33 & 0.06$^{+0.04}_{-0.02}$ & 7.16 & 5 & $\ll0.01$  \\
      ID48        & 0.9590 & 0.94$^{+1.06}_{-0.66}$ &  0.56 & 2 &    0.570 &  0.6 &  0.4 &  SMC &  72 & 12.16 & 11.10 & 0.94$^{+0.20}_{-0.06}$ & 4.69 & 4 & $\ll0.01$  \\
      ID49        & 0.9032 & 0.96$^{+0.16}_{-0.18}$ &  2.50 & 2 &    0.082 &  0.5 &  3.0 &   MW &  0 & 11.85 & 11.32 & 0.96$^{+0.12}_{-0.10}$ & 1.49 & 4 & 0.202  \\
      ID50        & 0.7610 & 0.66$^{+0.18}_{-0.10}$ &  1.08 & 2 &    0.339 &  0.3 &  5.0 &   MW &  0 & 11.52 & 11.58 & 0.64$^{+0.18}_{-0.08}$ & 0.94 & 4 & 0.442  \\
      ID51        & 0.9350 & 0.96$^{+0.38}_{-0.24}$ &  1.03 & 2 &    0.356 &  0.3 &  1.8 &  SMC & 45 & 12.00 & 10.41 & 1.04$^{+0.20}_{-0.28}$ & 1.10 & 4 & 0.355  \\
      ID52        & 1.2698 & 1.48$^{+0.18}_{-0.28}$ &  0.92 & 2 &    0.399 &  0.1 &  5.0 &   MW & 72 & 12.25 & 11.09 & 1.42$^{+0.20}_{-0.22}$ & 1.85 & 4 & 0.115 \\
      ID53        & 0.8517 & 0.72$^{+0.26}_{-0.22}$ &  0.07 & 3 &    0.976 &  0.5 &  5.0 &   MW &  0 & 11.53 & 10.54 & 0.66$^{+0.32}_{-0.12}$ & 0.17 & 5 & 0.973  \\
      \multicolumn{10}{r}{\emph{continued on next page}}
    \end{tabular}
  \end{center}
\end{table*}

\setcounter{table}{2}
\begin{table*}
  \vspace{-0.0cm}
  \scriptsize
  \caption{{\it Continued}}
  \vspace{-0.2cm}
  \begin{center}
    \begin{tabular}{lrrrrrrrrrrrrrrr}
      \hline
      \hline
      ID & $z_{\rm spec}$ & $z_{\rm phot}$ & $\chi_{\rm min}^{2}/\nu$ & $\nu$ & $P_{\chi^2}$ & Age   &  $\Theta$  & Ext. & $\theta_{\rm view}$ & $\log L_{\rm SB}$  & $\log L_{\rm AGN}$ &
      $z_{\rm phot}$ & $\chi_{\rm min}^{2}/\nu$ & $\nu$ & $P_{\chi^2}$ \\
      &              &               &                        &        &             & (Gyr)  &           &       & ($^{\circ}$)        & ($L_{\odot}$)      & ($L_{\odot}$) &
      &   &     &        \\
      \hline
      \hline
      ID54        & 1.2463 & 1.38$^{+0.28}_{-0.18}$ &  1.79 & 2 &    0.167 &  0.2 &  5.0 &   MW & 90 & 12.25 & 10.87 & 1.38$^{+0.24}_{-0.14}$ & 1.41 & 4 & 0.227  \\
      ID55        & 0.8493 & 0.90$^{+0.18}_{-0.36}$ &  0.46 & 2 &     0.634 &  0.2 &  2.8 &  LMC & 90 & 11.79 & 10.32 & 0.94$^{+0.18}_{-0.42}$ & 1.50 & 4 & 0.198  \\
      ID56        & 1.0174 & 1.04$^{+1.02}_{-0.30}$ &  1.33 & 2 &    0.265 &  0.4 &  3.0 &   MW & 45 & 11.76 & 10.15 & 1.24$^{+0.40}_{-0.36}$ & 1.09 & 4 & 0.359  \\
      ID57        & 0.2542 & 0.42$^{+0.04}_{-0.06}$ & 13.95 & 4 & $\ll0.01$ &  0.4 & 0.3 &  SMC &  - & 11.81 &     - & 0.42$^{+0.04}_{-0.06}$ & 13.95 & 4 & $\ll0.01$ \\
      ID58        & 0.8572 & 0.98$^{+0.20}_{-0.24}$ &  2.15 & 2 &     0.117 &  0.3 &  0.9 &  SMC & 45 & 12.35 & 11.02 & 0.96$^{+0.08}_{-0.08}$ & 2.21 & 4 & 0.065 \\
      ID59        & 2.2110 & 0.54$^{+0.52}_{-0.48}$ &  0.18 & 2 &     0.832 &  0.07 &  0.7 &  SMC & 90 & 11.16 & 9.88 & 0.54$^{+0.12}_{-0.04}$ & 1.93 & 4 & 0.103 \\
      \hline
      \hline
    \end{tabular}
  \end{center}
\end{table*}
\normalsize

\setcounter{figure}{4}
\begin{figure*}
  \hspace{-2.5cm}\includegraphics[height=11.5cm,width=16.0cm]
  {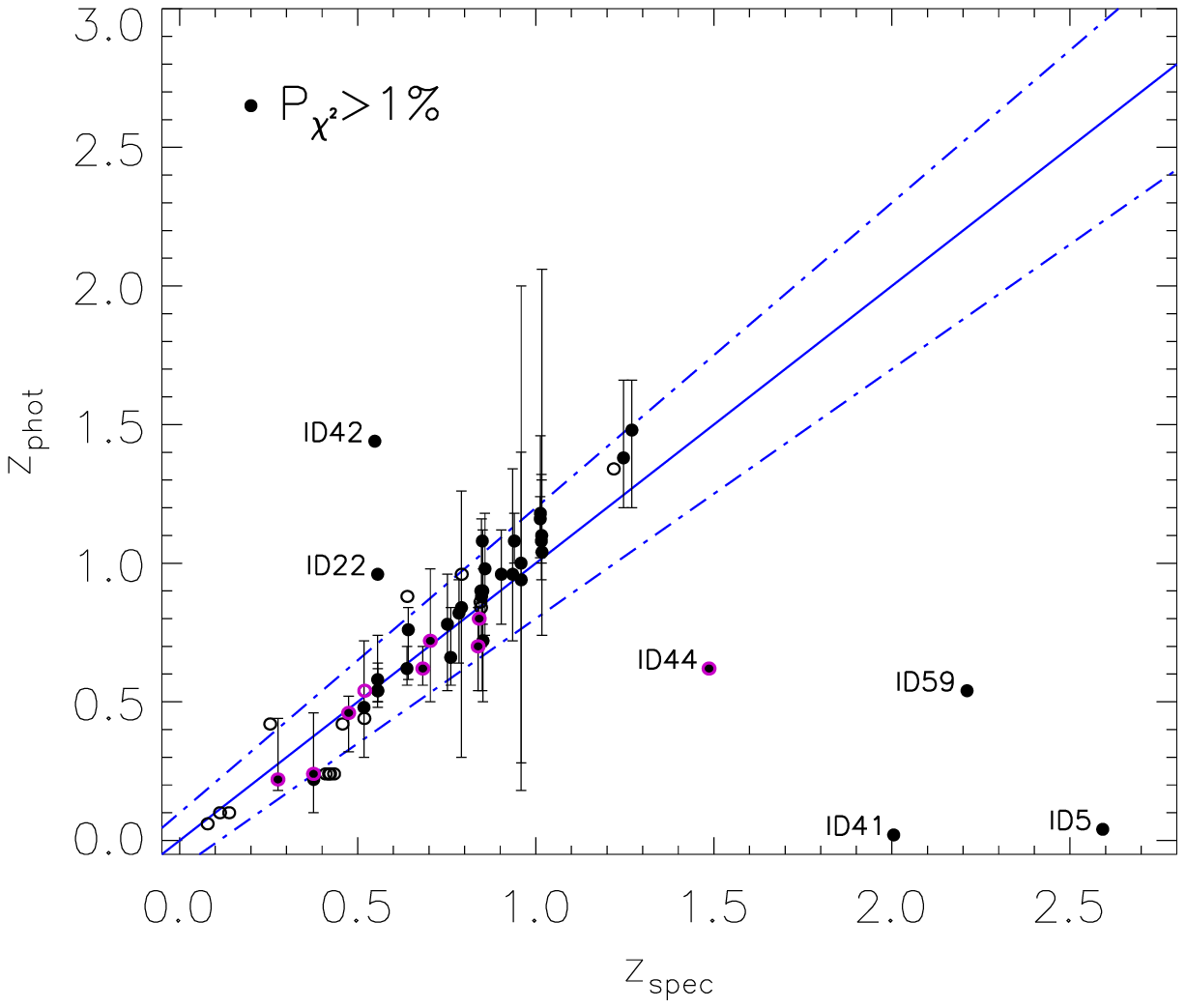}
  \vskip-0.8truecm\caption{Photometric versus spectroscopic redshift
    for the sources listed in Table~\ref{tab:realcat_info}. Filled
    circles mark the cases of a ``good'' fit to the SED (i.e.
    $P_{\chi^2}>1\%$) while the open circles correspond to a bad SED
    fit (i.e.  $P_{\chi^2}<1\%$). Sources with uncertain photometry at
    {\it AKARI} and {\it ISO} wavebands are circled in color. The
    dot-dashed lines delimit the region where $(z_{\rm phot}-z_{\rm
      spec})/(1+z_{\rm spec})<10\%$ while the solid line corresponds
    to the case $z_{\rm phot}=z_{\rm spec}$. Errors on photometric
    redshifts are in 99$\%$ confidence limit and, for reason of
    clarity, they are shown only for the objects with
    $P_{\chi^2}>1\%$, lying within the $10\%$-accuracy region.}
  \label{fig:zphot_vs_zspec_realcat}
\end{figure*}

ID42 is almost at the same redshift of ID22 and its spectrum has a
peculiar power-law shape above $8\,\mu$m. The {\it ISO} data suggest
an emission feature at 6.5$\,\mu$m that, if due to the restframe
3.3$\,\mu$m PAH feature, would suggest a redshift $z\sim1$ at variance
with the spectroscopic value. However the measurement is affected by
large uncertainties. By fitting the measured SED with the redshift
fixed to its spectroscopic value we get a minimum reduced $\chi^2$ of
$\sim3.5$. Examining the contribution of the individual wavebands to
the value of the minimum $\chi^2$ we find the AKARI and the
4.5$\,\mu$m {\it Spitzer} photometry to have the highest discrepancy
between observation and theoretical predictions. The fluxes at 4.5, 11
and 18 appear lower than expected.  These results indicate the
limitation of our reference SED templates in accounting for the
infrared spectrum of ID42. Therefore ID42 represents a challenge for
the SED models exploited here.

Failure in recovering the redshift of ID44 is quite certainly due to
flux ``contamination''. In fact, this object lies in a crowded field
(see Fig.~\ref{fig:PostageStamps}): there are two sources individually
detected by {\it Spitzer} (with comparable fluxes at 24$\,\mu$m) which
are blended at {\it AKARI} (and {\it ISO}) wavebands. The spectrum of
ID44 displays two bumps at 15 and 18$\,\mu$m that the SED fitting
interprets as redshifted PAH 7.7 and 11.3 $\mu$m features,
respectively, thus implying $z_{\rm phot}<1$.  Adopting the flux
measurements at 15 and 18$\mu$m as upper limits does not improve the
fit because the SED would assume a featureless power-law
shape. Therefore the only way to obtain a better photo-$z$ estimate
for this object is to use optical and radio photometry with
higher spatial resolution. \\

For sources at $z_{\rm spec}\gsim2$ the photometric redshift is found
to be systematically and significantly lower than the spectroscopic
value, i.e.  $z_{\rm phot}\le0.5$. This is not surprising if examine the
measured spectrum of the three objects (see Fig.~\ref{fig:SEDs}). The
SEDs of ID5 and ID41 are completely featureless, resembling a
power-law functional form, thus making the estimate of the redshift
impossible without the support of photometric data at other
wavelengths. On the other hand the SED of ID59 seems to indicate an
absorption at $\sim15\,\mu$m (apart from that the SED is very close to
a power-law). Our SED-fitting procedure interprets this feature as due
to the redshifted $9.7\,\mu$m silicate absorption in the spectrum of a
young starburst, providing $z_{\rm phot}\sim0.5$. In fact, at the
spectroscopic redshift of the source (i.e. $z_{\rm spec}\sim2.2$), the
$9.7\,\mu$m absorption falls above the 24$\,\mu$m waveband covered by
MIPS. It is therefore plausible that the observed absorption is due to
the transition between the rest-frame 3.3 and 6.2$\mu$m PAH
features. In that case the source would have a much higher intrinsic
luminosity than the value we obtained for $z_{\rm phot}=0.54$. An
independent constraint on the infrared luminosity of the source,
e.g. from measurements at submillimeter and/or decimetric radio
wavelengths, would be crucial for better recovering the redshift of
this source. We plan to test the addition of extra data on the
photo-$z$ estimates (and on SED model parameters) in a future paper. \\

\setcounter{figure}{5}
\begin{figure}
  \hspace{-2.0cm}\includegraphics[height=8.0cm,width=11.0cm]
  {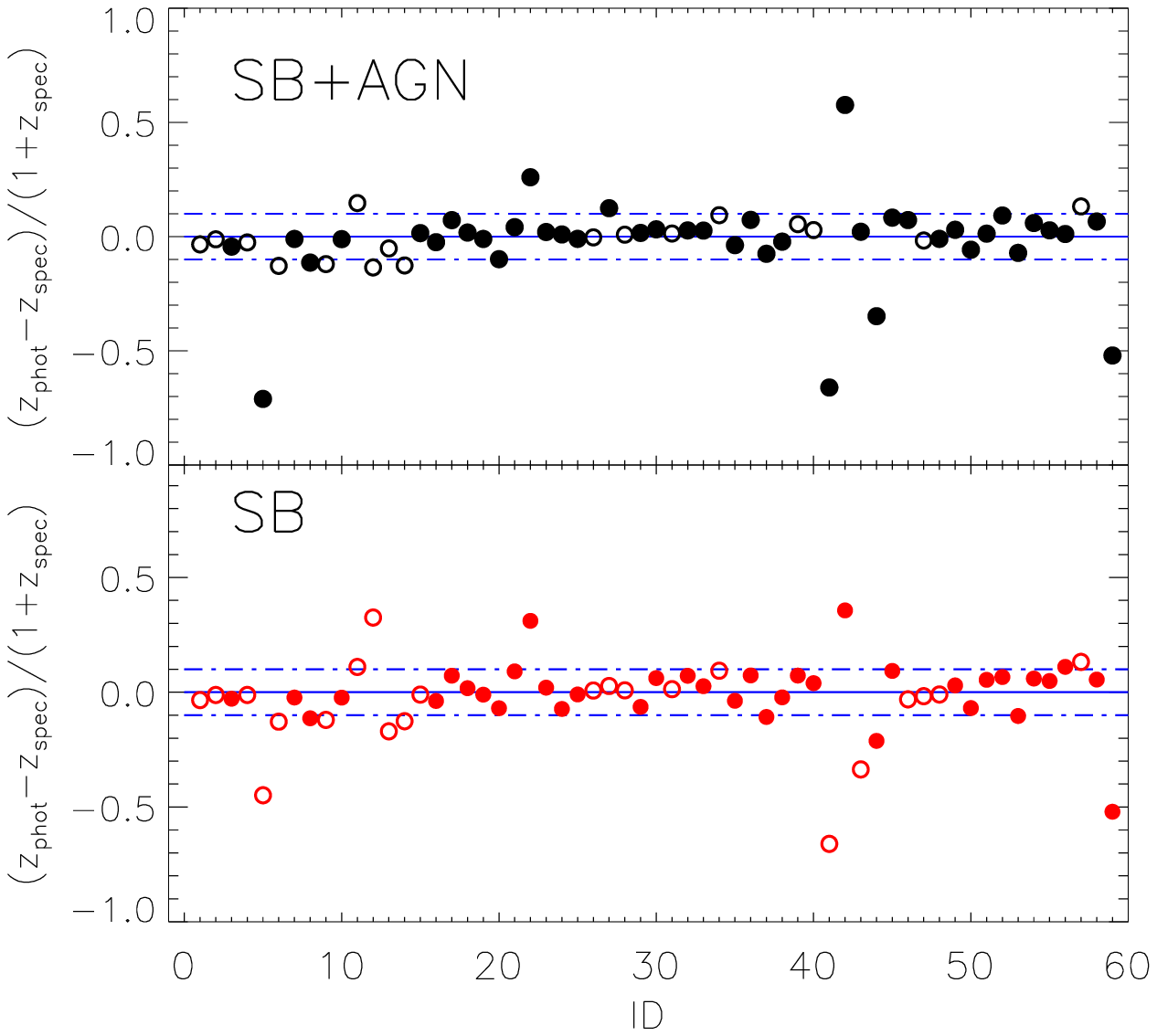}
  \vskip0.1truecm \caption{\label{fig:dz_comparison} Comparison of the
    photometric redshift accuracy achieved by fitting the infrared
    spectrum of the sources listed in Table~\ref{tab:realcat_info}
    with 2 SED components (i.e. starburst + AGN; upper panel) and with
    the starburst templates alone (lower panel). The meaning of the
    lines is the same as in Fig.~\ref{fig:zphot_vs_zspec_realcat}.}
\end{figure}

We note that the estimated 99$\%$ error on the photometric redshift is
particularly large for few sources lying in the redshift range
$0.8\lsim z_{\rm spec}\lsim1$, i.e. ID32, ID43, ID48, and ID56. This
follows from the shape of the SED of these objects. For ID32, ID43 and
ID48 the measured spectrum is close to a power law for
$\lambda\gsim8\mu$m, and this induces a degeneracy between the
redshift and the normalization of the AGN sed component ($f_2$) around
the minimum value of the $\chi^2$.  The case of ID56 is more
peculiar. At the redshift of the source (i.e. $z_{\rm spec}=1.017$),
the rest-frame 9.7$\,\mu$m silicate absorption enters the {\it AKARI}
18$\,\mu$m filters, which is the broadest among the {\it AKARI}
filters (FWHM$\sim$10$\,\mu$m); as a consequence the measured SED
appears relatively flat between $15$ and $24\,\mu$m. However, while
for objects like ID21, ID30, ID36, ID45, ID46 (all at $z=1$), a hint
of the silicate absorption at $18\,\mu$m is still visible, in the case
of ID56 the SED is featureless above 15$\,\mu$m. Consequently, the
observed spectrum can be still reasonably well fitted by a mixture of
starburst and AGN emission in a large redshift interval around
$z\sim1$. \\

According to our SED fitting results, sources with $P_{\chi^2}>1\%$
and $|\delta z|/(1+z_{\rm spec})\sim10\%$ manifest different levels of
AGN emission in their SED. In same cases the SED is well described by
a starburst template (e.g. ID17, ID35, ID38), while in other cases an
underlying AGN activity is found to contribute to the measured fluxes
either at all wavebands (e.g. ID24, ID42, ID48, ID52, ID58) or only at
longest wavelengths, i.e. 18 and 24$\,\mu$m (e.g. ID3, ID16). There
are also some very peculiar/interesting cases of sources at $z_{\rm
  spec}\sim1$ for which the flux above $\lambda\sim10\,\mu$m is
completely or significantly accounted for by the emission from an
edge-on viewed dusty torus (e.g. ID21, ID27, ID30, ID39, ID46).

It is worth noticing that our SED-fitting procedure is meant to
combine the SB and AGN SED components in such a way to minimize the
value of the $\chi^2$ in Eq.~\ref{eq:chisq} and not to maximize the
probability associated to the minimum value of the $\chi^2$. In fact,
while the value of $P_{\chi^2}$ depends on the number of degree of
freedom of the problem (and on the assumption that the quantity in
Eq.~\ref{eq:chisq} follows exactly a $\chi^2$-distribution), that of
the minimum $\chi^2$ does not. Consequently, there may be some sources
for which the best fit to their SED gives a smaller value for $P_{\chi^2}$ when the fit is performed by setting {\it a priori} $f_{2}$ to zero than when both $f_{1}$ and $f_{2}$ are taken as free parameters, even if the resulting minimum $\chi^2$ is smaller in the latter case.
In
Table~\ref{tab:SEDfit_results} we show, for comparison, the minimum
reduced $\chi^2$ and the associated
probability $P_{\chi^2}$ obtained when the measured SED is fitted with the starburst templates alone. \\
We note that there are several cases for which the value of
$P_{\chi^2}$ is slightly lowered when using two SED components and in
two cases it drops below the $1\%$ limit (ID39, ID40). On the other
hand, accounting for an AGN component improves the goodness of the fit
for many other objects in the sample, particularly when the measured
SED is close to a power-law (see e.g. ID5, ID41 and ID48). However,
for these extreme cases, a better fit doesn't imply a better redshift
accuracy which instead remains quite poor.  In
Fig.~\ref{fig:dz_comparison} we compare the photometric redshift
accuracy, $(z_{\rm phot}-z_{\rm spec})/(1+z_{\rm spec})$, achieved by
our two-SED fitting procedure (upper panel) with that obtained by
fitting the observed SED with the starburst templates alone (lower
panel). In both panels filled circles mark the sources with
$P_{\chi^2}>1\%$ while open circles indicate sources with
$P_{\chi^2}<1\%$. There are few cases in which, the redshift of the
source is better recovered when accounting for an AGN component: ID24,
ID29, ID43, however, we do not see on average a significant difference
in the achieved redshift accuracy between the two-SED fitting approach
and that based on the starburst templates alone.

For comparison, we also show in Fig~\ref{fig:SEDs} the result of the
best-fit obtained when $f_{2}=0$. There are
some sources, mainly at redshift $z\lsim0.5$, e.g. ID1, ID2, ID4,
ID12, ID13, ID47 and ID57, for which the fit to the SED is very poor
(i.e. $P_{\chi^2}\ll1\%$) independently on the adopted SED fitting
approach. Failure in reproducing the infrared spectrum of these
sources could be due to an underestimate of the fluxes at 11 and
18$\,\mu$m. In fact, AKARI photometry were obtained under the
assumption of point sources (see Pearson et al. 2008 for details),
while most of the objects listed above are significantly extended in
optical images (see Fig.~\ref{fig:PostageStamps}). Moreover, ID1 and
ID47 appear to be elliptical galaxies whose spectrum is not expected
to be well-recovered by the starburst SED templates exploited here.

\setcounter{figure}{6}
\begin{figure*}
  \begin{center}
    \hspace{-0.8cm}\includegraphics[height=11.5cm,width=16.0cm]
    {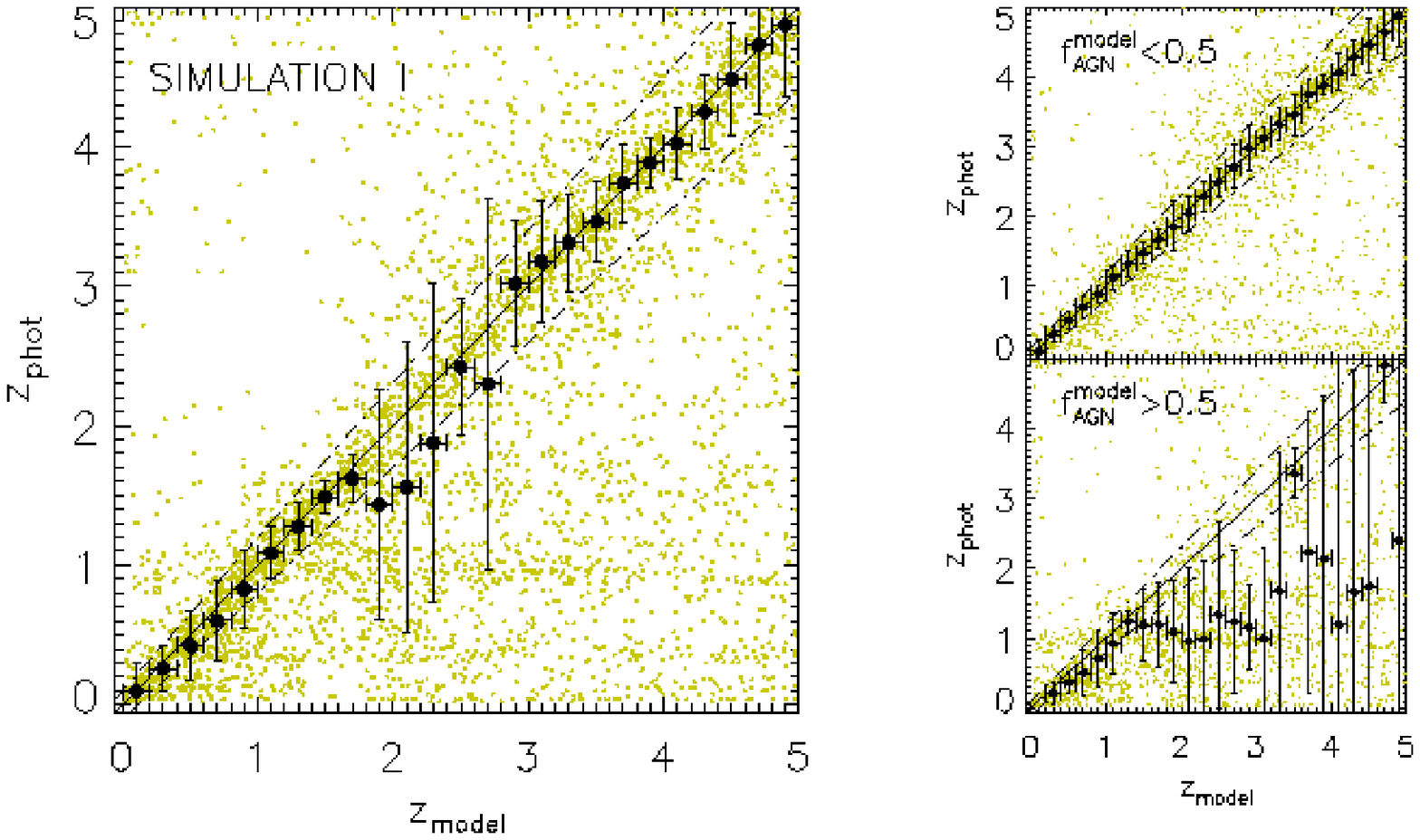}
    \vskip-1.4truecm \caption{Simulation of photometric redshift
      accuracy with the {\it AKARI} NEP Deep Survey. The simulated
      spectra have been derived from the same models used for
      constructing the reference SED templates, i.e. Takagi et
      al. (2003, 2004) for the starburst and Efstathiou $\&$
      Rowan-Robinson (1995) for the AGN. The filled circles with error
      bars are the mean value and the 1$\sigma$ dispersion provided by
      the Gaussian fit to the histogram of the $z_{\rm phot}$ values
      in intervals of 0.2 in $z_{\rm model}$. Only simulated spectra
      for which $P_{\chi^2}>1\%$ are taken into account (4919 over
      5000). The dot-dashed lines delimit the region where $(z_{\rm
        phot}-z_{\rm model})/(1+z_{\rm model})<10\%$, while the solid
      line corresponds to the case $z_{\rm phot}=z_{\rm model}$. The
      photometric redshift accuracy achieved for the whole sample is
      shown in the left-hand panel while that derived for the
      subsamples of spectra with $f_{\rm AGN}^{\rm model}<0.5$ and
      $f_{\rm AGN}^{\rm model}>0.5$ is shown in the right-hand
      panels.}\label{fig:zphot_vs_zmodel_simulcat_takagi}
  \end{center}
\end{figure*}

\setcounter{table}{3}
\begin{table}
  \scriptsize
  \caption{5$\sigma$ flux limits for the {\it AKARI} NEP-Deep Survey used in the simulations (from Wada et al. 2008).}
  \begin{center}
    \begin{tabular}{p{0.4cm}p{0.4cm}p{0.4cm}p{0.4cm}p{0.4cm}p{0.4cm}p{0.4cm}p{0.4cm}p{0.4cm}p{0.4cm}}
      \hline
      $\lambda_{c}$ ($\mu$m) &   2.43 &   3.16 &   4.14 &   7.3 &   9.1 &   10.7 &   15.7 &   18.3 &   23.0 \\
      \hline
      S$_{lim}$ (mJy)        &   14.2 &   11.0 &   8.0 &   48.9 &   58.5 &   70.9 &  117.0 &   121.4 &  275.8 \\
      \hline
    \end{tabular}
  \end{center}
  \label{tab:detect_limits}
\end{table}

\section{Simulations of the {\it AKARI} NEP Deep Survey }
\label{sec:simulations}

In order to investigate the photometric redshift accuracy achievable
with the {\it AKARI} NEP Deep Survey we have generated three
  different sets of 5000 simulated spectra, including contributions
  from both starburst and AGN.

In all the simulations, which are described in the next subsections,
the derived starburst and AGN components are linearly combined
together with a relative contribution to the bolometric luminosity
randomly selected between 0 and 1.  Redshifts are randomly assigned in
the range 0-5, assuming a uniform distribution, and the spectra are
then normalized to the 24$\,\mu$m fluxes generated from the source
count model of Lagache et al. (2003). A minimum 24$\,\mu$m flux of
100$\,\mu$Jy is assumed. The spectra are then convolved with the nine
AKARI/IRC filters, spanning the wavelength range 2-24$\,\mu$m (they are shown in
Fig.~\ref{fig:sed_vs_filters}).  We require the simulated spectra to
be detected in at least five of the observed bands, according to the
adopted 5$\sigma$ detection limits shown in
Table~\ref{tab:detect_limits} (from Wada et al. 2008). The
corresponding 1$\sigma$ limits are used to introduce Gaussian
fluctuations on the simulated fluxes and are adopted as the estimated
error on the resulting fluxes. When a simulated spectrum is undetected
at a certain waveband than both its flux and the accompanying error
are set equal to half the 5$\sigma$ detection limit at that waveband.

For all the three sets of simulations, the resulting fluxes are
  fitted using the reference SED templates presented in
  subsection~\ref{subsec:reference_templates}. A description of the
  SED models used to build the simulations and of the derived
  photometric accuracy is provided in the following subsections.

\setcounter{figure}{7}
\begin{figure*}
  \begin{center}
    \hspace{-0.8cm}\includegraphics[height=11.5cm,width=16.0cm]
    {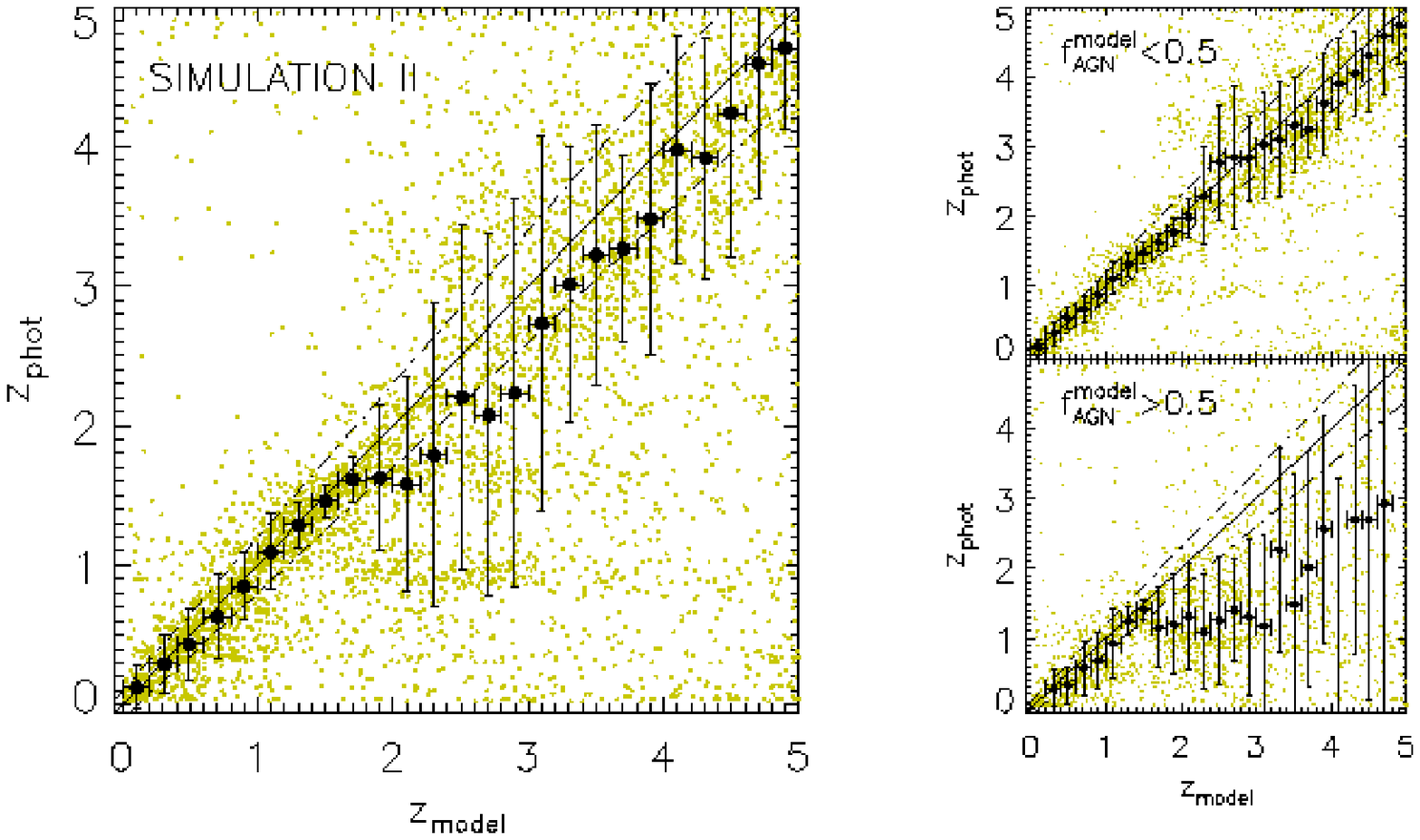}
    \vskip-1.4truecm \caption{Photometric redshift accuracy with the
      {\it AKARI} NEP Deep Survey derived from the set of simulated
      spectra generated using the models of Efstathiou et
      al. (2000, 2003) and of Efstathiou $\&$ Rowan-Robinson (1995)
      for the starburst/cirrus and AGN infrared emissions,
      respectively. The meaning of the symbols and of the lines is the
      same as in Fig.~\ref{fig:zphot_vs_zmodel_simulcat_takagi}. For
      4859 simulated spectra (out of 5000) the fit has provided
      $P_{\chi^2}>1\%$.}\label{fig:zphot_vs_zmodel_simulcat_efstat}
  \end{center}
\end{figure*}

\setcounter{figure}{8}
\begin{figure*}
  \begin{center}
    \hspace{-0.8cm}\includegraphics[height=11.5cm,width=16.0cm]
    {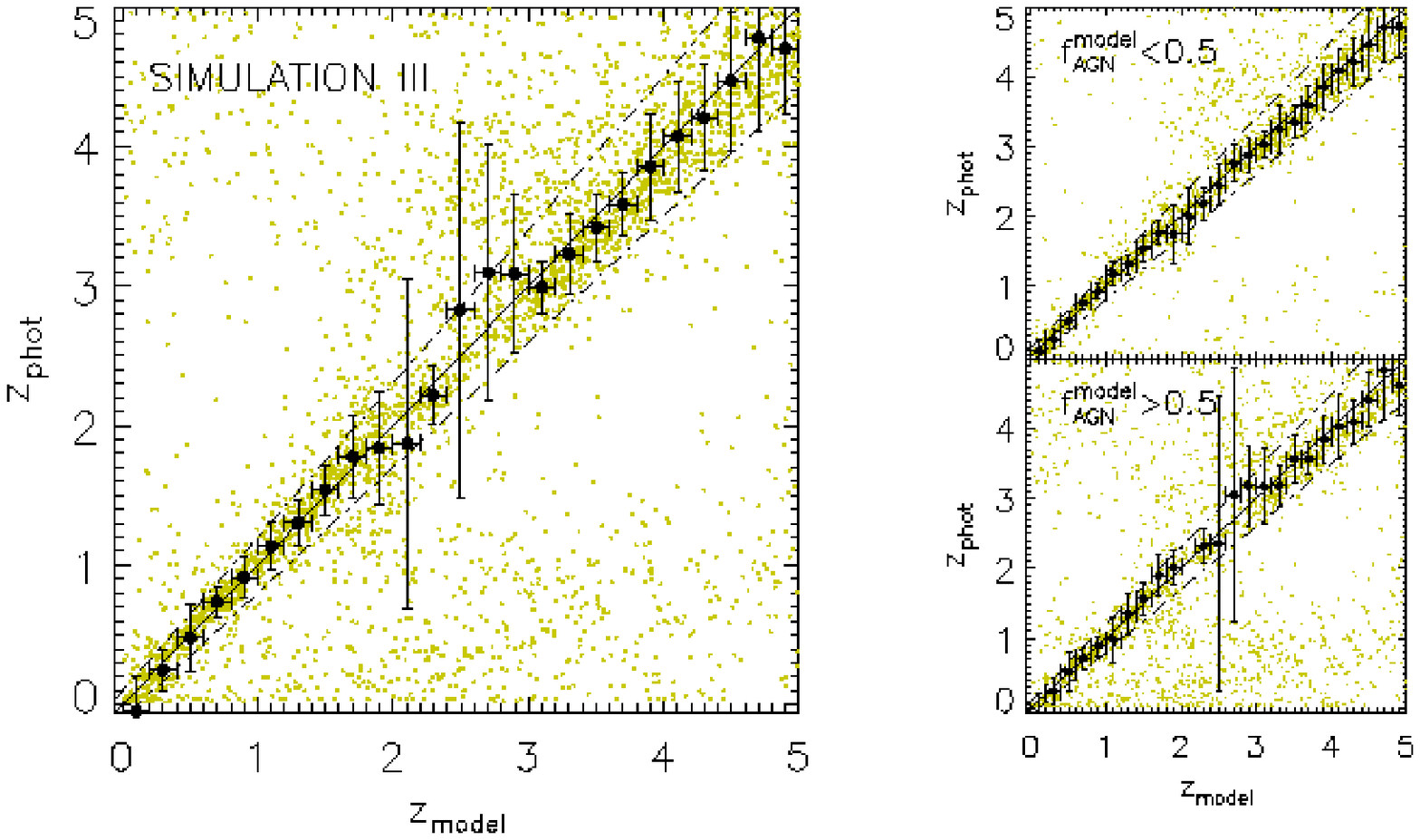}
    \vskip-1.4truecm \caption{Photometric redshift accuracy with the
      {\it AKARI} NEP Deep Survey derived for the set of simulated 
      spectra generated using the SWIRE SED templates. The meaning of
      the symbols and of the lines is the same as in
      Fig.~\ref{fig:zphot_vs_zmodel_simulcat_takagi}. The simulated
      spectra for which the fit has provided $P_{\chi^2}>1\%$ are 4060
      out of 5000.}\label{fig:zphot_vs_zmodel_simulcat_polletta}
  \end{center}
\end{figure*}

\subsection{Simulation I}
\label{subsec:simulations_takagi}

The first set of simulated flux measurements (5000 spectra in total)
have been obtained from the same SED models used to construct the
reference SED templates, i.e. TAH03 model for the starburst and ER95
for the AGN. Each SED component is derived by randomly selecting the
values of the SED parameters within the intervals specified in
subsection~\ref{subsec:reference_templates}.

The derived photometric redshifts, $z_{\rm phot}$, versus the
simulated redshifts, $z_{\rm model}$, are shown in the left-hand panel
of Fig.~\ref{fig:zphot_vs_zmodel_simulcat_takagi} where the filled
circles with error bars represent the mean value and the 1$\sigma$
dispersion obtained by fitting a Gaussian\footnote{The Gaussian fit
  has been performed using the {\sc sky\underline{~}stats} IDL
  routine.} to the histogram of the $z_{\rm phot}$ values in intervals
of 0.2 in $z_{\rm model}$. Only simulated spectra with
$P_{\chi^2}>1\%$ have been taken into account (they are represented by
the small dots in the same figure). On average, the accuracy
achievable on ($1+z$) is close to 10$\%$ (a limit indicated by the
dot-dashed lines in the same figure) below $z\sim2$ and above
$z\sim3$, while in the redshift interval $1.8\lsim z\lsim2.8$ the
dispersion on the photo-$z$ estimates is particularly high. This
behavior is mainly due to a significant number of points lying below
the 10$\%$-accuracy region. Interestingly, we found that these data
correspond to spectra with a dominant AGN component, and therefore
with almost featureless power-law shape. To prove this we have
splitted the whole set of simulated spectra into two subsamples
according to the input value of the AGN fraction, i.e. $f_{\rm
  AGN}^{\rm model}<0.5$ and $f_{\rm AGN}^{\rm model}>0.5$ and
calculated the photometric accuracy for each subsample. The results
are presented in the right-hand panels of
Fig.~\ref{fig:zphot_vs_zmodel_simulcat_takagi}. Power-law like spectra
have photometric redshifts which are systematically and significantly
lower than the input redshifts. Once they are removed from the sample
the achieved accuracy on $(1+z)$ gets close to 10$\%$ over the whole
range of redshifts probed by the simulation.

\subsection{Simulation II}
\label{subsec:simulations_efstat}

The second set of 5000 simulated spectra has been constructed using
the model of Efstathiou et al. (2000, hereafter ERS00) for the
starburst emission while the AGN component has been modelled using the
prescription of ER95, as done before. We have also included a third
SED component describing the emission from the diffuse interstellar
dust heated by old stars and/or quiescent star formation (infrared
cirrus), for which we have followed the model of
Efstathiou $\&$ Rowan-Robinson (2003, hereafter ER03). \\

According to ERS00 stars form primarily within optically thick giant
molecular clouds (GMCs) and their model provides prescriptions for the
evolution of the GMCs owing to the ionization-induced expansion of the
HII regions. The evolution of the stellar population within the GMC is
accounted for by using the Bruzual $\&$ Charlot (1993) stellar
population synthesis models with a Salpeter Initial Mass Function
(IMF) and a star mass range of 0.1-125 M$_{\odot}$.  The
absorption/emission properties of the dust within the GMCs are derived
according to the model of Siebenmorgen $\&$ Krugel (1992) which assume
three different populations of dust grains: large grains, small
graphite particles and PAHs.  Starburst galaxies are treated as an
ensemble of GMCs at different evolutionary stages.  The star formation
rate is modeled with an exponential functional form $\dot{M}\propto
e^{-t/\tau_{\rm SB}}$. Here we set the e-folding time of the
star-formation rate, $\tau_{\rm SB}$, to 20 Myr while the age of the
HII region phase, Age$_{\rm SB}$, and the initial optical depth of
GMCs, $\tau_{0}$, are treated as free parameters, with Age$_{\rm
  SB}=0-72$ Myr and $\tau_{0}=50-200$.

\setcounter{figure}{9}
\begin{figure*}
  \hspace{+0.5cm}\includegraphics[height=11.0cm,width=15.5cm]
  {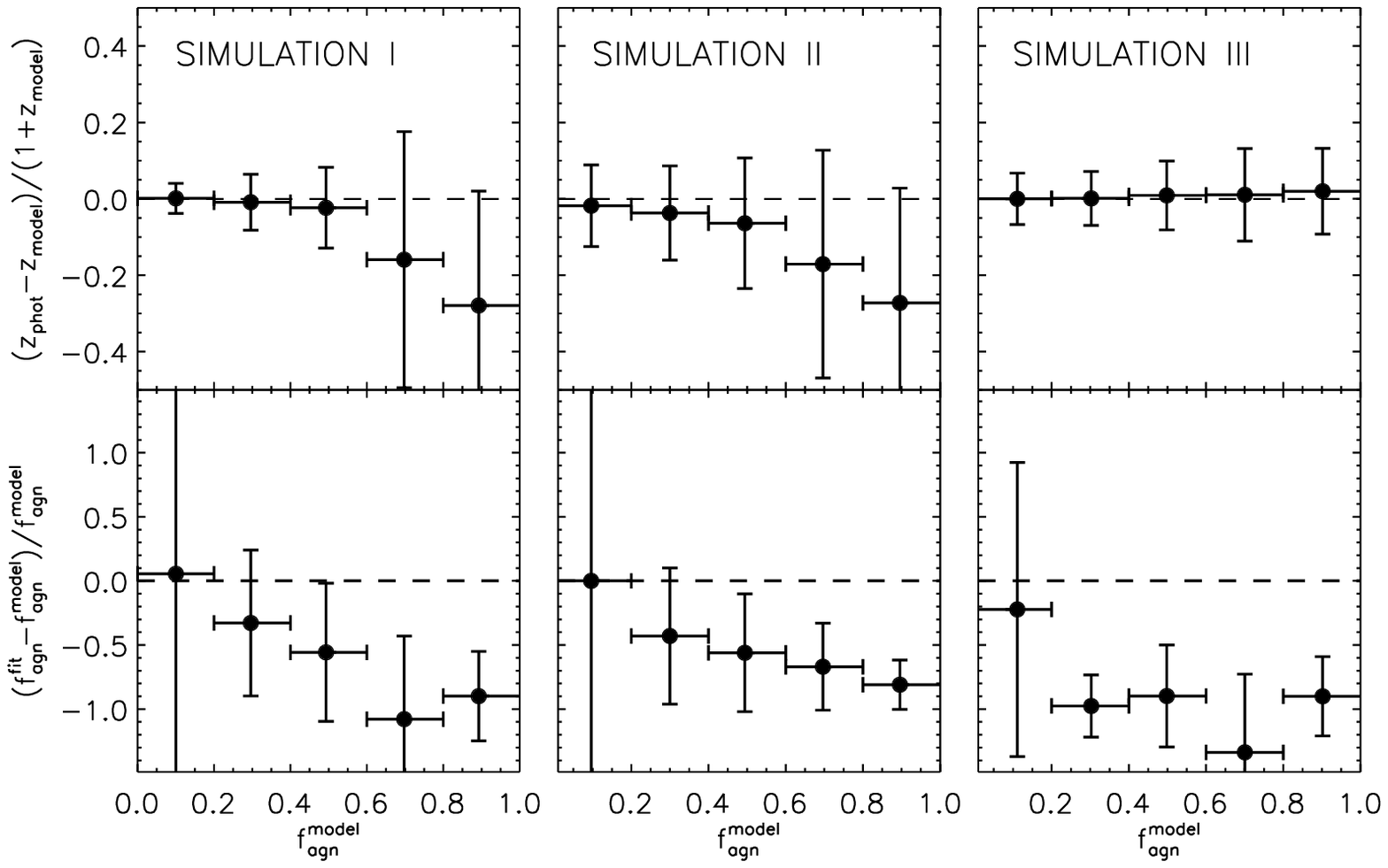}
  \vskip-1.0truecm\caption{{\it Upper panel}: photometric redshift
    accuracy, $\delta z/(1+z_{\rm model})$, as a function of the input
    value of the AGN fraction, $f_{\rm AGN}^{\rm model}$, derived from
    the three sets of simulations presented here. The filled circles
    with error bars are the mean value and the 1$\sigma$ dispersion
    provided by the Gaussian fit to the histogram of the $\delta
    z/(1+z_{\rm model})$ values in intervals of 0.2 in $f_{\rm
      AGN}^{\rm model}$. Only spectra with $P_{\chi^2}>1\%$ have been
    taken into account. {\it Lower panels}: corresponding accuracy
    achieved on the AGN fraction. Again, the filled circles with error
    bars are the mean value and the 1$\sigma$ dispersion provided by
    the Gaussian fit to the histogram of the $(f_{\rm AGN}^{\rm
      fit}-f_{\rm AGN}^{\rm model})/f_{\rm AGN}^{\rm model}$ values in
    intervals of 0.2 in $f_{\rm AGN}^{\rm model}$.}
  \label{fig:dz_vs_fagn_simulcat_ALL}
\end{figure*}

The infrared cirrus is characterized by lower optical depth and lower
temperatures of the dust ($\lsim$ 30 K), compared to what occurs in
starburst regions.  However, as in ERS00, stars are assumed to have
formed within GMCs but their evolution is followed well beyond the
complete dispersion of the GMCs (i.e. after $\sim72$ Myr), when the
starlight can be absorbed only by the general interstellar medium.
The input stellar radiation field is computed from the stellar
population synthesis model of Bruzual $\&$ Charlot (1993) while the
grain model of Siebenmorgen $\&$ Krugel (1992) is used again to derive
the absorption/emission properties of the dust.  The extinction
parameter $A_{V}$ regulates the proportion of the UV to near-IR light
that is absorbed by the ISM and re-emitted in the far-IR and sub-mm
bands.  The temperature of the interstellar dust is determined by the
intensity of the radiation field. ER03 characterize this in terms of
the ratio $\Phi$ of the bolometric intensity of the radiation field to
that in the stellar radiation field in the solar neighborhood. Since
the value of $\Phi$ influences only the far-infrared shape of the SED,
i.e. the position of the peak of the SED (the higher $\Phi$, the
hotter the dust and the lower the wavelength of the peak), the exact
choice of its value is not relevant for our purpose and therefore we
just set it to 1.  Before the complete evaporation of the GMC, the
non-spherical evolution of the molecular cocoon may allow a fraction
$f$ of the starlight to escape without any dust absorption from the
GMC. Here we set $f=1$ after 3 Myr as in ER03. The age of the cirrus
component, Age$_{\rm CIR}$, and the e-folding time of the
star-formation rate, $\tau_{\rm CIR}$, are treated as free parameters
and varied consistently with the redshift of the source (i.e. always
assuring that Age$_{\rm CIR}$ and $\tau_{\rm CIR}$ are lower than the
age of the Universe at the redshift of the source). The visual
extinction is chosen within the range $A_{V} = 0-3$.

Each simulated SED component, i.e. starburst, cirrus and AGN, is
derived by randomly selecting the values of the SED parameters within
the intervals mentioned before. The fractions of the bolometric
luminosity contributed by the AGN and by the starburst+cirrus are
selected at random between 0 and 1. The contribution from the
starburst+cirrus is then randomly divided between the two components
(i.e. starburst and cirrus); in this way we guarantee that the final
sample of simulated spectra include almost 50$\%$ of AGN dominated
SED, as in the previous sample.

The results on the photometric redshift accuracy are shown in
Fig.~\ref{fig:zphot_vs_zmodel_simulcat_efstat}, where the meaning of
the symbols and of the lines is the same as in
Fig.~\ref{fig:zphot_vs_zmodel_simulcat_takagi}. The achieved accuracy
on $(1+z)$ is close to $10\%$ below $z\sim1.8$ but decreases
significantly above $z\sim2$, particularly in the interval
$z\sim2-3$. Although AGN power-law like spectra are found to be
responsible for the significant fraction of data points lying below
the $10\%$-accuracy region (see right-hand panels in the same figure),
this effect alone does not account for the relatively low
  accuracy achieved in the interval $2\lsim z\lsim3$. In this redshift
  range the only valuable redshift indicators are the 1.6$\,\mu$m bump
  and the PAH 3.3$\,\mu$m emission feature. The simulated spectra
  including a cirrus component are more challenging for our reference
  SED templates to reproduce. In fact the latter are meant for fitting
  relatively young starburst galaxies only, with (eventually) an
  additional AGN component. It is also worth noting that for $2\lsim
  z\lsim3$ the 1.6$\,\mu$m bump is shifted to wavelengths
  $4.8\lsim\lambda\lsim6.4$, where the {\it AKARI} coverage is quite
  poor (see Fig.~\ref{fig:sed_vs_filters}). This contributes to the
  observed decrease in the photo-z accuracy in that redshift
  interval.

\subsection{Simulation III}
\label{subsec:simulations_polletta}

The third simulation is built from the empirical spectra used to fit
  the SED of different types of sources in the {\it Spitzer} Wide-Area
  Infrared Extragalactic(SWIRE) survey (see Polletta et al. 2008 and
  references therein). The SWIRE template
  library\footnote{http://cass.ucsd.edu/SWIRE/mcp/templates/swire$\underline{~}$templates.html}
  contains 25 templates: 3 ellipticals, 7 spirals, 6 starbursts, 7
  AGNs (3 type 1 AGNs, 4 type 2 AGNs), and 2 composite
  (starburst+AGN). The elliptical, spiral and starburst templates were
  generated using the the GRASIL code (Silva et al. 1998). The
  ellipticals correspond to three different ages: 2 ,5 and 13 bilion
  years. The 7 spirals range from early to late types (S0-Sdm). The
  starburst templates correspond to the SEDs of NGC6090, NGC6240, M82,
  Arp220, IRAS22491-1808, and IRAS20551-4250. In all of the spirals
  and starburst templates the spectral region between 5 and
  12$\,\mu$m, where PAH broad emission and silicate absorption
  features are observed, was replaced using observed infrared spectra
  from the PHT-S spectrometer on the {\it ISO} and from IRS on {\it
    Spitzer}.

  AGN templates corresponding to Seyfert 1.8 and Seyfert 2 galaxies
  were obtained by combining models, broadband photometric data (NED)
  and {\it ISO} PHT-S spectra of a random sample of 28 Seyfert
  galaxies. Three other AGN templates represent optically selected
  QSOs with different values of infrared/optical flux ratios, derived
  by combining the Sloan Digital Sky Survey (SDSS) quasar composite
  spectrum and rest-frame IR data of a sample of 35 SDSS/SWIRE quasars
  and then assuming three different IR SEDs. Of the remaining AGN
  templates (type 2 QSOs) one was obtained by combining the observed
  optical/near-IR spectrum of the red quasar FIRST J013435.7-093102
  and the rest-frame IR data from the quasars in the Palomar-Green
  sample with consistent optical SEDs. The other type 2 QSO template
  corresponds to the model used to fit the SED of a heavily obscured
  type 2 QSO, SWIRE$\underline{~}$J104409.95+585224.8 (Polletta et
  al. 2006)

  The composite (AGN+SB) templates are empirical templates created to
  fit the SEDs of the heavily obscured BAL QSO Mrk 231 (Berta 2005)
  and the Seyfert 2 galaxy IRAS 19254-7245 South (Berta et
  al. 2003). They both contain a powerful starburst component,
  responsible for their large infrared luminosities, and an AGN
  component that contributes to the mid-IR luminosities.

  Elliptical, spirals, starburst and AGN+SB templates were randomly
  selected within the available sample. If the SED did not already
  include an AGN component (i.e. it was not one of the two SWIRE
  composite spectra) than an AGN templates was chosen at random from
  the SWIRE sample and added to the SED, with a relative contribution
  to the bolometric luminosity randomly selected from a uniform
  distribution between 0 and 1.

  The results of the photometric redshift accuracy are presented in
  Fig.~\ref{fig:zphot_vs_zmodel_simulcat_polletta}. Again we observe a
  large scatter in the photo-z estimates in the interval $2\lsim
  z\lsim3$ where the main infrared features are shifted outside the
  range of wavelengths covered by {\it AKARI} and the 1.6$\,\mu$m
  feature is not well sampled because of the lack of coverage around
  $\lambda\sim5-6\,\mu$m. Despite this, the accuracy achieved on
  $(1+z)$ is better than 10$\%$ up to $z=5$.

\subsection{Discussion}

\setcounter{figure}{10}
\begin{figure}
  \begin{center}
    \includegraphics[height=5.8cm,
    width=8.0cm]{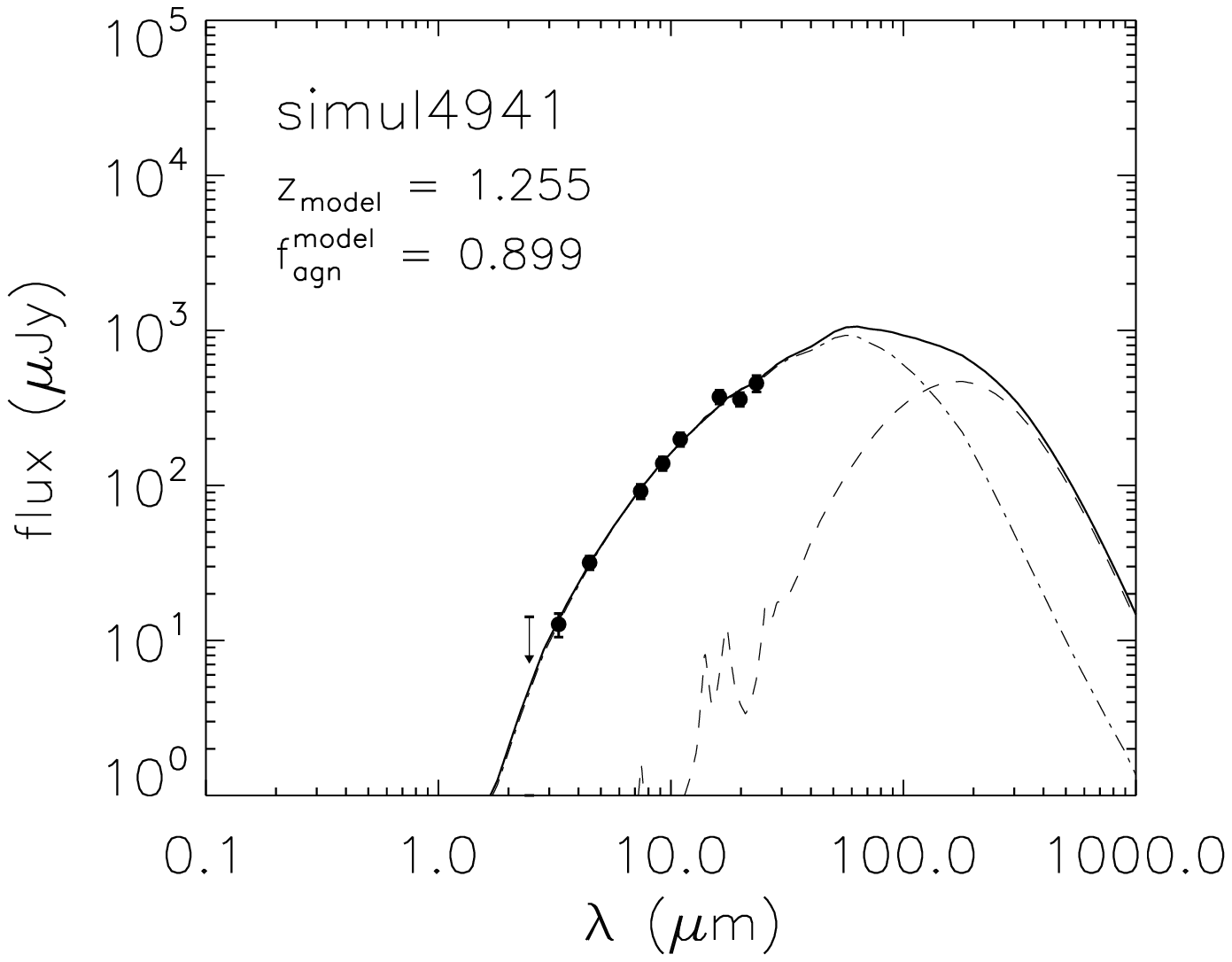}
    \includegraphics[height=5.8cm,
    width=8.0cm]{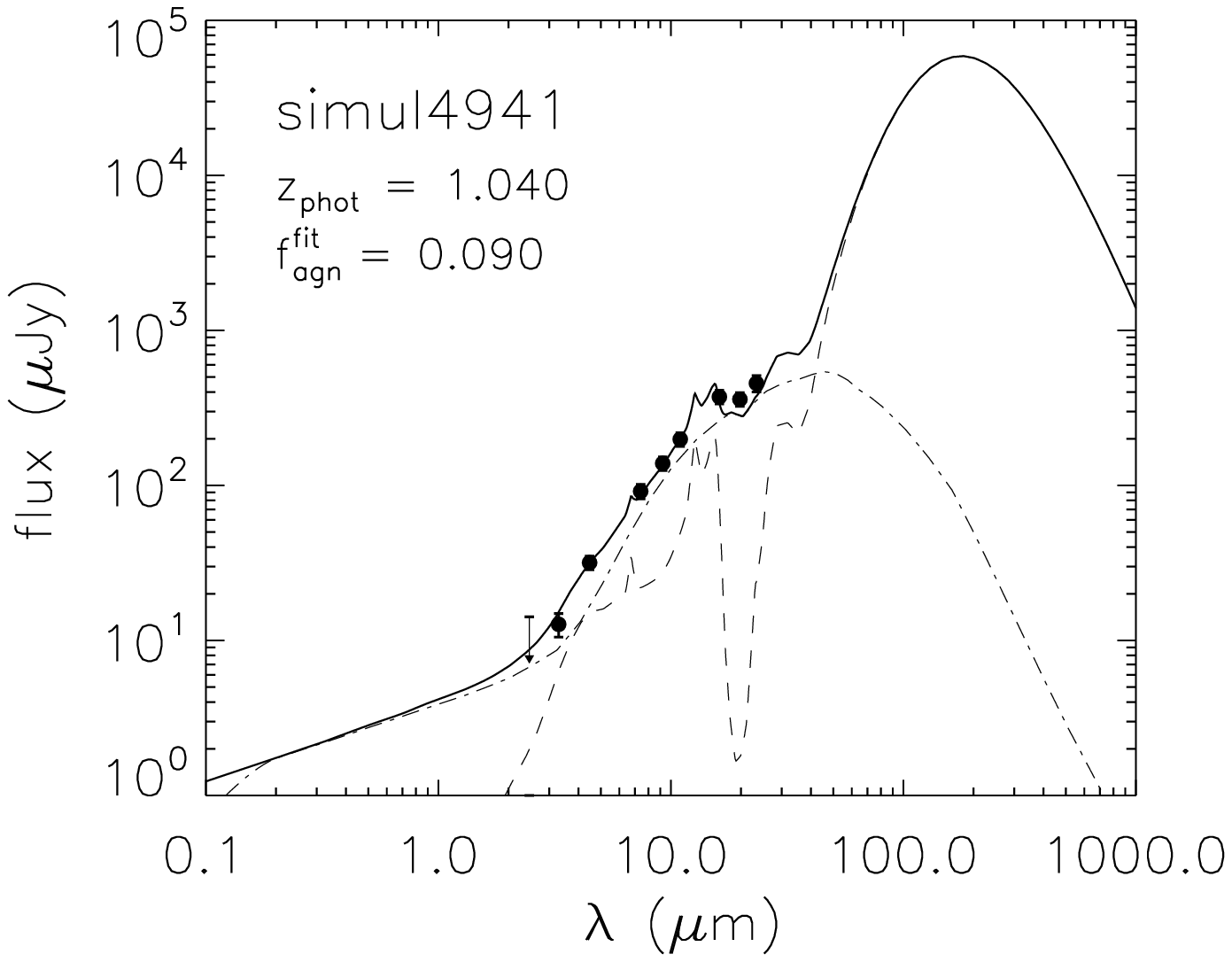}
    \vskip0.2truecm \caption{Example of a simulated SED with a
      dominant AGN component ($f_{\rm AGN}^{\rm model}>0.8$) for which
      the best-fit provides low AGN fraction ($f_{\rm AGN}^{\rm
        fit}<0.2$). The input SED model is shown in the upper panel:
      the solid line is the total SED and includes contributions from
      a starburst (dashed line) and an AGN (dot-dashed), as derived
      from the models of Takagi et al. (2003) and Efsathiou $\&$
      Rowan-Robinson (1995), respectively. The SED best-fit is shown
      in the lower panel. The meaning of the lines is the same as
      above.}\label{fig:sed_example}
  \end{center}
\end{figure}

The results of the simulations suggest that for $z_{\rm
  model}\lsim1.8$, i.e. when the photo-$z$ estimate is based on the
passage of the PAH features through the {\it AKARI} filters, the
achieved accuracy on redshift does not depend significantly on the
precise details of the underlying starburst model and the estimated
redshift can be considered reliable, irrespective of the AGN
fraction. At higher redshifts instead, featureless AGN-dominated
spectra make the recovery of redshift extremely challenging from
infrared data alone. The effect of the AGN infrared emission on the
photo-$z$ accuracy is made even clearer in the top panels of
Fig.~\ref{fig:dz_vs_fagn_simulcat_ALL} where the quantity $(z_{\rm
  phot}-z_{\rm model})/(1+z_{\rm model})$ is shown as a function of
the input value of the AGN fraction for the three sets of simulations
presented here. The dispersion on ($1+z$) increases from $\lsim10\%$
to $\gsim30\%$ as the AGN fraction passes from 0.1 to 0.5, while for
$f_{\rm AGN}^{\rm model}\gsim0.5$ the photometric estimate of the
redshift becomes totaly unreliable, at least according to the results
based on the first two sets of simulations. On the other hand, the
recovery of the AGN fraction itself becomes very challenging when
$f_{\rm AGN}^{\rm model}\gsim0.2$ as demonstrated in the lower panel
of the same figure. An example of such a case is shown in
Fig.~\ref{fig:sed_example}.

These results indicates that infrared data alone are a valuable tool
for redshift estimate but they are not very efficient in constraining
the AGN fraction, at least when the infrared SED is AGN dominated. In
such cases the support of data at other wavelengths (e.g. optical and
near-infrared, radio, X-ray) is needed to recover the real nature of
the infrared emission.

We note however that AGN power-law like spectra can be easily singling
out by fitting a line to the observed $\log S-\log\lambda$ diagram
($S$ being the measured flux and $\lambda$ the observing wavelength)
and analyzing the probability associated to the corresponding
$\chi^2$, $P_{\chi^2}^{\rm pl}$. In
Fig.~\ref{fig:zphot_vs_zmodel_simulcat_powerlawfit} we show the
photometric redshift accuracy achieved in the three sets of
simulations after dividing the spectra into those with
$P_{\chi^2}^{\rm pl}>1\%$ (i.e. power-law shape) and those with
$P_{\chi^2}^{\rm pl}<1\%$. In all cases, the criterion of selection of
the SED based on fitting a line to the observed $\log S-\log\lambda$
diagram acts exactly in the same way as the one relying on the input
value of the AGN fraction (see right-hand panels of
Figs~\ref{fig:zphot_vs_zmodel_simulcat_takagi} to
\ref{fig:zphot_vs_zmodel_simulcat_polletta}). Therefore we suggest the
following rules as a practical guideline for interpreting photometric
redshifts in {\it AKARI} NEP Deep Survey when using the methods
illustrated here: if the observed spectrum is poorly described by a
power-law than the photometric redshift can be considered accurate up
to $z_{\rm phot}\sim5$. When instead the observed spectrum is very
close to a power-law, it is high probable that the redshift of the
source has been significantly underestimated and it should be
rejected.

\setcounter{figure}{11}
\begin{figure*}
  \hspace{-0.0cm}\includegraphics[height=12.5cm,width=17.0cm]
  {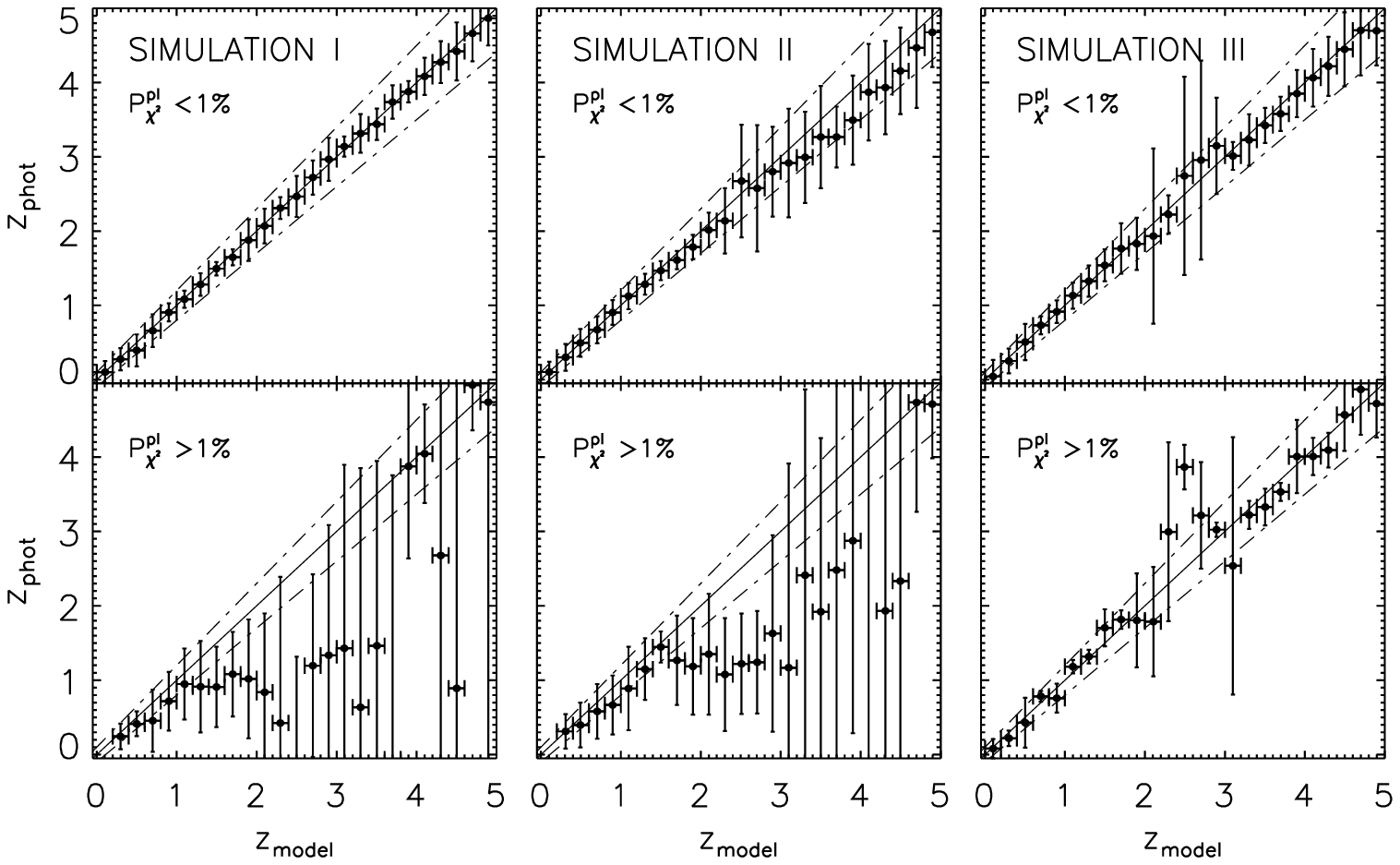}
  \vskip-1.7truecm\caption{Simulated photometric redshift accuracy
    with the {\it AKARI} NEP Deep Survey for power-law like spectra
    ($P_{\chi^2}^{\rm pl}>1\%$, bottom panels) and for spectra which
    are poorly fitted by a power-law ($P_{\chi^2}^{\rm pl}<1\%$, top
    panels). The panels refers from left to right to simulations based
    on the Takagi et al. (2003) model for the starbursts, the
    Efstathiou et al. (2000, 2003) models for the starburst/cirrus and
    the SWIRE sed templates (Polletta et al. 2008).}
  \label{fig:zphot_vs_zmodel_simulcat_powerlawfit}
\end{figure*}

\section{Conclusions}
\label{sec:conclusions}

We have tested the reliability of the photometric redshift estimates
based on infrared photometry alone on a sample of 59 galaxies with
spectroscopic redshifts $0.5\lsim z_{\rm spec}\lsim1.5$ drawn from the
GOODS-N field, combining together infrared data from literature
(i.e. {\it Spitzer} and {\it ISO}) and new {\it AKARI} 11 and
18$\,\mu$m photometric data. Our SED fitting procedure allows for the
contribution to the infrared emission from both starburst and AGN,
described according to the models of Takagi et al. (2003) and
Efstathiou $\&$ Rowan-Robinson (1995), respectively.  Three different
sets of simulations derived from both theoretical SED models and empirical
spectra have been used to tested the photo-z accuracy
achievable with the {\it AKARI} NEP Deep Survey. \\
The main conclusions are summarized as follows.
\begin{itemize}
\item The SEDs of 42 out of our 59 sources are well fitted
  (i.e. $P_{\chi^2}>1\%$) by our starburst+AGN SED reference
  templates. For all the sources in the sample bar 7 the achieved
  accuracy on $(1+z)$ is close to or better than $10\%$. Even in the
  case of a ``bad'' fit (i.e. $P_{\chi^2}<1\%$) the mean features in
  the spectrum are recognized and the redshift is still reasonably
  well recovered. Sources at $z\gsim2$ (three in total) display in
  general a power-law infrared spectrum, whose lack of any evident
  feature make the estimate of redshift quite difficult on the basis
  of infrared data alone.  In such cases data at longer wavelengths,
  near or beyond the peak associated to the dust emission or at
  decimetric radio wavelengths, are needed to remove the degeneracies
  in redshifts.
\item Simulations show that the infrared data alone produced by the
  {\it AKARI} NEP Deep Survey will provide photometric redshifts with
  a typical accuracy of $|z_{\rm phot}-z_{\rm spec}|/(1+z_{\rm
    spec})\sim10\%$ (1$\,\sigma$) at $z\lsim2$, in agreement with our
  findings for the spectroscopic sample. At higher redshifts the PAH features are shifted outside the wavelength range covered by {\it
    AKARI} and the 1.6$\,\mu$m stellar bump is then exploited as a redshift indicator; the accuracy achievable in this case on $(1+z)$ is
  $\sim10-15\%$, provided that the AGN contribution to the
  infrared emission is subdominant.
 
Although our reference SED templates allow for an AGN
  contribution to the infrared emission, the AGN fraction is poorly
  constrained when the SED is AGN dominated and therefore does not
  exhibit any evident features.  In these cases the support of
  photometric data at other wavelengths (e.g. X-ray, submillimeter or
  decimeter radio wavelengths) can help to better constrain the
  relative contributions of starburst and AGN to the observed SED.
\end{itemize}

\section*{ACKNOWLEDGMENTS}
We are grateful to the anonymous referee for helpful comments that
improved the paper. We wish to thank Myung Gyoon Lee, Micol Bolzonella, Mattia Vaccari,
Giulia Rodighiero and Simon Dye for usefull suggestions and stimulating discussions. This work
was supported by STFC grant PP/D002400/1.

\end{document}